\documentclass[ALICE,manyauthors]{cernphprep}

\usepackage[sort&compress,numbers,square,comma]{natbib}
\usepackage{doi}
\usepackage{hyperref}
\usepackage{lineno}

\usepackage[utf8x]{inputenc}
\usepackage{graphicx}
\usepackage[margin=1.0in]{geometry}
\usepackage{xcolor}
\usepackage{xspace}
\usepackage{multirow}
\usepackage{rotating} 
\usepackage{upgreek}

\newcommand {\pp}        {\ensuremath{\mathrm {p\kern-0.05em p}}}                                                                  \newcommand {\PbPb}      {\ensuremath{\mathrm{Pb\mbox{--}Pb}}}
\newcommand {\PbPbinbf}  {\ensuremath{\mathrm{\mbox{\bf{Pb}--\bf{Pb}}}}}
\newcommand {\CuCu}      {\ensuremath{\mathrm{Cu\mbox{--}Cu}}}
\newcommand {\AuAu}      {\ensuremath{\mathrm{Au\mbox{--}Au}}}
\newcommand {\nucnuc}    {\ensuremath{\mathrm{A\mbox{--}A}}}

\newcommand {\mumu}      {\ensuremath{\mu^+\mu^-}}
\newcommand {\mumuMB}    {\ensuremath{\mu \mu \mbox{-} \mathrm{MB}}}

\newcommand {\ccbar}     {\ensuremath{{\mathrm c}\bar{{\mathrm c}}}}
\newcommand {\qcbar}     {\ensuremath{\bar{{\mathrm c}}}}
\newcommand {\qc}        {\ensuremath{\mathrm c}}
\newcommand {\jpsi}      {\ensuremath{\mathrm{J}\kern-0.02em/\kern-0.05em\psi}}
\newcommand {\psip}      {\ensuremath{\psi\mathrm{(2S)}}}
\newcommand {\chic}      {\ensuremath{\chi_{c}}}
\newcommand {\psiratio}  {\ensuremath{\left[ \psip / \jpsi \right]}}

\newcommand {\pt}        {\ensuremath{p_{\mathrm{\textsc{t}}}}}

\newcommand {\meanpt}    {\ensuremath{\langle p_{\mathrm{\textsc{t}}} \rangle}}
\newcommand {\meanptsq}  {\ensuremath{\langle p_{\mathrm{\textsc{t}}}^{2} \rangle}}
\newcommand {\pttrig}    {\ensuremath{p_{\mathrm{\textsc{t}}}^{\mathrm{trig}}}}
\newcommand {\y}         {\ensuremath{y}}

\newcommand {\cent}[2]   {\ensuremath{\mathrm{#1\mbox{--}#2\%}}}    
\newcommand {\sqrtSnn}           {\ensuremath{\sqrt{s_{_{\mathrm{NN}}}}}}

\newcommand {\sqrtSnnE}[2][TeV]  {$\sqrtSnn = #2\,\mathrm{#1}$}

\newcommand {\sqrtS}             {\ensuremath{\sqrt{s}}\,}
\newcommand {\sqrtSE}[2][TeV]    {$\sqrtS = #2\,\mathrm{#1}$} 
\newcommand {\Raa}       {\ensuremath{R_\mathrm{AA}}}
\newcommand {\raa}       {\ensuremath{r_\mathrm{AA}}}
\newcommand {\Taa}       {\ensuremath{\langle T_\mathrm{AA} \rangle}}

\newcommand {\Npart}     {\ensuremath{\langle N_{\mathrm{part}} \rangle}}

\newcommand {\gev}       {\ensuremath{\,\mathrm{GeV}}}
\newcommand {\tev}       {\ensuremath{\,\mathrm{TeV}}}
\newcommand {\mevc}      {\ensuremath{\,\mathrm{MeV}\kern-0.05em/\kern-0.02em c}}
\newcommand {\mevcc}     {\ensuremath{\,\mathrm{MeV}\kern-0.05em/\kern-0.02em c^2}}
\newcommand {\gevc}      {\ensuremath{\,\mathrm{GeV}\kern-0.05em/\kern-0.02em c}}
\newcommand {\gevcc}     {\ensuremath{\,\mathrm{GeV}\kern-0.05em/\kern-0.02em c^2}}
\newcommand {\gevcsq}    {\ensuremath{\,\mathrm{GeV}^2\kern-0.05em/\kern-0.02em c^2}}
\newcommand {\gevfmcube} {\ensuremath{\,\mathrm{GeV}\kern-0.05em/\kern-0.02em \mathrm{fm}^3}}
\newcommand {\Ae}        {\ensuremath{A\kern-0.1em\times\kern-0.1em\varepsilon}}
\newcommand {\Fnorm}     {\ensuremath{F_{\mathrm{norm}}}}
\newcommand {\Tc}        {\ensuremath{T_{\mathrm{c}}}}

\begin{document}

\begin{titlepage}
%
%
%

\PHyear{2015}      
\PHnumber{157}     
\PHdate{26 June 2015}   

%
%
\title{Differential studies of inclusive J/$\psi$ and $\psi$(2S) production  \\ at forward rapidity in \PbPbinbf\  collisions  at  $\mathbf{\sqrt{{\textit s}_{_{NN}}}}$ = 2.76\,TeV }
\ShortTitle{\jpsi\ and  \psip\ production in \PbPb\ collisions at \sqrtSnnE{2.76}}   

\Collaboration{ALICE Collaboration\thanks{See Appendix~\ref{app:collab} for the list of collaboration members}}
\ShortAuthor{ALICE Collaboration} 

\begin{abstract}
The production of \jpsi\  and \psip\ was studied with the ALICE detector  in  \PbPb\ collisions at the LHC. 
The measurement was  performed at forward rapidity ($2.5 < \y < 4 $) down to zero transverse momentum (\pt)  in the dimuon decay channel. 
 Inclusive \jpsi\ yields  were extracted in different  centrality classes and the centrality dependence of the average \pt\ is presented.
The \jpsi\ suppression, quantified with the nuclear modification factor (\Raa), was measured  as  a function of  centrality, transverse momentum and rapidity. 
Comparisons with similar measurements at lower collision energy and theoretical models indicate that the \jpsi\ production is  the result of an interplay between color screening and recombination mechanisms in a deconfined partonic medium, or at its hadronization.
Results on the \psip\ suppression are provided via the ratio of \psip\ over \jpsi\ measured in \pp\ and \PbPb\ collisions.   
\end{abstract}
\end{titlepage}
\setcounter{page}{2}

\section{Introduction}

At high temperature, lattice quantum chromodynamics predicts the existence of a  deconfined phase of  quarks and gluons where chiral symmetry is restored~\cite{Karsch:2001cy}.    
This state of matter is known as the Quark Gluon Plasma (QGP)~\cite{Shuryak:1980tp}, and its  characterization is the goal of ultra-relativistic heavy-ion collision studies. 

Among the probes used to investigate the QGP and quantify its properties, quarkonium states are one of the most prominent and have generated a large amount of results both theoretical and experimental.
According to the  color-screening model~\cite{Matsui:1986dk,Karsch:1990wi}, measurement of the in-medium dissociation probability of the different quarkonium states could provide an estimate of the  system temperature.
Dissociation is expected to take place when the medium reaches or exceeds the critical temperature for the phase transition (\Tc), depending on the binding energy of the quarkonium state. 
In the charmonium (\ccbar) family,  the strongly bound \jpsi\ could survive significantly above \Tc\  (1.5--2 \Tc) whereas \chic\  and \psip\ melting  should occur near \Tc\ (1.1--1.2 \Tc)~\cite{Satz:2005hx,Satz:2013xja}.
The determination of the in-medium quarkonium properties remains a challenging theoretical task. Intense and persistent investigations  on the theory side are ongoing~\cite{Mocsy:2013syh}.
Shortly after quarkonium suppression was suggested as a strong evidence of QGP formation,  
the first ideas  of charmonium  enhancement via recombination of  \qc\ and  \qcbar\  appeared~\cite{matsui:1987,Svetitsky:1987gq}. 
Since then, the \jpsi\ enhancement mechanism has been more formalized and quantitative  predictions~\cite{BraunMunzinger:2000px,Thews:2000rj,Gorenstein:2000ck,Zhao:2011cv,Liu:2009nb} were made. 
Since the charm quark density produced in hadronic collisions increases with energy~\cite{Abelev:2012vra},  recombination mechanisms are predicted
to give rise to a sizable \jpsi\ production at LHC energies, which is likely to partially compensate or exceed the \jpsi\ suppression due to color-screening in the QGP. 
The observation of \jpsi\ enhancement in nucleus-nucleus collisions via recombination would constitute an evidence for deconfinement and hence for QGP formation. 
In addition, information for the characterization of the QGP can come from the study of the \psip\  meson, a state which is less strongly bound and not affected by higher mass charmonium decays with respect to the \jpsi. In the pure melting scenario, the relative production of \psip\ with respect to \jpsi\ is expected to be very small at the LHC~\cite{Karsch:1990wi}, which is not the case if recombination occurs~\cite{Andronic:2009sv,Chen:2013wmr}.

\jpsi\ suppression was observed experimentally in the most central heavy-nucleus collisions at the SPS~\cite{Alessandro:2004ap,Arnaldi:2007zz}, RHIC~\cite{Adare:2006ns,Adare:2011yf,Adamczyk:2012ey,Adamczyk:2013tvk}
and LHC~\cite{Aad:2010aa,Chatrchyan:2012np,Abelev:2012rv,Abelev:2013ila,Adam:2015rba}, ranging  from a center-of-mass energy per nucleon pair (\sqrtSnn) of about 17\gev\ to 2.76\tev .  
The \psip\ suppression was measured at the SPS~\cite{Alessandro:2006ju} and the LHC~\cite{Khachatryan:2014bva}.   
The interpretation of these results is not straightforward as they are also subject to other effects, not all related to the presence of a QGP.
A fraction of \jpsi\ originates from the strong and electromagnetic feed-down of the \chic\ and \psip. Therefore, a melting of these higher mass states before they can decay into the \jpsi\ will lead to an effective suppression of the \jpsi\ yield already for a medium that does not reach the \jpsi\ dissociation temperature.
Assuming charmonium states are initially produced with the same relative abundancies in  \PbPb\ collisions as  in \pp\ collisions, the  \chic\  and \psip\ melting would result in a reduction of the \jpsi\ yield of about 40\%~\cite{Faccioli:2008ir}. 
In addition, a non-prompt \jpsi\ and \psip\ component from the weak decay of beauty hadrons also contributes to the inclusive measurements. 
Since the beauty hadrons decay outside the QGP volume, this contribution is not sensitive to the color-screening of charmonia.
Finally,  a fraction of the  \jpsi\ and \psip\ suppression can be ascribed to cold nuclear matter (CNM) effects, also present in proton--nucleus collisions~\cite{Abelev:2013yxa,Abelev:2014zpa}.
The CNM effects group together the nuclear absorption of the charmonia,  the modification of the parton distribution functions (PDF) in the nuclei that  leads to a reduction (shadowing) or an enhancement (anti-shadowing) of the \ccbar\ pair production, and the energy loss of charm quarks in the nucleus.

 Numerous studies of \jpsi\ production in different collision systems at different energies  are now available.  Comparisons between experiments and to theoretical models can be made over wide kinematic ranges in rapidity and transverse momentum.
We already published the centrality, transverse momentum (\pt) and rapidity (\y) dependence of the \jpsi\ nuclear modification factor in \PbPb\ collisions at  \sqrtSnnE{2.76}~\cite{Abelev:2012rv,Abelev:2013ila}. In this paper, those results are extensively compared to available theoretical models and lower energy data. New results on the \jpsi\ \meanpt\ and \meanptsq\  versus centrality, and on the centrality (\pt) dependence of the \jpsi\ suppression for various \pt\ (centrality) ranges are also presented. Furthermore, we show results on \psip\ in \PbPb\ collisions, measured via the \psiratio\ ratio, as a function of centrality.

The remainder of this paper is organized as follows: the experimental apparatus and the data sample are  presented in sections~2 and 3. Section~4 gives the definition of the observables used in the analysis. The analysis procedure is then  described in sections~5 and 6.   Systematic uncertainties  are discussed in section~7.  The \jpsi\ results  are given in sections~8 and 9 while section~10 is dedicated to the \psip\ results.  Finally, section~11 presents our conclusions. 

\section{The ALICE detector}
The ALICE detector is described in detail in~\cite{Aamodt:2008zz}. 
At forward rapidity  ($2.5 < \y\ < 4$) the production of quarkonium  states is studied in the muon spectrometer via their \mumu\ decay channels down to zero \pt.  
In the ALICE reference frame, the positive $z$ direction is along the counter-clockwise beam direction.   
The muon spectrometer covers a negative pseudo-rapidity ($\eta$) range and consequently a negative \y\ range. 
However, due to the  symmetry of the \PbPb\ system, the results are presented with a positive \y\ notation, while keeping the negative sign for $\eta$.

The muon spectrometer consists of a ten-interaction-lengths (4.1 m) thick absorber, which filters the muons, in front of five tracking stations comprising two planes of cathode pad chambers each. The third station is located inside a dipole magnet with  a 3 Tm field integral. 
The tracking apparatus is completed by a Muon Trigger system (MTR) composed of four planes of resistive plate chambers  downstream from a seven-interaction-lengths (1.2 m) thick iron wall,  which absorbs secondary hadrons escaping from the front absorber and low-momentum  muons coming mainly from charged pion and kaon decays. 
A small-angle conical absorber protects the tracking and trigger chambers against secondary particles  produced by the interaction of large rapidity  primary  particles with the beam pipe. 
Finally, a rear absorber protects the trigger chambers from the background generated by beam-gas interactions downstream from the spectrometer.

In addition, the Silicon Pixel  Detector (SPD),  scintillator arrays (V0) and Zero Degree Calorimeters (ZDC) were used in this analysis. 
The SPD consists of two cylindrical layers  covering  $|\eta|  <  2.0$  and $|\eta| < 1.4$  for the inner and outer ones, respectively,  and  provides the coordinates of the primary vertex of the collision. 
The V0 counters, two arrays of 32 scintillator tiles each, are located on both sides of the nominal interaction point and cover  $2.8 < \eta < 5.1$ (V0-A) and $-3.7 < \eta < -1.7$ (V0-C). 
The ZDC are located on either side of the interaction point at $z \approx \pm $114 m
and detect spectator nucleons at zero degree with respect to the LHC beam axis. 
The V0 and ZDC detectors provide triggering information and event characterization.

\section{Data sample}
\label{sec:datasample}

The data sample analysed in this paper corresponds to \PbPb\ collisions at \sqrtSnnE{2.76}. 
These collisions were delivered by the LHC during 190 hours of stable beam operations spread over three weeks in November and December 2011.

The Level-0 (L0) minimum bias (MB) trigger was defined as the coincidence of  signals in V0-A  and V0-C detectors synchronized with the passage of two crossing lead bunches.  
This choice for the  MB condition  provides a high triggering efficiency ($> 95\%$) for hadronic interactions. 
To improve the trigger purity, a threshold on the energy deposited in the neutron ZDC rejects the contribution from electromagnetic dissociation processes at the Level-1 (L1) trigger level.  
Beam induced  background is further reduced at the offline level by timing cuts on the signals  from  the V0  and the ZDC. 

The charmonium analysis was carried out on a data sample, where in addition to the MB prerequisite, a trigger condition of at least one or two reconstructed muon candidate tracks in the MTR (trigger tracks) was required in each event.
The MTR logic allows for programming several L0 trigger decisions based on (i) the detection of one or two muon trigger tracks, (ii) the presence of opposite-sign or like-sign trigger track pairs and (iii) a lower threshold on the approximate transverse momentum (\pttrig) of the muon candidates. The latter selection is performed by applying 
a cut on the maximum deviation of the trigger track from an infinite momentum track originating at the nominal interaction point.
Due to the finite spatial resolution of the trigger chambers, this does not lead to a sharp cut in \pt, and the corresponding \pttrig\ threshold is defined in simulation as the \pt\ value for which the muon trigger probability is 50\%. 
The following muon-specific L0 triggers were used:
\begin{itemize}
\item Single muon low \pt\ ($\pttrig =  1\gevc$): MSL 
\item Opposite-sign dimuon low \pt\ ($\pttrig = 1\gevc$ on each muon): MUL
\item Like-sign dimuon low \pt\ ($\pttrig = 1\gevc$  on each muon): MLL
\end{itemize}  
A data sample of $17.3 \cdot 10^{6}$ \PbPb\ collisions was collected  with the  \mumuMB\ trigger, defined as the  coincidence of the MB and MUL conditions.
A scaling factor \Fnorm\ is computed for each run --- corresponding to a few hours maximum of continuous data taking --- in order  to normalize the number of \mumuMB\  triggers to the number of equivalent MB triggers. 
It is defined as the ratio, in a MB data sample, between the total number of events and  the number of events fulfilling the \mumuMB\ trigger condition.
It should be noted that the MB sample used in this calculation was recorded in parallel to the \mumuMB\ triggers.
The  \Fnorm\ value, $30.56 \pm 0.01 (\rm{stat.}) \pm 1.10 (\rm{syst.})$, is given by the 
average  over all runs  weighted by the  statistical uncertainties.  
A small fraction of opposite-sign dimuons were misidentified by the trigger algorithm as like-sign pairs. 
Although for the \jpsi\ it amounts to less than 1\% when considering the full sample, it increases up to 4\% at high \pt\ in peripheral collisions.
In this analysis, the missing fraction of opposite-sign dimuons was recovered by extracting the number of produced \jpsi\ and \psip\ from the union of the MUL and MLL data sample (MUL$\cup$MLL). This is different from the selection applied in the former paper~\cite{Abelev:2013ila}, where only the MUL data sample was used.
On the other hand, the efficiency of the trigger algorithm to determine the sign of the muon pairs does not impact the normalization of the collected data sample to the number of equivalent MB events described above. 
This was cross-checked by computing the normalization factor of the MUL$\cup$MLL data sample, resulting in less than 1\% difference in the extracted number of equivalent MB events.

The  integrated luminosity  corresponding  to the analysed data sample  is $\mathcal{L}_{\rm int} =   N_{\mumuMB} \cdot \Fnorm / \sigma_{\PbPb}  = 68.8 \pm 0.9 (\mathrm{stat.}) \; \pm 2.5 (\mathrm{syst.} \; \Fnorm ) \,  ^{+5.5}_{-4.5}  (\mathrm{syst.} \; \sigma_{\PbPb})  \, \upmu\mathrm{b}^{-1}$  using  an inelastic \PbPb\ cross section $\sigma_{\PbPb} = 7.7 \pm 0.1 \, ^{+0.6}_{-0.5}\,  \mathrm{b}$~\cite{ALICE:2012aa}.

\section{Definition of observables}
\label{sec:observables}

The centrality determination is based on a fit of the V0 signal amplitude distribution as described  in~\cite{Abelev:2013qoq}. 
Variables  characterizing  the collision  such as the average number of participant nucleons (\Npart) and the average nuclear overlap function (\Taa) for each  centrality class are given in Tab.~\ref{tab:taa}. 
In this  analysis a cut corresponding to the most central 90\%  of the inelastic nuclear cross section was applied as for these events the MB trigger is fully efficient and the residual contamination from electromagnetic processes is negligible.
\begin{table}[ht!]
\begin{center}
\begin{tabular}{ccc|ccc}
\hline
\hline
Centrality    & \Npart\              &  \Taa\ (mb$^{-1}$)  & Centrality    & \Npart\              &  \Taa\ (mb$^{-1}$)\\
\hline
\cent{0}{10}\     & 356.0$\pm$3.6            &  23.44$\pm$0.76     &\cent{0}{20}\     & 308.1$\pm$3.7            &  18.91$\pm$0.61\\
\cent{10}{20}\    & 260.1$\pm$3.8            &  14.39$\pm$0.45     &\cent{0}{40}\     & 232.6$\pm$3.4            &  12.88$\pm$0.42\\
\cent{20}{30}\    & 185.8$\pm$3.3            &  8.70$\pm$0.27      &\cent{0}{90}\     & 124.4$\pm$2.2            &  6.27$\pm$0.21\\
\cent{30}{40}\    & 128.5$\pm$2.9            &  5.00$\pm$0.18      &\cent{20}{40}\    & 157.2$\pm$3.1            &  6.85$\pm$0.23\\
\cent{40}{50}\    &  84.7$\pm$2.4            &  2.68$\pm$0.12      &\cent{20}{60}\    & 112.8$\pm$2.6            &  4.42$\pm$0.16\\
\cent{50}{60}\    &  52.4$\pm$1.6            &  1.317$\pm$0.071    &\cent{40}{60}\    & 68.6$\pm$2.0             &  1.996$\pm$0.097\\
\cent{60}{70}\    &  29.77$\pm$0.98          &  0.591$\pm$0.036    &\cent{40}{90}\    & 37.9$\pm$1.2             &  0.985$\pm$0.051\\
\cent{70}{80}\    &  15.27$\pm$0.55          &  0.243$\pm$0.016    &\cent{50}{90}\    & 26.23$\pm$0.84           &  0.563$\pm$0.033 \\
\cent{80}{90}\    &   7.49$\pm$0.22          &  0.0983$\pm$0.0076  &\cent{60}{90}\    & 17.51$\pm$0.59           &  0.311$\pm$0.020\\
\hline
\hline
\end{tabular}
\end{center}
\caption{\label{tab:taa} The average number of participant nucleons \Npart\ and the  average value of the nuclear overlap function \Taa\ with their associated systematic uncertainties  for the  centrality classes, expressed in percentages of the nuclear cross section~\cite{Abelev:2013qoq}, used in these analyses.}
\end{table}                        

For each centrality class $i$, the measured number of \jpsi\  ($N_{\jpsi}^{i}$) is normalized  to the equivalent number of minimum bias events ($N_{\mathrm{events}}^{i}$). 
To obtain $N_{\mathrm{events}}^{i}$, one simply multiplies the number of \mumuMB\ triggered events  by the \Fnorm\ factor scaled by the width of the centrality class.
Corrections  for the  branching ratio of the dimuon decay channel ($\mathrm{BR}_{\jpsi \rightarrow \mumu} = 5.93\pm0.06\%$) and for the acceptance times efficiency ($A\times \epsilon^{i}$) of the detector are then applied.
The \jpsi\ yield ($Y_{\jpsi}^{i}$) in a centrality class $i$  is  given by 
\begin{eqnarray}
\frac{\mathrm{d}^{2} Y^{i}_{\jpsi}}{\mathrm{d} \pt \mathrm{d} \y} =  \frac{\mathrm{d}^{2} N_{\jpsi}^{i} / \mathrm{d} \pt \mathrm{d} \y}{\mathrm{BR}_{\jpsi \rightarrow \mumu} \cdot N_{\rm{events}}^{i}\cdot A \times \epsilon^{i} (\pt,\y) }.
\label{eq:jpsiyield}
\end{eqnarray}
It is then combined with the inclusive \jpsi\ cross section measured in \pp\ collisions at the same energy to form the nuclear modification factor \Raa\ defined as
\begin{eqnarray}
\Raa^{i}(\pt,\y) =   \frac{ \mathrm{d}^{2} Y^{i}_{\jpsi} / \mathrm{d} \pt \mathrm{d} \y }{ \Taa^{i}  \cdot \mathrm{d}^{2} \sigma_{\jpsi}^{\pp} / \mathrm{d} \pt \mathrm{d} \y }.\label{eq:Raa}
\end{eqnarray}
The  \pt\ and \y\ integrated \jpsi\ cross section is
$\sigma_{\jpsi}^{\pp}(\pt < 8\,\gevc,\ 2.5<y<4) =   3.34 \pm 0.13 (\rm{stat.}) \linebreak[1] \pm 0.24  (\rm{syst.})  \pm0.12 \linebreak[1] (\rm{luminosity}) \linebreak[1] ^{+0.53}_{-1.07} (\rm{polarization}) \upmu\mathrm{b}$~\cite{Abelev:2012kr}.

The ALICE measurements reported here refer to inclusive \jpsi\ yields, i.e.~include prompt \jpsi\  (direct \jpsi\ and feed-down from \psip\ and \chic) and non-prompt  \jpsi\  (decay of B-mesons).
Contrary to prompt \jpsi,  \jpsi\  from  B-meson decays  do not directly probe the hot and dense medium created in the \PbPb\ collisions. 
Beauty hadron decays occur outside the QGP, so the non-prompt \jpsi\ \Raa\ is instead related to the energy loss of the beauty quarks in the medium. 
Although the prompt \jpsi\ \Raa\ cannot be directly measured with the ALICE muon spectrometer, it can be evaluated via 
\begin{eqnarray}  \centering
\Raa^{\mathrm{prompt}}   =    \frac{ \Raa  -  F_{\rm B} \cdot \Raa^{\mathrm{non\mbox{-}prompt}} }{ 1 -F_{\mathrm{B}} } 
\label{eq:promptraa}
\end{eqnarray}  
where  $F_{\mathrm{B}}$ is the fraction of non-prompt to inclusive \jpsi\  measured in \pp\ collisions, and $\Raa^{\mathrm{non\mbox{-}prompt}}$ is the nuclear modification factor of \jpsi\ from B-meson decays  in \PbPb\ collisions.
The  non-prompt and prompt \jpsi\ differential cross sections  as a function of \pt\ and \y\ were measured by LHCb in \pp\ collisions at  \sqrtSE{2.76 \; \mathrm{and} \; 7}~\cite{Aaij:2012asz,Aaij:2011jh} in a kinematic range overlapping with that of the ALICE muon spectrometer.  
Therefore, one can extract the \pt\ and \y\ dependence of  $F_{\rm B}$ from these data and use it in Eq.~\ref{eq:promptraa}.
A reliable determination of  
$\Raa^{\mathrm{non\mbox{-}prompt}}$ presents further complications.
We have thus chosen two extreme hypotheses,  independent of centrality, corresponding to the absence of medium effects on beauty hadrons  ($\Raa^{\mathrm{non\mbox{-}prompt}} = 1$)  or to a complete suppression  ($\Raa^{\mathrm{non\mbox{-}prompt}} = 0$), to evaluate conservative limits on $\Raa^{\mathrm{prompt}}$.

An excess of \jpsi\ compared to the yield expected assuming a smooth evolution of the \jpsi\ hadro-production and nuclear modification factor was observed in peripheral \PbPb\ collisions at very low \pt~\cite{Adam:2015gba}. This excess might originate from the  photo-production of \jpsi. This contribution is negligible in \pp\ collisions --- from LHCb measurement at \sqrtSE{7}~\cite{Aaij:2013jxj}, it is $\mathcal{O}(10^{-3})\%$ --- but it is enhanced by a factor $\mathcal{O}(10^{4})$ in \PbPb\ collisions, thus reaching the order of magnitude of the observed excess. The \jpsi\ coherent photo-production has been measured in ultra-peripheral \PbPb\ collisions~\cite{Abelev:2012ba}. It is centered at very low \pt, with $\sim 98\%$ of these \jpsi\ below 0.3 \gevc. An incoherent photo-production component is also observed in ultra-peripheral \PbPb\ collisions. About 30\%  of this contribution has a $\pt<0.3$ \gevc, the rest being mainly located in the \pt\ range 0.3--1 \gevc. The influence of possible photo-production mechanisms on the inclusive \jpsi\ \Raa\ presented in this paper has been evaluated by repeating the analysis placing a low \pt\ threshold on the \jpsi\ of 0.3 \gevc.
Assuming that the observed excess in peripheral \PbPb\ collisions is indeed due to the photo-production of \jpsi, and that the relative contribution of the incoherent over coherent components is the same as the one estimated in ultra-peripheral collisions, then this selection would remove about 75\% of the full photo-production contribution. Numerical values of \Raa\ with the low \pt\ threshold at 0.3 \gevc\ are given in the  Appendix~\ref{app:datatables}.  All the figures and values presented in the paper refer to the inclusive  \jpsi\ \Raa\ but  estimates of the difference between the inclusive and hadronic (without \jpsi\ photo-production) \jpsi\ \Raa, are indicated where appropriate.

The results for the \psip\ analysis are given in terms of the ratio of their production cross sections (or, equivalently, of their production yields), expressed as 
\begin{equation}
\psip / \jpsi = \frac{N^{\mathrm{i}}_{\psip}}{N^{\mathrm{i}}_{\jpsi}} \cdot \frac{(\Ae^{\mathrm{i}})_{\jpsi}}{(\Ae^{\mathrm{i}})_{\psip}}. 
\end{equation}
When forming such a ratio the normalization factor $N_{\mathrm{events}}^{\mathrm{i}}$  cancels out, as do most of the systematic uncertainties on \Ae\ corrections.
The double ratio $\psiratio_{\PbPb}/ \psiratio_{\pp}$ is used in order to directly compare the relative abundances of \psip\ and \jpsi\ in nucleus-nucleus and pp collisions.

\section{Signal extraction}
\label{sec:signal_extraction}

After a description of the muon selection procedure, we present here the two methods used to extract the \jpsi\ and \psip\ signals. 
The first one is directly based on fits of the \mumu\  invariant mass distribution while the second one makes use of the event mixing technique to subtract the combinatorial background.

\subsection{Muon reconstruction}
\label{sec:muonrec}
The muon reconstruction starts with the exclusion of parts of the detector that show problems during data taking such as high voltage trips, large electronic noise, pedestal determination issues. This selection is performed on a run-by-run basis to account for the time evolution of the apparatus. After pedestal subtraction, the adjacent well-functioning pads of both cathodes of each tracking chamber having collected a charge are grouped to form pre-clusters. 
These pre-clusters might be the superposition of several clusters of charges deposited by several particles crossing the detector close to each others.
The number of clusters of charges contributing to the pre-cluster and their approximate location are determined with a Maximum Likelihood - Expectation Maximization (MLEM) algorithm. It assumes that the charge distribution of each single cluster follows a two-dimensional integral of the Mathieson function~\cite{Mathieson:1988}. If the estimated number of clusters is larger than 3, the pre-cluster is split into several groups of 1, 2 or 3 clusters selected with the minimum total coupling to all the other clusters into the pre-cluster. Each group of clusters is then fitted using a sum of Mathieson functions, taking the MLEM results as a seed, to extract the precise location of where the particles crossed the detector.
The overall spatial resolution is around 200 (550) $\mathrm{\upmu m}$ in average in the (non-)bending direction.

The track reconstruction starts from the most downstream stations, where the multiplicity of secondary particles is smallest, by forming pairs of clusters in the two chambers of station 5(4), 
and deriving the parameters and associated errors of the resulting muon track candidates. The candidates are then extrapolated to the station 4(5), validated if at least one compatible cluster is found in the station and duplicate tracks are removed. 
 The procedure continues extrapolating the tracks to stations 3, 2 and 1, validating them by the inclusion of at least one cluster per station. The selection of compatible clusters is based on a 5$\sigma$ cut on a $\chi^2$ computed from the cluster and track local positions and errors. If several compatible clusters are found in the same chamber, the track is duplicated to consider all the possibilities and for each of them the track parameters and associated errors are recomputed using a Kalman filter. 
At each of the tracking steps, the track candidates, whose parameters indicate that they will exit the geometrical acceptance of the spectrometer in the next steps are removed. 
 At the end of the procedure, the quality of the track is improved by adding/removing clusters based on a 4$\sigma$ cut on the local $\chi^2$ and fake tracks sharing clusters with others in the three outermost stations with respect to the interaction point are removed. The choice of the $\chi^2$ cuts is a compromise between maximizing the tracking efficiency ($< 1$--2\% muon rejection) and minimizing the amount of fake tracks (negligible background for this analysis). Finally, muon track candidates are extrapolated to the interaction vertex measured by the SPD taking into account the energy loss and the multiple Coulomb scattering in the front absorber.

An accurate measurement of the tracking chamber alignment is essential to reconstruct the tracks with enough precision to identify resonances in the $\mumu$ invariant mass spectrum, especially the \psip\ for which the signal-to-background ratio is low. The absolute position of the chambers was first measured using photogrammetry before the data taking. Their relative position was then precisely determined using a modified version of the MILLEPEDE package~\cite{Blobel:2002ax}, combining several samples of tracks taken with and without magnetic field. The small displacement of the chambers when switching on the dipole was measured by the Geometry Monitoring System (an array of optical sensors fixed on the chambers) and taken into account. The resulting alignment precision is $\sim 100 \, \mathrm{\upmu m}$, leading to a reconstructed \jpsi\ invariant mass resolution of about 70 \mevcc, and about 10\% higher for the \psip. The resolution is dominated by the energy loss fluctuation and multiple Coulomb scattering of the muons in the front absorber. 
More details on the muon spectrometer performances are given in~\cite{Abelev:2014ffa}.

In this analysis, the muon track candidates also have to fulfill the following requirements.   
First, the reconstructed track must match a trigger track with  a \pttrig\ above the threshold set in the MTR for triggering the event (1\gevc\ in this analysis). The trigger track is reconstructed from the average position of the fired strips on the two trigger stations, as computed by the trigger algorithm. The matching is based on a 4$\sigma$ cut on a $\chi^2$ computed from the tracker and trigger track parameters and errors  including the angular dispersion due to the multiple Coulomb scattering of the muon in the iron wall.
Second, the transverse radius coordinate of the track at the end of the front absorber must be in the range  $17.6 <  R_{\mathrm{abs}} <  89.5$ cm. 
Muons exiting the absorber at small and large angles, thus outside the $R_{\mathrm{abs}}$ cut range, have crossed a different amount of material, either the beam shield or the envelope of the absorber, affecting the precision of the energy loss and multiple Coulomb scattering corrections.   
Third, in order to  remove muon candidates  close to the edge of the spectrometer acceptance, a cut on the track  pseudo-rapidity  $-4 <  \eta <  -2.5$ is applied.

\subsection{\jpsi\ signal}
\jpsi\ candidates are formed by combining pairs of opposite-sign tracks reconstructed within  the geometrical acceptance of the muon spectrometer. 
The aforementioned cuts at the single muon track level remove most of the hadrons escaping from or produced in the front absorber, as well as a large fraction of low \pt\ muons from pion and kaon decays, secondary muons produced in the front absorber, and fake tracks.
The \jpsi\ peak becomes visible in the \mumu\ invariant mass spectra even before any background subtraction.
At the dimuon level only cuts on rapidity  ($2.5 < \y < 4$) and  transverse momentum  ($ \pt  < 8$ \gevc) are applied.
The \jpsi\ raw yields  are extracted by using two different methods.

In the first method, the opposite-sign dimuon invariant mass distribution is fitted with a sum of two functions. 
The signal is described by a double-sided Crystal Ball function (CB2).  
This function is an extension of the Crystal Ball function~\cite{CBdef}, i.e.~a Gaussian with a power-law tail in the low mass range, with an additional independent power-law tail in the high mass range.
The CB2 function reproduces very well the \jpsi\ line shape in the Monte Carlo (MC) simulations.
The underlying continuum is described by a variable width Gaussian function.
This function  is  built on a Gaussian form, whose width  is dependent on the invariant mass of the dimuon.
It was checked that including or excluding a \psip\ contribution in the fitting procedure has a negligible effect on the extracted \jpsi\ yield within the present statistical and signal-extraction-related systematic uncertainties. Since the significance of the \psip\ signal in the centrality, \pt\ and \y\ intervals used for the \jpsi\ analysis is too small to extract its contribution, we do not include it in the fit for this analysis.
During the fitting procedure, the width of the \jpsi\ peak is kept as a free parameter as it cannot be reproduced perfectly in simulations, and its value varies from 65 to 76~\mevcc\ (larger than those from MC by about 5--10\%).  
The pole mass is also kept free although the differences observed between data and simulation are at the per mille level.
The tail parameters cannot be constrained by the fit. Therefore they are fixed to values obtained from an embedding simulation (described in section~\ref{sec:embedding}) and adjusted for each \pt\ and \y\ interval under study in order to account for the observed dependence on the \jpsi\ kinematics.
On the contrary, the \jpsi\ shape does not show a dependence on centrality, hence the CB2 tail parameters tuned on a centrality integrated MC sample are used in all the bins.
\begin{figure}[htbp]
\begin{center}
\includegraphics[width=1\linewidth,keepaspectratio]{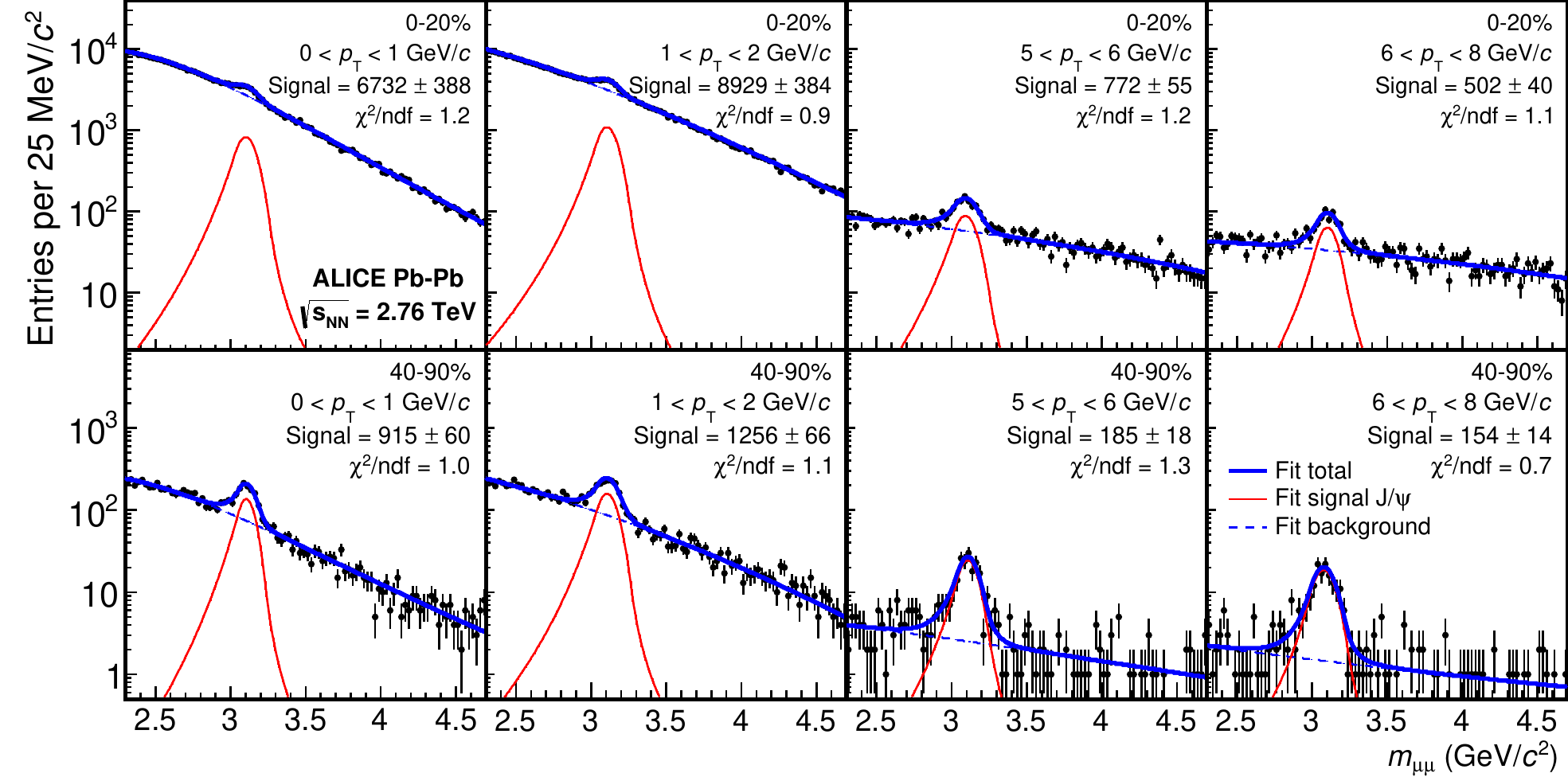}
\caption{Fit to the opposite-sign dimuon invariant mass distribution in the \cent{0}{20}\ (upper row) and \cent{40}{90}\  (lower row) centrality classes, for $2.5 < \y\  <  4$, in various \pt\ intervals.}
\label{fig:InvMassPerf}
\end{center}
\end{figure}
Figure~\ref{fig:InvMassPerf} presents  fits of the opposite-sign dimuon invariant mass ($ m_{\mu\mu} $) distributions for different \pt\ ranges in central (top row) and peripheral (bottom row) collisions. 
The signal-to-background ratio (S/B)  and the significance 
(S/$\sqrt{\mathrm{S+B}}$) 
of the signal are  evaluated within 3 standard deviations with respect to the \jpsi\ pole mass.
The  S/B varies from 0.2 to 6.5 when going from the most  central collisions to the most peripheral ones. 
Integrated over centrality and \y\ (\pt), the S/B ranges from 0.2 (0.2) to 1.2 (0.6) with increasing \pt\ (\y).   
In all the centrality, \pt\ or \y\ intervals considered in this analysis, the significance is always larger than 8.

\begin{figure}[htbp]
\begin{center}
\includegraphics[width=1\linewidth,keepaspectratio]{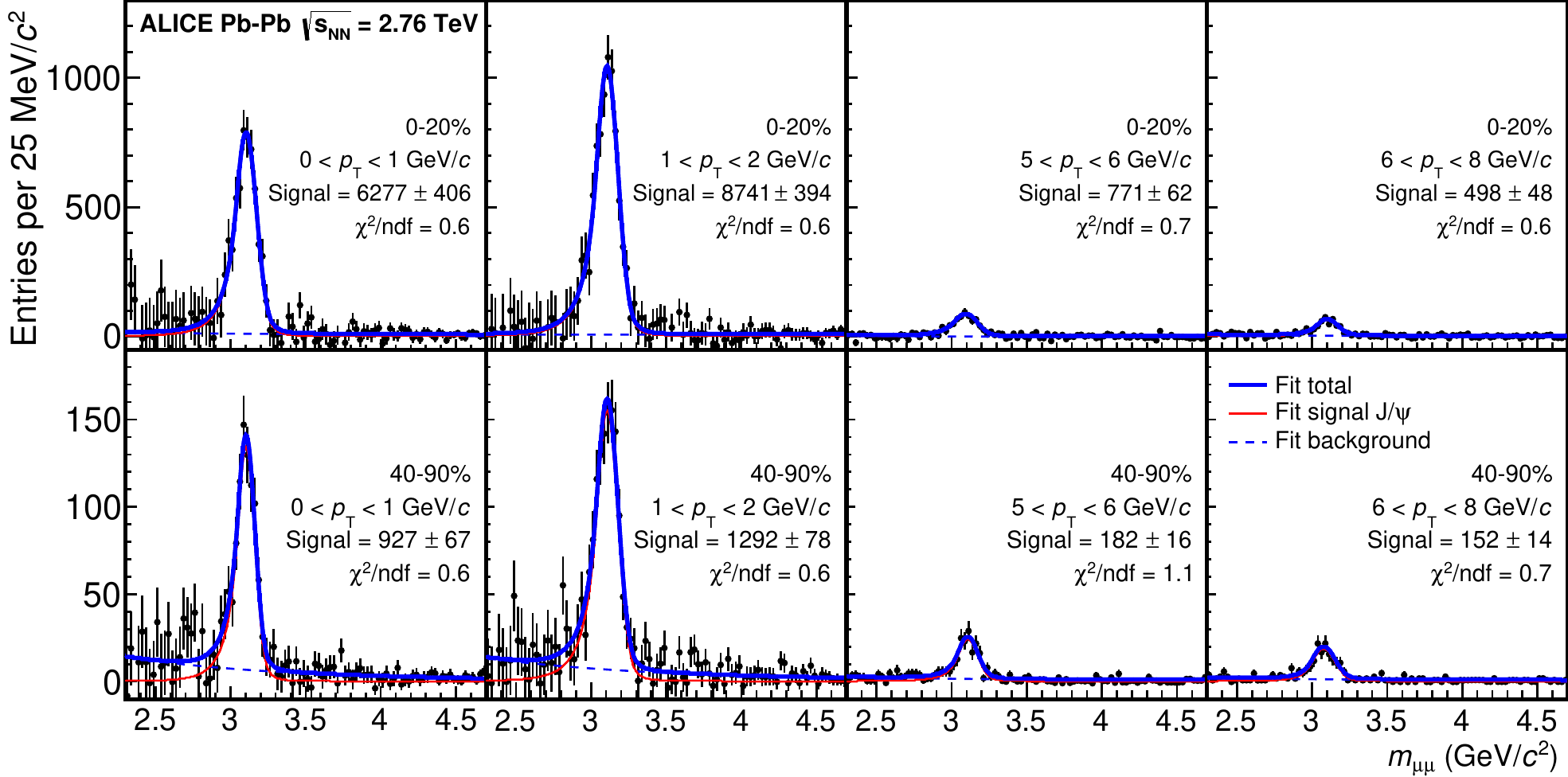}
\caption{Fit to the opposite-sign dimuon invariant mass distribution after background subtraction in the \cent{0}{20}\ (upper row) and \cent{40}{90}\ (lower row) centrality classes, for $2.5 < \y\  <  4$, in various \pt\ intervals.}
\label{fig:InvMassVsCent_wmixing}
\end{center}
\end{figure}

In the second method,  the combinatorial background is  subtracted using an event-mixing technique.  
The opposite-sign muon pairs from mixed-events  are formed by combining muons from single muon low \pt\ (MSL) triggered events. 
In order to limit the effect of efficiency fluctuations between runs and to take into account the dependence of muon multiplicity and kinematic distributions on the collision centrality, events in the same run and in the same centrality class are mixed together.  
The mixed-event spectra are normalized to the data using the combination of the measured like-sign pairs such as 
\begin{equation}
\label{eq:normmixing}
\int   \frac{\mathrm{d} N_{+-}^{\mathrm{mixed}}}{\mathrm{d} m_{\mu\mu}} \; {\rm d} m_{\mu\mu} = \int 2 R \sqrt{ \frac{\mathrm{d} N_{++}}{\mathrm{d} m_{\mu\mu}}  \frac{\mathrm{d} N_{--}}{\mathrm{d} m_{\mu\mu}} } \;  {\rm d} m_{\mu\mu} 
\end{equation}
where $N_{+-}$,  $N_{++}$ and $N_{--}$ are the number of opposite-sign, positive like-sign and negative like-sign muon pairs. 
The  $R$ factor in  Eq.~\ref{eq:normmixing} is defined by 
\begin{equation}
\label{eq:rfactor}
 R = \frac{ \frac{\mathrm{d} N_{+-}^{\mathrm{mixed}}}{\mathrm{d} m_{\mu\mu}}}{  2  \sqrt{ \frac{\mathrm{d} N_{++}^{\mathrm{mixed}}}{\mathrm{d} m_{\mu\mu}} \frac{\mathrm{d} N_{--}^{\mathrm{mixed}}}{\mathrm{d} m_{\mu\mu}} } } 
\end{equation}
and is introduced in order  to correct for differences in acceptance between like-sign and opposite-sign muon pairs. 
Above a dimuon invariant mass of 1.8 \gevcc, the $R$ factor is equal to unity with deviations smaller than 1\%. 
The accuracy of the mixed-event technique was assessed by comparing the distributions of like-sign muon pairs from mixed-events to the same-event ones, which  agree within 1\%  over the  mass, \pt\ and \y\ ranges under study.
This agreement justifies  the use of the normalization given by Eq.~\ref{eq:normmixing}, which implies that the correlated signal in the like-sign dimuon spectra is negligible with respect to the combinatorial background. 
The mass spectra of the opposite-sign  mixed-event  pairs  are then subtracted from the data. 
The resulting background-subtracted spectra are fitted following the same procedure as in the first method, except that the variable width Gaussian function is replaced by an exponential function to account for residual background.
Figure~\ref{fig:InvMassVsCent_wmixing} shows fits of the background-subtracted  opposite-sign dimuon invariant mass distributions for different \pt\ ranges in central (top row) and peripheral (bottom row) collisions. 

\subsection{\psip\ signal}

The invariant mass spectra used to extract the \psiratio\ ratio are obtained in the same way as described in the previous section, implementing the same cuts applied at the muon and dimuon levels. 
In order to improve the significance of the \psip\ signal, a wider centrality and \pt\ binning than the one used for the \jpsi\ analysis was adopted, and the analysis is  performed integrated over the full rapidity domain $2.5 < \y\  <  4$.
The fits to the invariant mass spectra are performed by modeling the \psip\ signal  with a CB2 function. 
Given the very low S/B ratio, the normalization is chosen as the only free parameter for \psip. 
The tails of the CB2 function describing the \psip\ are fixed to those extracted for the \jpsi.
The position of the \psip\ pole mass is  fixed to the one of the \jpsi, shifted by the corresponding $\Delta m = m_{\psip} - m_{\jpsi}$  value taken from the PDG~\cite{Agashe:2014kda}.  
The width of the \psip\ is  fixed to the one of the \jpsi\ scaled by the ratio $\sigma_{\psip} / \sigma_{\jpsi}$ estimated from MC simulations. 
\begin{figure}[htbp]
\begin{center}
\includegraphics[width=0.75\linewidth,keepaspectratio]{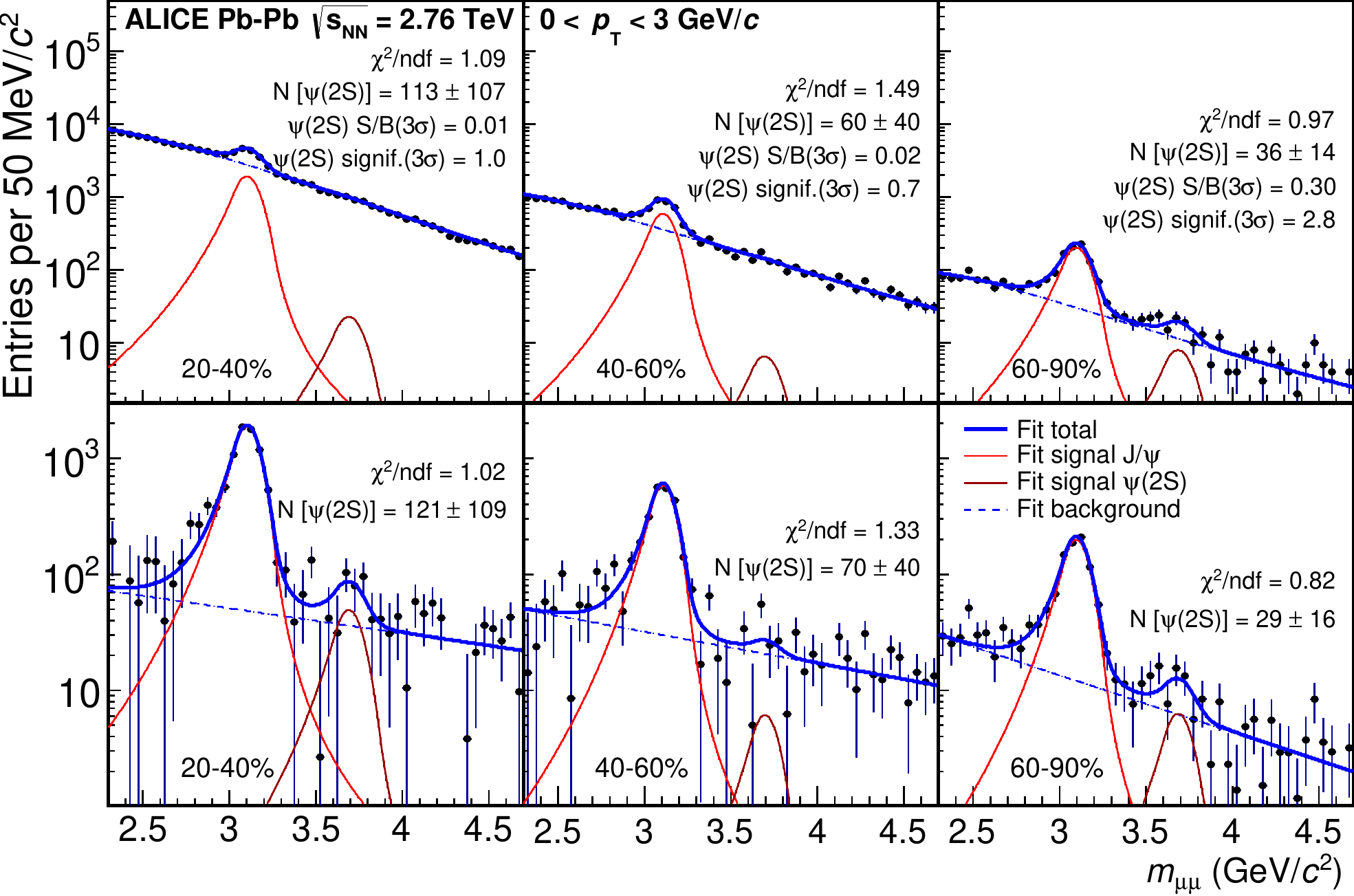}
\caption{Opposite-sign dimuon invariant mass distribution  for the  \cent{20}{40}, \cent{40}{60}\ and  \cent{60}{90}\  centrality classes,  for $2.5 < \y\  <  4$  and  $ \pt < 3\gevc$, before (top row) and after background subtraction (bottom row) via event mixing.   
In these intervals the \psip\ signal is extracted whereas  in all other centrality and \pt\ intervals, only the 95\% confidence level upper limits are provided.}
\label{fig:InvMassPerf2}
\end{center}
\end{figure}

Fits of the invariant mass spectra showing the \psip\ are visible in Fig.~\ref{fig:InvMassPerf2} for the  $ \pt < 3\gevc $ interval in  the  centrality classes \cent{20}{40}, \cent{40}{60}
 and \cent{60}{90}. 
For the other intervals in centrality and \pt\ the \psip\ signal could not be extracted, i.e.~the ratio \psiratio\  is consistent with zero.
In these cases, only the  95\% confidence level upper limit is computed.

\section{Acceptance and efficiency correction}
\label{sec:embedding}

In the \jpsi\  analysis, embedding  simulations are used  to compute the centrality, \pt\ and \y\ dependences of  the acceptance times efficiency factor (\Ae). 
The Monte Carlo embedding technique consists of adding the detector response from a simulated signal event (charmonium in this case) to a real data event, and then performing the reconstruction as for real events. 
This has the advantage of providing the most realistic background conditions, which is necessary for \PbPb\ collisions where high multiplicities are reached:
at $\eta = 3.25$, $\mathrm{d} N_{\mathrm{ch}} / \mathrm{d} \eta \approx$  1450 for the \cent{0}{5}\ most central events~\cite{Abbas:2013bpa}.
This leads to a large  detector occupancy, which can reach about 3\% in the most central collisions and  alter the track reconstruction efficiency.

Monte Carlo \jpsi\ were embedded in MB triggered events recorded in parallel to  the  opposite-sign dimuon triggered events. 
Only one \jpsi\ was simulated per event  at the position of the real event primary vertex reconstructed by the SPD.
The shapes of the input MC \pt\ and \y\ distributions were tuned  to match the measured distribution in \PbPb\ collisions (see discussion in section~\ref{sec:sysuncertMC}).   
The muons from the \jpsi\ decay were then transported through a simulation of the ALICE detector using GEANT3~\cite{GEANTref}. 
The detector simulated response was then merged with that of a real \PbPb\ event and the result was processed by the normal reconstruction chain.
Embedding simulations  were  performed  on a run-by-run basis to account for the time-dependent status of the tracking chambers. The residual misalignment of the detection elements, whose amplitude is evaluated by analyzing the residual distance between the clusters and the tracks in data, was also taken into account. For the trigger chambers, the efficiency maps measured in data were used in the simulations.

\begin{figure}[!b]
\begin{center}
\begin{tabular}{cc} 
\hspace{-1cm}
\includegraphics[width=0.54\linewidth,keepaspectratio]{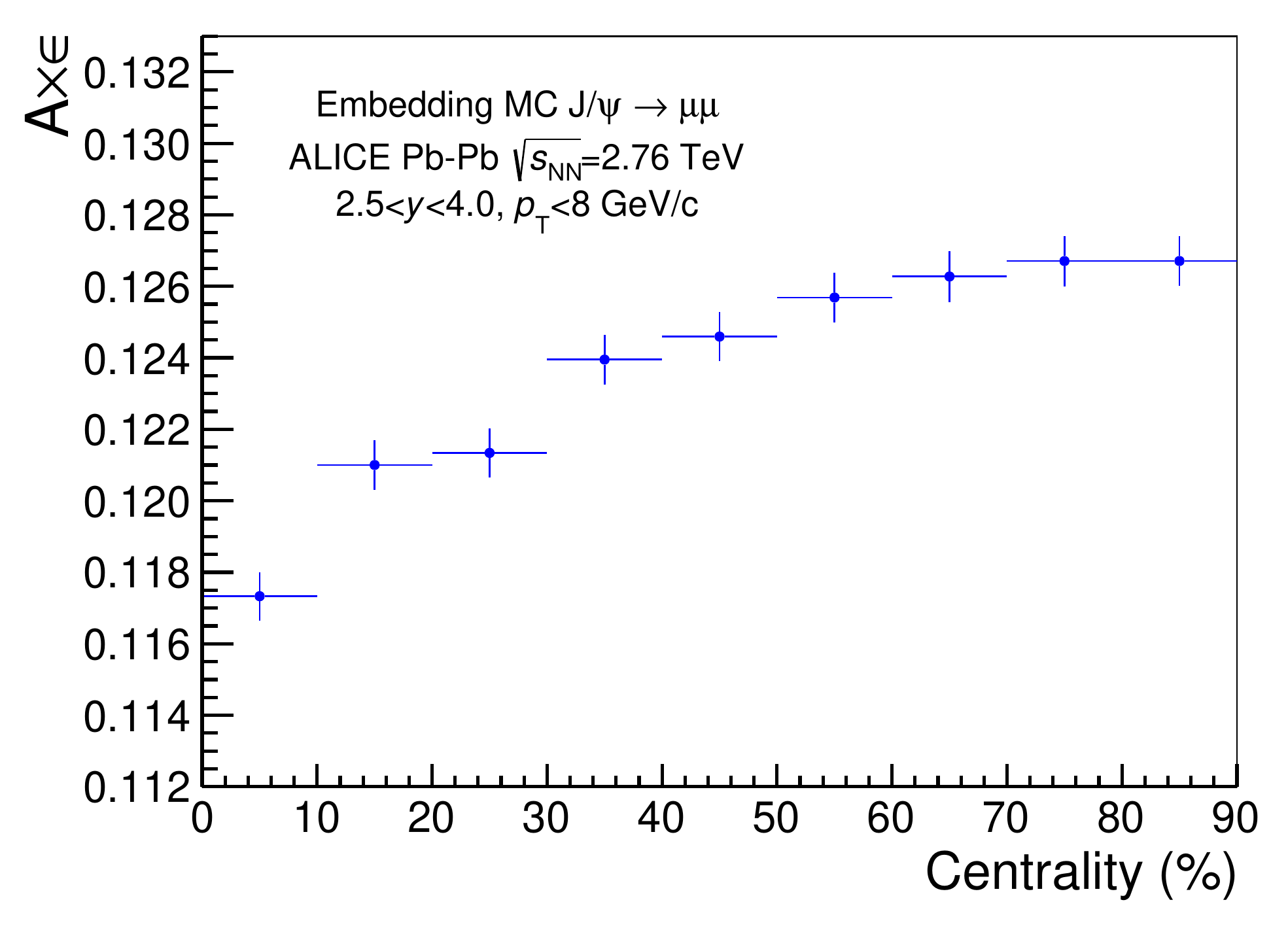}
\hspace{-0.1cm}
\includegraphics[width=0.54\linewidth,keepaspectratio]{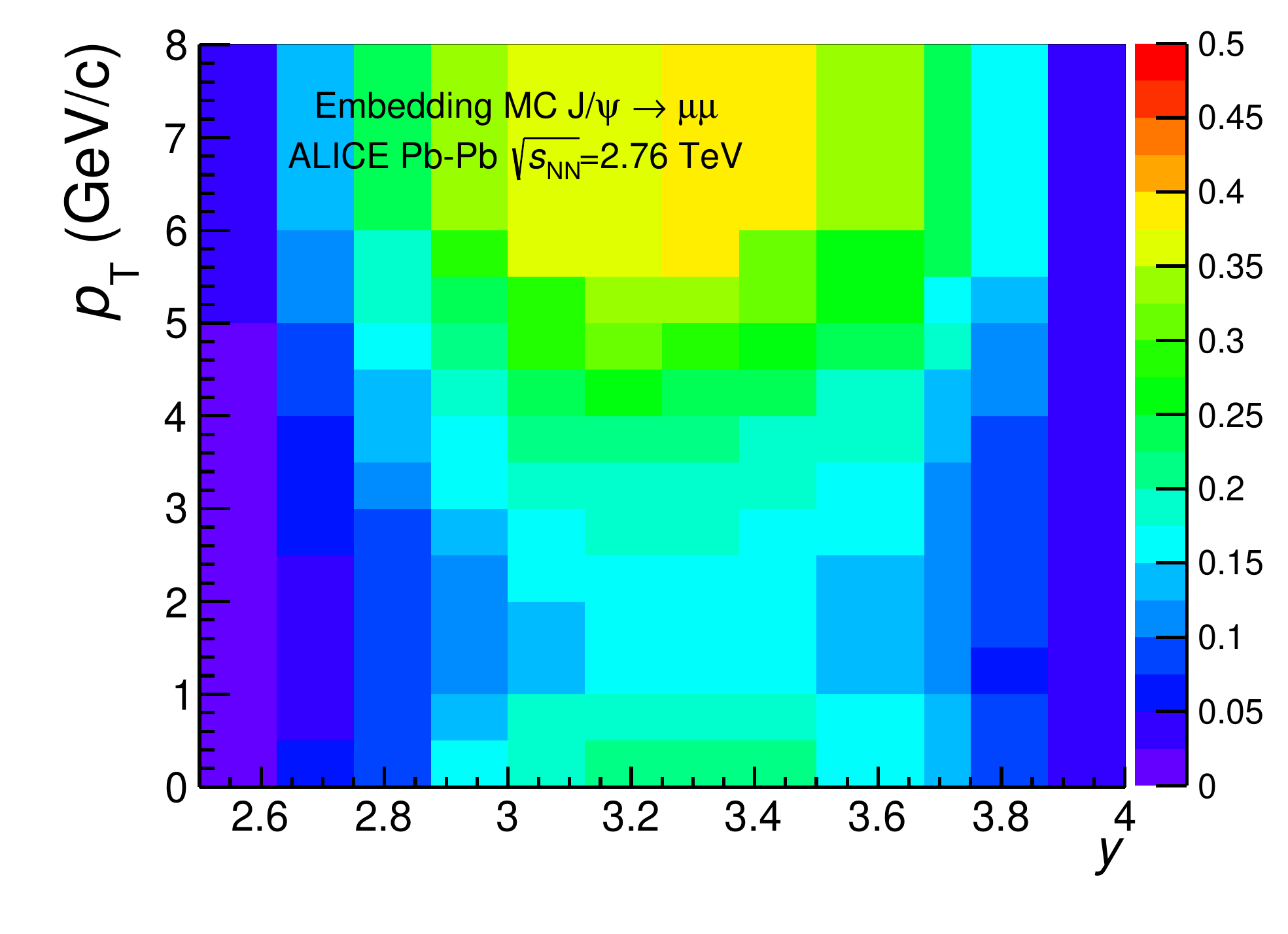}
\end{tabular}
\caption{The \jpsi\ acceptance times efficiency, shown as a function of centrality (left) and as a function of \pt\ and \y\, for the centrality class \cent{0}{90} (right). The vertical error bars in the left panel represent the statistical uncertainties.}
\label{fig:effEmbCent}
\end{center}
\end{figure}
The left panel of Fig.~\ref{fig:effEmbCent} shows the \jpsi\ \Ae\ as a function of collision centrality  in the rapidity domain $2.5 < y < 4$ and in the \pt\ range $  \pt < 8\gevc$.
We observe a relative decrease of 8\% of the \jpsi\ reconstruction efficiency from the \cent{80}{90}\ centrality class  to the \cent{0}{10}\ centrality class.
This decrease is mostly due to a drop of about 3\% of the single muon trigger efficiency in the most central collisions  whereas the decrease of the single muon tracking efficiency is only on the order of 1\%.
When considering specific \pt\ or \y\ intervals, a maximum relative variation of $\sim$ 30\% of the \Ae\ decrease with centrality is observed.
The right panel of Fig.~\ref{fig:effEmbCent} shows the  \pt\ versus \y\   dependence of \Ae. 
The  rapidity dependence  of  \Ae\   reflects the geometrical acceptance of the muon pairs with a maximum  centered at the middle of the rapidity interval and a decrease towards the edges of the acceptance.
The \pt\ dependence of \Ae\ is non-monotonic, with a minimum at $\pt \approx 1.8$ \gevc\ corresponding to  \jpsi\ kinematics for which one of the decay muons does not fall into  the muon spectrometer acceptance.

For the \psip\ resonance, the embedding technique was not used. 
Since, in this case, only the ratio \psiratio\ is extracted, the \Ae\ correction factors for both resonances were evaluated through pure signal MC simulations, assuming that the dependence of the efficiency as a function of the centrality is the same for \jpsi\ and \psip, and  therefore cancels out in the ratio. 
The effect of possible differences in the centrality dependence of \Ae\ was studied and included as a source of systematic uncertainty.

\section{Systematic uncertainties}
\label{sec:sysuncert}

In the following, each source of systematic uncertainty is detailed.
Most of them  affect the \jpsi\  and \psip\ results identically and vanish in the \psiratio\ ratio.
Systematic uncertainties specific to the \psip\ analysis are explicitly  mentioned.

\subsection{Signal extraction}

The systematic uncertainty on the signal extraction results from several fits of the invariant mass spectra, where signal line shape parameters, background description and fit range are varied as detailed below. In each centrality, \pt\ and \y\ intervals, the raw yield and the statistical uncertainty are given by the average of the results obtained from the different fits. The corresponding systematic uncertainty is defined as the RMS of these results. It was also checked that every individual result differs from the mean value by less than three RMS.

The \jpsi\ line shape is well described by the CB2 function, whose pole mass and width are constrained by the data while the tail parameters have to be fixed to values extracted from the embedding simulation. Alternatively, another set of tails was extracted from \pp\ data, where a large statistics and a better S/B were available. 
In this case, the \pt\ and \y\ dependence of the tail parameters could not be determined with sufficient precision, so the same values were used for all \pt\ and \y\ intervals.
In the event mixing approach,  
the  influence of different normalizations of the opposite-sign mixed-event spectrum  to the opposite-sign same-event spectrum was investigated.
We have tested a normalization performed on a run-by-run basis or after merging of all the runs, and a normalization  based on the integral of the invariant mass spectrum in the intermediate mass region ($1.5<m_{\mu\mu}<2.5$ \gevcc).
None of these tests showed deviations larger than 1\% in the number of extracted \jpsi, and thus were not included in the tests used to extract the systematic uncertainty on the signal extraction.
The fit range of the invariant mass spectra was also varied considering  a narrow  ($2.3 < m_{\mu\mu}< 4.7$\gevcc) and a wide ($2 < m_{\mu\mu} < 5$\gevcc) interval. Finally, all the combinations of signal line shape, background description (with or without using the event-mixing technique) and fit range are performed to account for possible correlations.

The same procedure as above was applied when the \psip\ signal was included in the fit function for the specific centrality and \pt\ intervals presented in this analysis. To account for the fact that the \psip\ width was fixed to the one of the \jpsi\ scaled by the ratio $\sigma_{\psip} / \sigma_{\jpsi}$ estimated from MC simulations, all the fits were repeated varying the scaling factor by $\pm$10\%. This variation accounts for the fluctuations observed in \pp\ data when fitting the invariant mass spectra leaving the width of the \psip\ free or fixing it as described above.

The systematic uncertainty on the signal extraction varies within  the 1--4\% range depending on the centrality class.  
Considering the  \pt\ intervals 0--2, 2--5 and 5--8\gevc\ used for the \Raa\  multi-differential studies, we obtain systematic uncertainties in the ranges  1--4\%, 1--4\% and 1--3\%, respectively.  
As a function of \pt, the systematic uncertainty on the signal extraction varies from 1\% to 4\%;  for the centrality  \cent{0}{20}, \cent{20}{40}\ and \cent{40}{90}, the values are in the ranges 1--5\%, 1--4\% and 1--2\%, respectively.  
As a function of \y, the systematic uncertainty on the signal extraction varies from 1\% to 4\%. 
Concerning the \psip\ analysis, in the intervals where the signal was extracted, the systematic uncertainty is 14\%, 45\% and 24\%  for centrality ranges \cent{60}{90}, \cent{40}{60}\  and \cent{20}{40}\  for $ \pt < 3\gevc$.

\subsection{Monte Carlo input parametrization}
\label{sec:sysuncertMC}

The estimation of \Ae\ factors depends on the charmonium \pt\ and \y\ shapes used as input distributions in the MC simulation. In order to evaluate the sensitivity of the results on this choice, several MC simulations were performed, each one including modified \pt\ and \y\ distributions.
For the \jpsi, the modification of the shapes was done in order to take into account the possible correlation between \pt\ and \y\ (as observed by LHCb in \pp\ collisions~\cite{Aaij:2011jh}) and the correlation between \pt\ (\y) and the centrality of the collision (as reported in this paper).
A systematic uncertainty of 3\% is found for \Ae\ integrated over \pt\ and \y\ and is taken as correlated as a function of the centrality. The \pt\ (\y) dependence of this uncertainty varies in the range 0--1\% (3--8\%). The larger effect seen in the \y\ dependence occurs at the low and high limits, where the acceptance falls steeply.

The same procedure was followed for the \psip, assuming that the correlations between \pt\ and \y\ and with the centrality are of the same magnitude as those observed for the \jpsi. 
A systematic uncertainty of 2\% is evaluated for the \psiratio\ ratio in the $  \pt < 3\gevc$  interval.

\subsection{Centrality dependence of the \psiratio\ \Ae }

The embedding technique was not used for the \psip\ MC simulations as we have assumed the same \Ae\ dependence as a function of the centrality for the \psip\ and the \jpsi. 
In order to evaluate the systematic uncertainty introduced by this assumption, a  conservative $\pm$30\% variation of the \Ae\ loss as a function of centrality was applied to the \psip. 
This corresponds to the maximum variation of the  \Ae\ loss between peripheral and central collisions observed for the \jpsi\ in different \pt\ and \y\ intervals.
The effect on the $\left( \Ae \right)_{\jpsi} / \left( \Ae \right)_{\psip}$ ratio is 1\% or lower in all the centrality classes considered.
Since this effect is much smaller than the systematic uncertainty on the signal extraction, it is neglected.

\subsection{Tracking efficiency}

The tracking algorithm, as described in section~\ref{sec:muonrec}, does not require all the chambers to have fired to reconstruct a track. This redundancy of the tracking chambers can be used to measure their individual efficiencies from data, and since such efficiencies are independent from each other, we can combine them to assess the overall tracking efficiency. 
This evaluation of the tracking efficiency is not precise enough to be used to directly correct the data, because only the mean efficiency per chamber can be computed with the statistics available in each run. However, by comparing the result obtained from data with the same measurement performed in simulations, we can control the accuracy of these simulations and assess the corresponding systematic uncertainty on the \Ae\ corrections.

A 9\% relative systematic uncertainty is obtained for  the \jpsi\  by comparing the measured tracking efficiency in simulations and in peripheral \PbPb\ collisions. This uncertainty is constant and fully correlated as a function of centrality. 
From low to high \pt\ (\y), the systematic uncertainty varies from 9\% to 7\% (7\% to 6\% with a maximum of 12\% at $\y \simeq 3.25$). On top of that, a small difference was observed in the centrality dependence of this measurement between data and embedding simulations. This results in an additional  1\% systematic uncertainty in the \cent{0}{10}\ centrality class and 0.5\% in \cent{10}{20}.

Another systematic uncertainty can arise from correlated dead areas located in front of each other in the same station, which cannot be detected with the  method detailed above. A dedicated study has shown that this effect introduces a 2\% systematic uncertainty, fully correlated as a function of centrality and predominantly uncorrelated as a function of \pt\ and \y.

In the \psiratio\ ratio the systematic uncertainty on the tracking efficiency largely cancels out  because the \psip\ and \jpsi\ decay muons have similar \pt\ and \y\ distributions and, therefore, cross about the same regions of the detector. 
Since the possible remaining systematic uncertainty is much smaller than that on the signal extraction, it is neglected in this analysis.

\subsection{Trigger efficiency}
The systematic uncertainty on the \jpsi\ \Ae\ corrections related to the trigger efficiency has two origins: the intrinsic efficiency of the trigger chambers and the response of the trigger algorithm. 
The first part was determined from the uncertainties on the trigger chamber efficiencies measured from data and applied to simulations. 
Propagating these efficiencies in \jpsi\ simulations results in a 2\% systematic uncertainty on the \Ae\ corrections, fully correlated as a function of centrality and mainly uncorrelated as a function of \pt\ and \y.
The effect of the systematic uncertainty on the shape of the trigger response as a function of the muon \pt\ was determined by weighting MC \jpsi\ decay muons with different trigger response functions obtained in data and simulations. These functions were defined as the fraction, versus \pt, of the single muons passing a 0.5\gevc\ \pttrig\ threshold that also satisfy the 1\gevc\ \pttrig\ threshold used in this analysis. The resulting systematic uncertainty on the \jpsi\ \Ae\ correction integrated over \pt\ and \y\ is 1\%. As a function of \pt, it amounts to 3\% for $ \pt < 1\gevc$ and 1\% elsewhere. As a function of \y, a 1\% uncorrelated systematic uncertainty was  obtained. These uncertainties are fully correlated as a function of centrality.

The systematic uncertainty on the modification of the trigger response  as a function of centrality, i.e.~for increasing multiplicity, was assessed by changing the detector response (space size of the deposited charge) to the passage of particles in embedding simulations. The corresponding uncertainties on the \jpsi\ \Ae\ corrections are 1\% in the \cent{0}{10}\ and  \cent{10}{20}\ centrality classes, and 0.5\% in \cent{20}{30}\ and  \cent{30}{40}.

As for the case of tracking efficiency, this source of  systematic uncertainty largely cancels out in the \psiratio\ ratio and is neglected.

\subsection{Matching efficiency}

The systematic uncertainty on the matching efficiency between the tracking and the trigger tracks is 1\%. 
It is given by the differences observed between data and simulations when applying different $\chi^{2}$ cuts on the matching between the track reconstructed in the tracking chambers and the one reconstructed in the trigger chambers.  
This uncertainty is fully correlated as a function of the centrality and largely uncorrelated as a function of \pt\ and \y.

Also in this case, the effect on the \psiratio\ ratio is negligible.

\subsection{\pp\ reference}

The statistical and systematic uncertainties on the measurement of the  \jpsi\ differential cross section  in  \pp\  collisions at \sqrtSE{2.76} are available in~\cite{Abelev:2012kr}. 
The statistical uncertainty is combined with that of the \PbPb\ measurement when calculating the \Raa\ as a function of \pt\ and \y, but is considered as a fully correlated systematic uncertainty as a function of the centrality.
The correlated and uncorrelated part of the systematic uncertainty on the \pp\ reference as a function of \pt\ and \y\ are both fully correlated as a function of the centrality.

The \psip\ statistics in the \sqrtSE{2.76} \pp\ data sample are too low to be used for the normalization of the $\psiratio_{\PbPb}$ ratio.
For this reason, \pp\ results obtained at higher energy (\sqrtSE{7})~\cite{Abelev:2014qha} were used, thus introducing an additional source of systematic uncertainty.
An interpolation procedure, as the one described in~\cite{Abelev:2014zpa}, was applied   in order to extract the $\psiratio_{\pp}$ ratio at \sqrtSE{2.76}. 
The discrepancy between the result of this interpolation in the kinematic range  $ \pt < 3\gevc$   $ 2.5 < \y < 4$  and the value obtained at \sqrtSE{7} is 10\%: this relative difference is included in the systematic uncertainty on the \pp\ reference.

\subsection{Normalization}
The systematic uncertainty on the normalization is the one attached to the scaling factor \Fnorm\ and  amounts to 4\%. 
This value corresponds to one standard deviation of the distribution of the \Fnorm\ computed for each run  used in the analysis. 
This systematic uncertainty is fully correlated as a function of the centrality, \pt\ and \y.

\subsection{Others}
Systematic uncertainties on the nuclear overlap function \Taa\ are available in Tab.~\ref{tab:taa}. 
Another systematic uncertainty on the definition of the centrality classes arises from the V0 amplitude cut,  which corresponds to 90\% of the hadronic cross section~\cite{Abelev:2013qoq}. 
A maximum uncertainty of 5\% is obtained in the centrality class (\cent{80}{90}) vanishing with increasing centrality or in wider centrality classes.

Systematic uncertainties due to the unknown polarization of the \jpsi\ are not propagated and  we assume that \jpsi\ production is unpolarized both in \pp\ and in \PbPb\ collisions. 
In \pp\ collisions at \sqrtSE{7}, \jpsi\ polarization measurements at mid-rapidity ($\pt > 10 \gevc$) and forward-rapidity ($\pt > 2 \gevc$)  are  compatible with zero~\cite{Chatrchyan:2013cla,Abelev:2011md,Aaij:2013nlm}.
In \PbPb\ collisions,  \jpsi\ mesons produced from initial parton--parton hard scattering are expected to have the same polarization as in \pp\ collisions and those produced from charm quarks recombination in the medium are expected to be unpolarized.

\subsection{Summary}
The systematic uncertainties related to the \jpsi\ analysis are summarized in Tab.~\ref{tab:syst}. 
Concerning the \psip\ analysis, most of the systematic uncertainties cancel out in the \psiratio\ ratio and the main contributors are the signal extraction (14--45\%) and the pp reference (10\%).

\begin{table}[htp]
\centering
\small
\begin{tabular}{c l | c c | c c | c}
\hline
\hline
\multicolumn{2}{c|}{\multirow{2}{*}{Sources}}		& \multicolumn{2}{c|}{Centrality} 							& \multicolumn{2}{c|}{\pt}   								&  {\y}~\cite{Abelev:2013ila} \\
        			&						& $ \pt<8  \gevc$~\cite{Abelev:2013ila} 			& \pt\ bins					& \cent{0}{90}~\cite{Abelev:2013ila}   				& centrality bins				&   \\
\hline 
\multicolumn{2}{l|}{Signal extraction}	 & 1--3    & 1--4   & 1--4 & 1--5 & 1--4 \\
\multicolumn{2}{l|}{MC parametrization} & 3$^{\mathrm{*}}$ & 1--3$^{\mathrm{*}}$ & 0--1 & 0--1					& 3--8 \\
\multicolumn{2}{l|}{Tracking efficiency}      		& 0--1 and 11$^{\mathrm{*}}$ 	& 0--1 and 9--11$^{\mathrm{*}}$ 	& 9--11 and 1$^{\mathrm{*}}$   	& 9--11 and 0--1$^{\mathrm{*}}$	& 8--14 and 1$^{\mathrm{*}}$ \\
\multicolumn{2}{l|}{Trigger efficiency}        		& 0--1 and 2$^{\mathrm{*}}$		& 0--1 and 2$^{\mathrm{*}}$		& 2--4 and 1$^{\mathrm{*}}$    	& 2--4 and 0--1$^{\mathrm{*}}$	& 2 and 1$^{\mathrm{*}}$ \\
\multicolumn{2}{l|}{Matching efficiency} & 1$^{\mathrm{*}}$ & 1$^{\mathrm{*}}$ &  1 & 1  & 1  	\\
\multirow{2}{*}{$\sigma_{\jpsi}^{\pp}$} 	& stat.    	& 4$^{\mathrm{*}}$ & 5--12$^{\mathrm{*}}$ &  6--21  & 6--21 & 7--11  \\
							& syst.    	& 8$^{\mathrm{*}}$  & 7$^{\mathrm{*}}$ & 5--6 and 6$^{\mathrm{*}}$ & 5--6 and 6$^{\mathrm{*}}$ & 5--6 and 6$^{\mathrm{*}}$ \\
\multicolumn{2}{l|}{\Fnorm}		& 4$^{\mathrm{*}}$ & 4$^{\mathrm{*}}$ & 4$^{\mathrm{*}}$ & 4$^{\mathrm{*}}$ & 4$^{\mathrm{*}}$  \\
\multicolumn{2}{l|}{\Taa} & 3--8         				& 3--6      				& 3$^{\mathrm{*}}$          		& 3--5$^{\mathrm{*}}$		& 3$^{\mathrm{*}}$   \\	
\multicolumn{2}{l|}{Centrality limits}		   	& 0--5         				& 0--3					& 0    					& 0--2$^{\mathrm{*}}$		& 0   	\\ 
\multicolumn{2}{l|}{B.R.}         				& n/a  & 	n/a			  				& n/a       					& 1$^{\mathrm{*}}$							&  n/a \\
\hline
\hline
\end{tabular}
\normalsize
\caption{\label{tab:syst}Summary of the systematic uncertainties (in \%) entering the \jpsi\ yield and/or \Raa\ calculation as a function of centrality, \pt\ and \y. Numbers with an asterisk correspond to the systematic uncertainties fully correlated as a function of the given variable.}
\end{table}

\section{Inclusive \jpsi\ mean transverse momentum}

\begin{figure}[htbp]
\begin{center}
\begin{tabular}{c} 
	\includegraphics[width=0.54\linewidth,keepaspectratio]{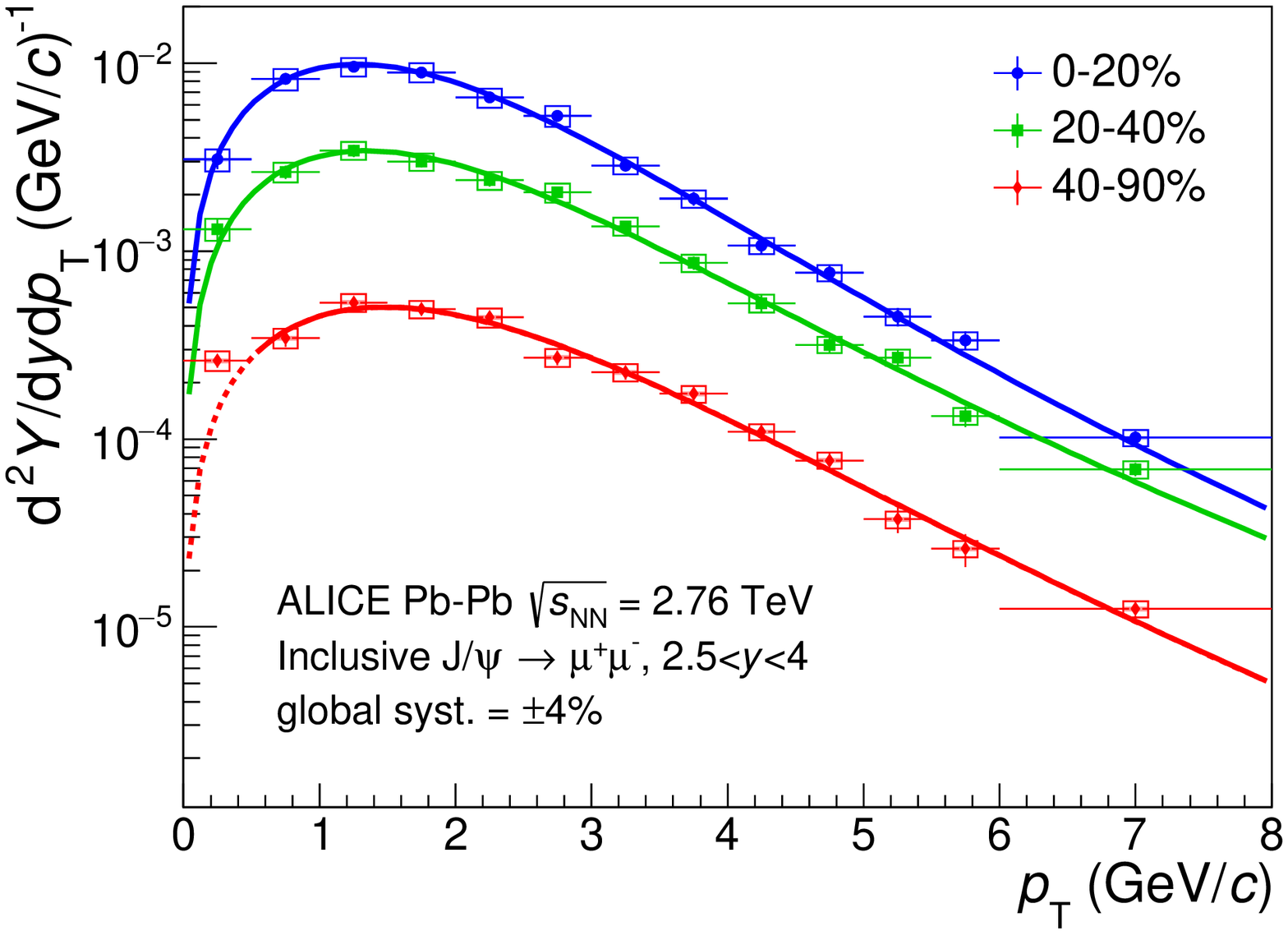}
\end{tabular}
\caption{Differential yields of inclusive \jpsi\ in \PbPb\ collisions at \sqrtSnnE{2.76}  as a function of \pt\ for three centrality classes. Solid lines correspond to the results from the fit described in the text.}  
\label{fig:dndptdy}
\end{center}
\end{figure}

The \pt\ dependence of the \jpsi\ yields per MB collision, defined by Eq.~\ref{eq:jpsiyield}, was studied for three centrality classes (\cent{0}{20}, \cent{20}{40}\ and \cent{40}{90}) and is displayed in Fig.~\ref{fig:dndptdy}.
The statistical uncertainties appear as vertical lines.
The systematic uncertainties uncorrelated as a function of \pt\ are shown as open boxes, while the ones fully correlated as a function of \pt\ but uncorrelated as a function of centrality are shown as shaded areas (mostly hidden by the points).
The global systematic uncertainty, fully correlated as a function of centrality and \pt, is quoted directly in the figure.
Numerical values for the \jpsi\  yields can be found in  Appendix~\ref{app:datatables}.
The inclusive \jpsi\  mean transverse momentum was computed by fitting the \pt\ distribution of inclusive \jpsi\ yields with the function 
\begin{eqnarray}
 f(\pt) = C \times   \frac{\pt}{\left( 1 + (\pt/p_{\mathrm{0}})^{2} \right)^{\mathrm{n}}},
\label{eq:meanpt}
\end{eqnarray}
where $C$, $p_{0}$ and $n$ are free parameters. 
This function is commonly used to reproduce the \jpsi\ \pt\ distribution in hadronic collisions, see for instance~\cite{Yoh:1978id,Adare:2006kf,Bossu:2011qe}. 
Fit results for the three centrality classes are displayed as full lines in the figure.
An excess over this function is revealed in the lowest \pt\ interval (corresponding to $ 0 < \pt < 500\mevc$) for peripheral \PbPb\ collisions.
It could be caused by a residual contribution from \jpsi\ coherent photo-production, which was measured in ultra-peripheral collisions~\cite{Abelev:2012ba}.
A quantitative measurement of this contribution in hadronic collisions is reported in~\cite{Adam:2015gba}.
Thus, in the most peripheral centrality class  (\cent{40}{90})  the fit was performed for $\pt > 500\mevc$ and extrapolated down to zero (dotted line).
In the \cent{0}{20}\ and  \cent{20}{40} centrality classes, no  \jpsi\ excess was observed and fits were performed down to zero \pt.
As a cross-check, the same procedure as for the peripheral centrality class was tested and the obtained results are fully compatible within uncertainties.

Values of the mean transverse momentum (\meanpt) and mean squared transverse momentum (\meanptsq) obtained  from the fits are given in Tab.~\ref{tab:meanpt} as a function of centrality.  
The statistical (systematic) uncertainty is extracted by fitting the \pt\ distribution considering only the statistical (\pt-uncorrelated systematic) uncertainty of the measurement. 
For comparisons, the \meanpt\ and \meanptsq\ results from PHENIX were recomputed with the function defined by Eq.~\ref{eq:meanpt}, adjusted in the measured \pt\ range and extrapolated  to \pt\ = 8\gevc\ to match our \pt\ range.
\begin{table}[t]
\begin{center}
\begin{tabular}{ccccc}
\hline
\hline
\pt\ range  &  \y\ range  &  Centrality  & \meanpt\ $\pm$ stat. $\pm$ syst.   & \meanptsq\ $\pm$ stat. $\pm$ syst. \\
(\gevc)            &   &  &  (\gevc)  & (\gevcsq)  \\ 
\hline
\multicolumn{5}{c}{ \PbPb\ \sqrtSnnE{2.76} }   \\ 
 0--8  & 2.5--4    &  \cent{0}{20}\   & 1.92 $\pm$ 0.02 $\pm$ 0.03   &  5.17 $\pm$ 0.12 $\pm$ 0.16      \\
 0--8  & 2.5--4    &  \cent{20}{40}\  & 2.04 $\pm$ 0.02 $\pm$ 0.04   &  5.83 $\pm$ 0.11 $\pm$ 0.17      \\
 0.5--8  & 2.5--4  &  \cent{40}{90}\  & 2.22 $\pm$ 0.03 $\pm$ 0.04   &  6.72 $\pm$ 0.14 $\pm$ 0.20       \\
\hline
\multicolumn{5}{c}{ \pp\ \sqrtSE{2.76}~\cite{Abelev:2012kr_err}}\\ 
 0--8  & 2.5--4       &  n/a         & 2.28 $\pm$ 0.04 $\pm$ 0.03      &  7.06 $\pm$ 0.26 $\pm$ 0.13      \\
\hline
\multicolumn{5}{c}{ \pp\ \sqrtSE{0.2}~\cite{Adare:2006kf}}\\ 
 0--7  & 1.2--2.2  &  n/a         & 1.61 $\pm$ 0.01  $\pm$ 0.012    &  3.60 $\pm$ 0.06        $\pm$ 0.07 \\
\hline
\multicolumn{5}{c}{ \AuAu\ \sqrtSnnE{0.2}~\cite{Adare:2011yf}}\\ 
 0--5  & 1.2--2.2  &  \cent{0}{20}\    & 1.94 $\pm$ 0.18             &  5.79 $\pm$ 1.33      \\
 0--6  & 1.2--2.2  &  \cent{20}{40}\   & 1.87 $\pm$ 0.07             &  4.78 $\pm$ 0.34      \\
 0--6  & 1.2--2.2  &  \cent{40}{60}\   & 1.74 $\pm$ 0.04             &  4.19 $\pm$ 0.27      \\
 0--6  & 1.2--2.2  &  \cent{60}{92}\   & 1.61 $\pm$ 0.05             &  3.87 $\pm$ 0.27      \\
\hline
\multicolumn{5}{c}{ \CuCu\ \sqrtSnnE{0.2}~\citep{Adare:2008sh}}\\ 
 0--5  & 1.2--2.2  &  \cent{0}{20}\    & 1.68 $\pm$ 0.04  $\pm$  0.02    &  3.79 $\pm$ 0.25   $\pm$ 0.11 \\
 0--5  & 1.2--2.2  &  \cent{20}{40}\   & 1.69 $\pm$ 0.04  $\pm$  0.02    &  3.71 $\pm$ 0.18       $\pm$ 0.08 \\
 0--5  & 1.2--2.2  &  \cent{40}{60}\   & 1.68 $\pm$ 0.05  $\pm$  0.02    &  3.91 $\pm$ 0.30      $\pm$ 0.11 \\
 0--5  & 1.2--2.2  &  \cent{60}{94}\   & 1.66 $\pm$ 0.10  $\pm$  0.04    &  4.13 $\pm$ 0.64       $\pm$ 0.24 \\
\hline
\hline
\end{tabular}
\end{center}
\caption{\label{tab:meanpt} Values of \meanpt\ and \meanptsq\  at various energies and colliding systems. 
The statistical and systematic uncertainties are quoted separately, except for PHENIX measurements in \AuAu\ collisions where the quadratic sum is given.
If the measurement is not available or not used in the range $0 < \pt < 8\gevc $, the fit function is extrapolated down to 0 and up to 8\gevc\ to compute \meanpt\ and \meanptsq. 
}
\end{table}                        
These results are also given in Tab.~\ref{tab:meanpt} along with the measurement in \pp\ collisions at \sqrtSE{2.76} with updated uncertainties~\cite{Abelev:2012kr_err}.

\begin{figure}[htbp]
\begin{center}
\begin{tabular}{cc} 
\hspace{-1cm}
	\includegraphics[width=0.54\linewidth,keepaspectratio]{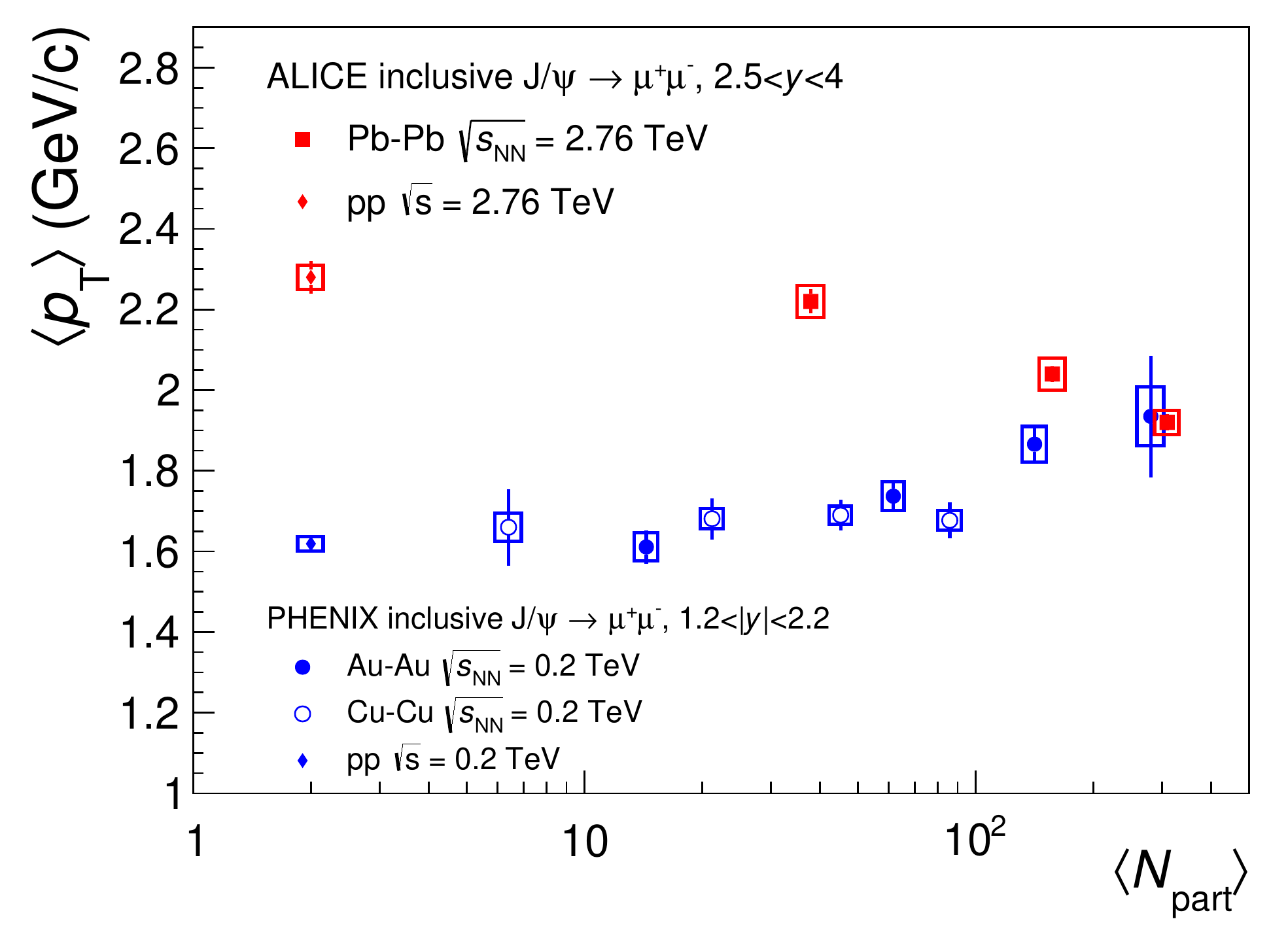} & 
\hspace{-0.5cm}
	\includegraphics[width=0.54\linewidth,keepaspectratio]{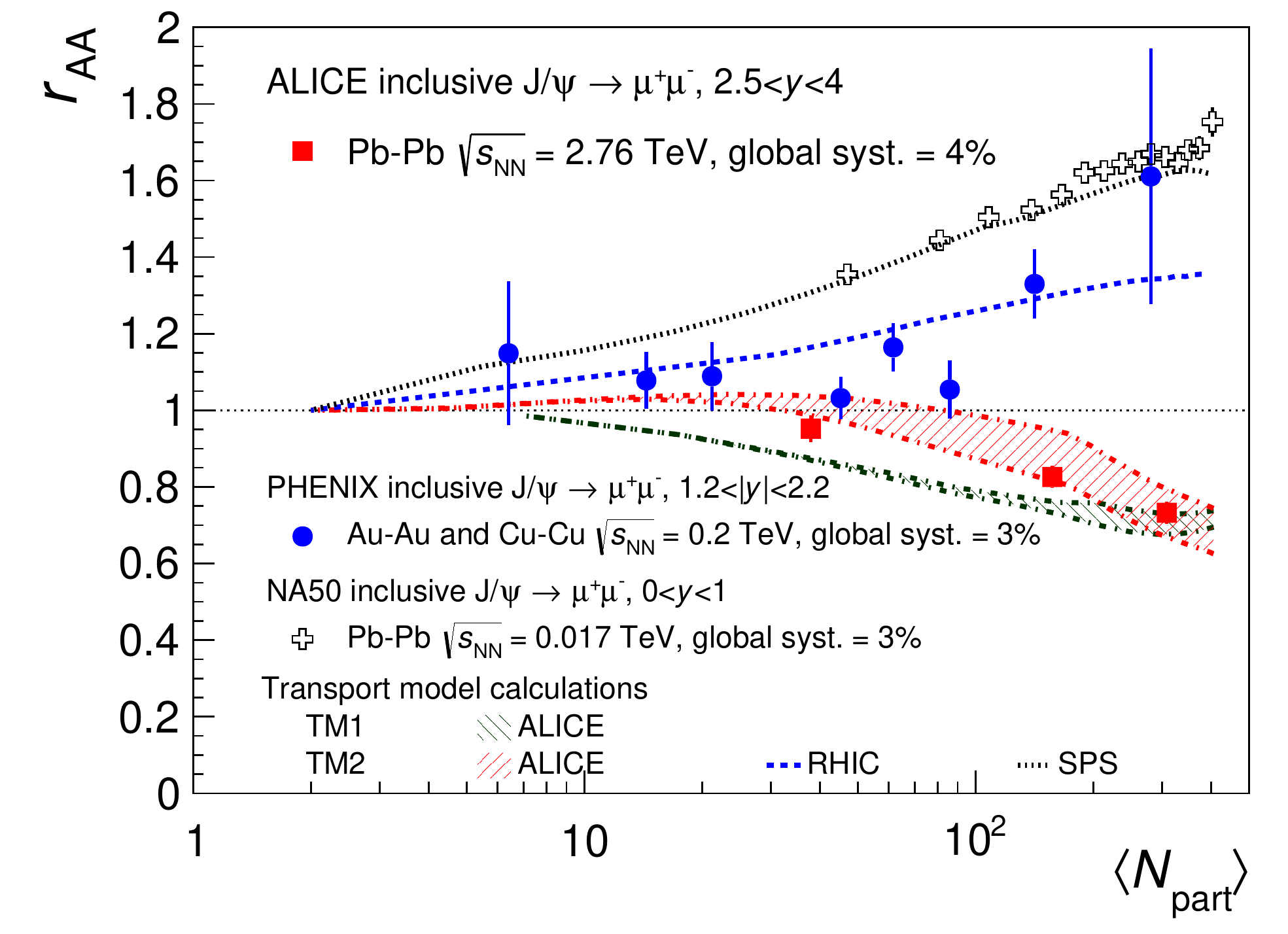} \\ 
\end{tabular}
\caption{Mean transverse momentum \meanpt\ measured by  ALICE~\cite{Abelev:2012kr,Abelev:2012kr_err} and PHENIX~\cite{Adare:2006kf,Adare:2008sh,Adare:2011yf} as a function of the number of participant nucleons (left). \raa\ measured by  NA50~\cite{Abreu200185}, PHENIX and ALICE and compared to model calculations~\cite{Zhao:2011cv,Zhou:2014kka}, as a function of the number of participant nucleons (right).}  
\label{fig:meanpt1}
\end{center}
\end{figure}

The \meanpt\ of inclusive \jpsi\ measured in \pp\ and \PbPb\ collisions at  \sqrtSnnE{2.76} is shown in Fig.~\ref{fig:meanpt1} (left side) as a function of \Npart.
The error bars (open boxes) represent the statistical (systematic) uncertainties.
A clear downward trend in \meanpt\ is observed when going from \pp\ to the most central \PbPb\ collisions.
The  \meanpt\ decrease from peripheral (\cent{40}{90}) to central (\cent{0}{20}) collisions is significant, the two values being  separated by more than $5\sigma$.     
These results are compared to the ones obtained  by PHENIX in \pp, \CuCu\ and \AuAu\ collisions at \sqrtSnnE{0.2}. 
There is no evidence for a decreasing trend, contrary to what is observed in the ALICE measurement.

In order to compare the evolution of $\meanptsq_{\nucnuc}$ at different energies, one can form the variable \raa\ defined as 
\begin{eqnarray}
\raa =  \frac{\meanptsq_{\nucnuc}}{\meanptsq_{\pp}}.
\label{eq:raa}
\end{eqnarray}
This variable  was measured over the  wide range of energies and colliding systems covered by NA50 and PHENIX experiments. 
The comparison with the ALICE results is done in Fig.~\ref{fig:meanpt1} (right side).
A very different \Npart\ dependence is seen, especially  when comparing \PbPb\ collisions at the SPS and the LHC. 
At the SPS energy of \sqrtSnnE{0.017}~\cite{Abreu200185}, the increase of the \jpsi\ \meanptsq\ with the centrality of the  collision was attributed to the  Cronin effect~\cite{Cronin:1974zm}, interpreted as an extra \pt\ kick due to multiple scatterings of the initial partons producing  the \jpsi. 
At the LHC, a clear decrease of   \raa\ is observed as a function of \Npart.
This behavior could  be related to the onset of recombination phenomena and to the thermalization of charm quarks. 
Theoretical calculations~\cite{Zhao:2011cv,Zhou:2014kka},  based on transport models (described in the next section) are able to reproduce the \raa\ at SPS, RHIC and LHC energies.  
They correlate the specific dependence of \raa\ on collision centrality with the increased importance of recombination effects in the \jpsi\ production mechanism at the LHC.

\section{Nuclear Modification Factor}

Some of the  \Raa\ results presented here were already published in~\cite{Abelev:2013ila} and are shown again in this section, where they are compared with model calculations and with results from previous experiments. They include the centrality dependence of \Raa\ (Fig.~\ref{fig:raacent}), the \pt\ dependence of \Raa\ for the full centrality range  \cent{0}{90} and for the centrality class \cent{0}{20}\ (Fig.~\ref{fig:raapt} top row) and the rapidity dependence of \Raa\ (Fig.~\ref{fig:raarap}).
The new results shown in this section include the  centrality dependence of \Raa\ for three \pt\ intervals (Fig.~\ref{fig:raacentptranges}) and the \pt\ dependence of \Raa\ for the centrality classes \cent{20}{40}\ and \cent{40}{90}\ (Fig.~\ref{fig:raapt} bottom row).
These new results were obtained using a slightly different trigger selection, as explained in section~\ref{sec:datasample}.
The consistency of the results obtained with the two selections was verified.

\subsection{Centrality dependence of \Raa}
\begin{figure}[htbp]
\begin{center}
\begin{tabular}{cc} 
\hspace{-1cm}
\includegraphics[width=0.54\linewidth,keepaspectratio]{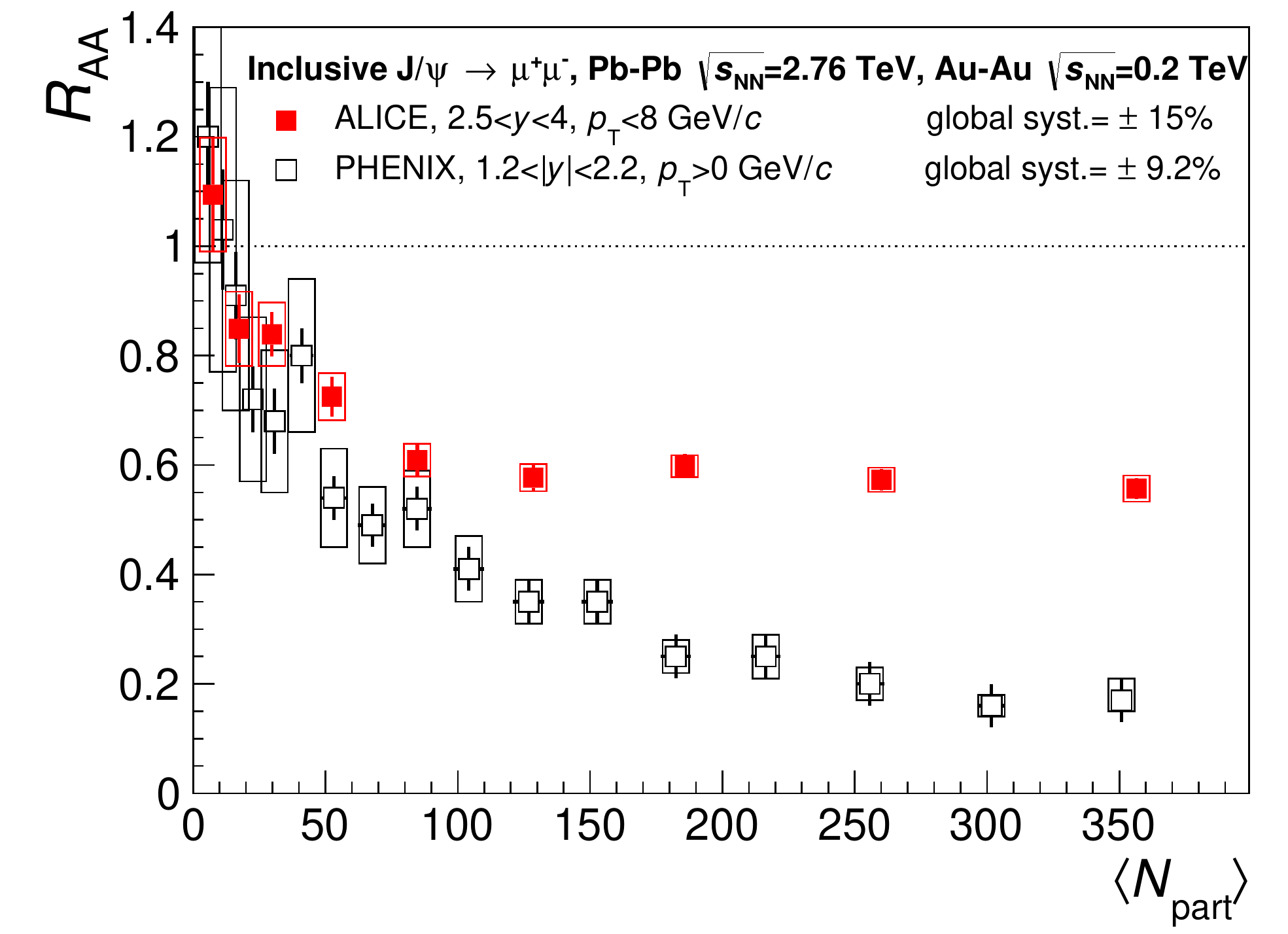}
\hspace{-0.4cm}
\includegraphics[width=0.54\linewidth,keepaspectratio]{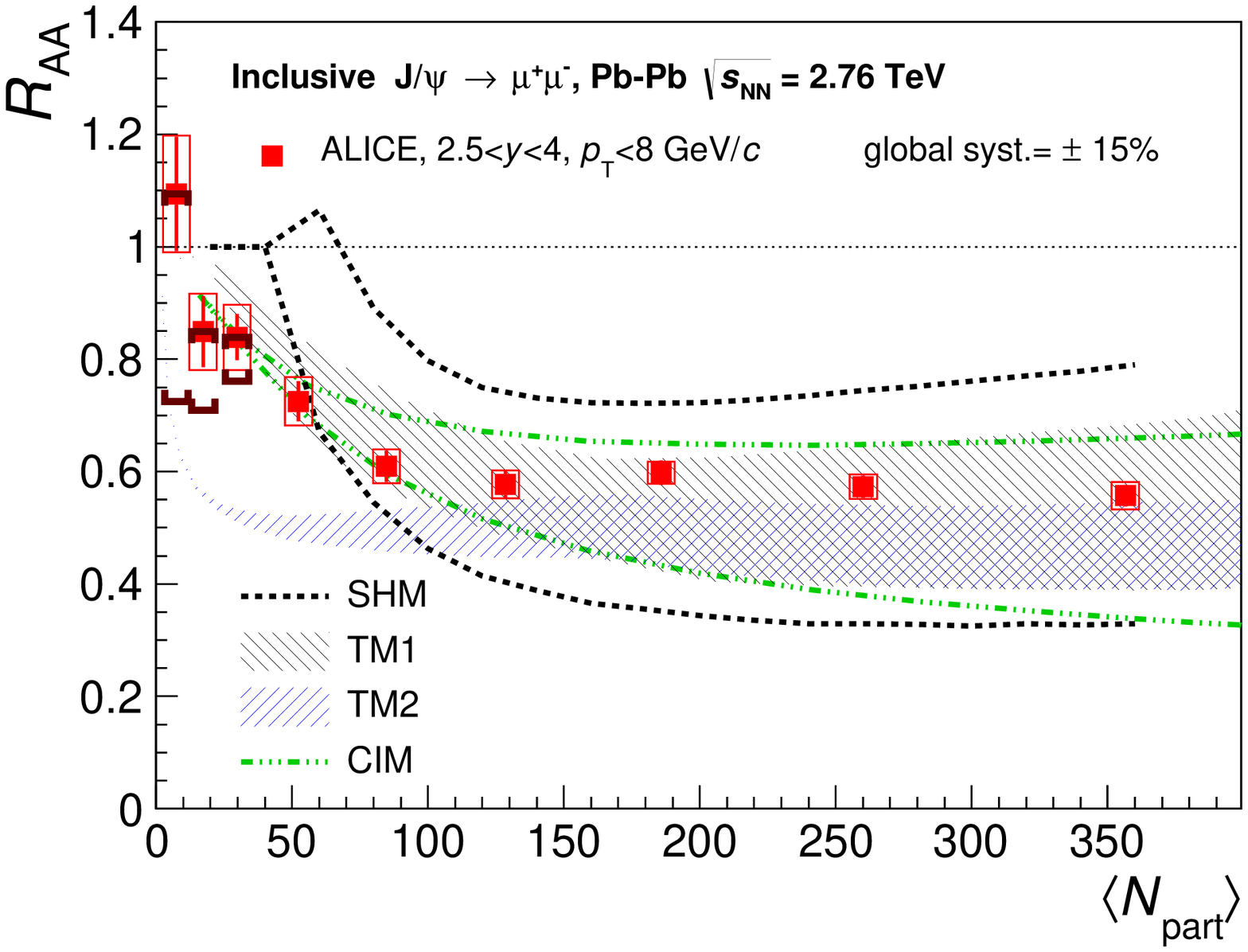}
\end{tabular}
\end{center}
\caption{\label{fig:raacent}  Inclusive \jpsi\ \Raa\ as a function of the number of participant nucleons measured in \PbPb\ collisions at \sqrtSnnE{2.76}~\cite{Abelev:2013ila}, compared to the PHENIX measurement in \AuAu\ collisions at \sqrtSnnE{0.2}~\cite{Adare:2011yf} (left) and  to theoretical models~\cite{Andronic:2011yq,Zhao:2011cv,Zhou:2014kka,Ferreiro:2012rq}, which all include a \jpsi\ regeneration component (right). The brackets shown in the three most peripheral centrality classes on the right figure quantify the possible range of variation of the hadronic \jpsi\ \Raa\ for two extreme hypotheses on the photo-production contamination in the inclusive measurement, see text for details.}
\end{figure}

Our measurement of the inclusive \jpsi\ \Raa\ at \sqrtSnnE{2.76} in the range $2.5 < \y  < 4$ and $  \pt < 8  \gevc$  is shown in Fig.~\ref{fig:raacent} as a function of  \Npart. 
Statistical  (uncorrelated systematic)  uncertainties  are represented by vertical error bars (open boxes).
A global correlated systematic uncertainty affecting all the values by the same amount is quoted in the legend.
The same  convention is applied  in the following figures, unless otherwise specified. 
The  \jpsi\ \Raa\  in the centrality class \cent{0}{90}\ (corresponding to $\Npart\ \sim 124 $, see Tab.~\ref{tab:taa}) is $\Raa^{\cent{0}{90}} = 0.58 \pm  0.01 \rm{(stat.)} \pm 0.09 \rm{(syst.)}$,  indicating a clear \jpsi\ suppression.
This suppression is significantly less pronounced than that observed at lower energy in PHENIX in a similar kinematic range, as previously discussed in~\cite{Abelev:2012rv,Abelev:2013ila}.
For \Npart\ larger than  70, corresponding to the 50\% most central \PbPb\ collisions, the \jpsi\ \Raa\ is consistent with being constant, within uncertainties. 
Such behavior was not  observed in heavy ion collisions at lower energies (SPS, RHIC), where \Raa\ is continuously decreasing as a function of centrality.

\begin{figure}[h!]
\begin{center}
\begin{tabular}{cc} 
\hspace{-1cm}
\includegraphics[width=0.54\linewidth,keepaspectratio]{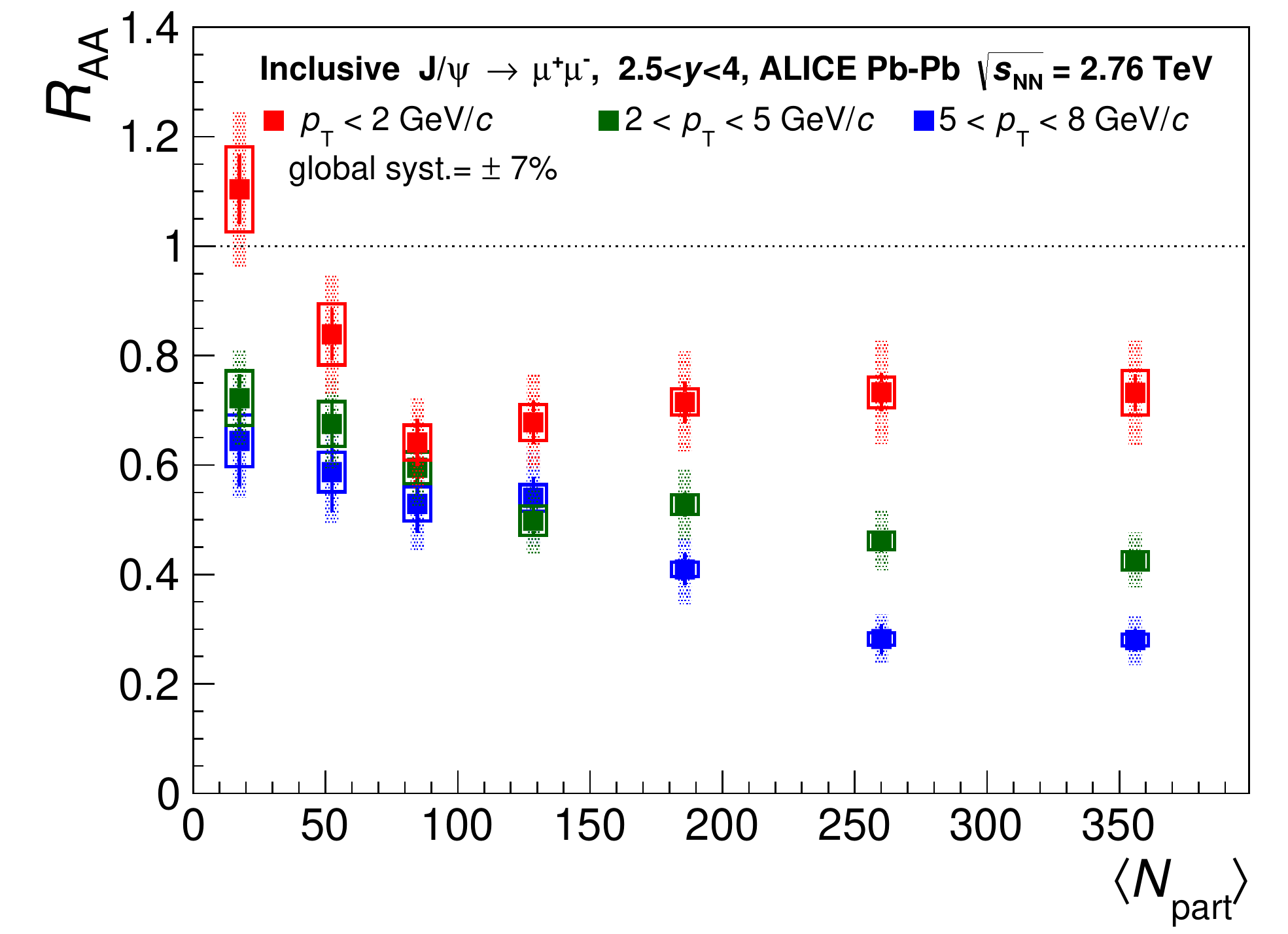}
\hspace{-0.4cm}
\includegraphics[width=0.54\linewidth,keepaspectratio]{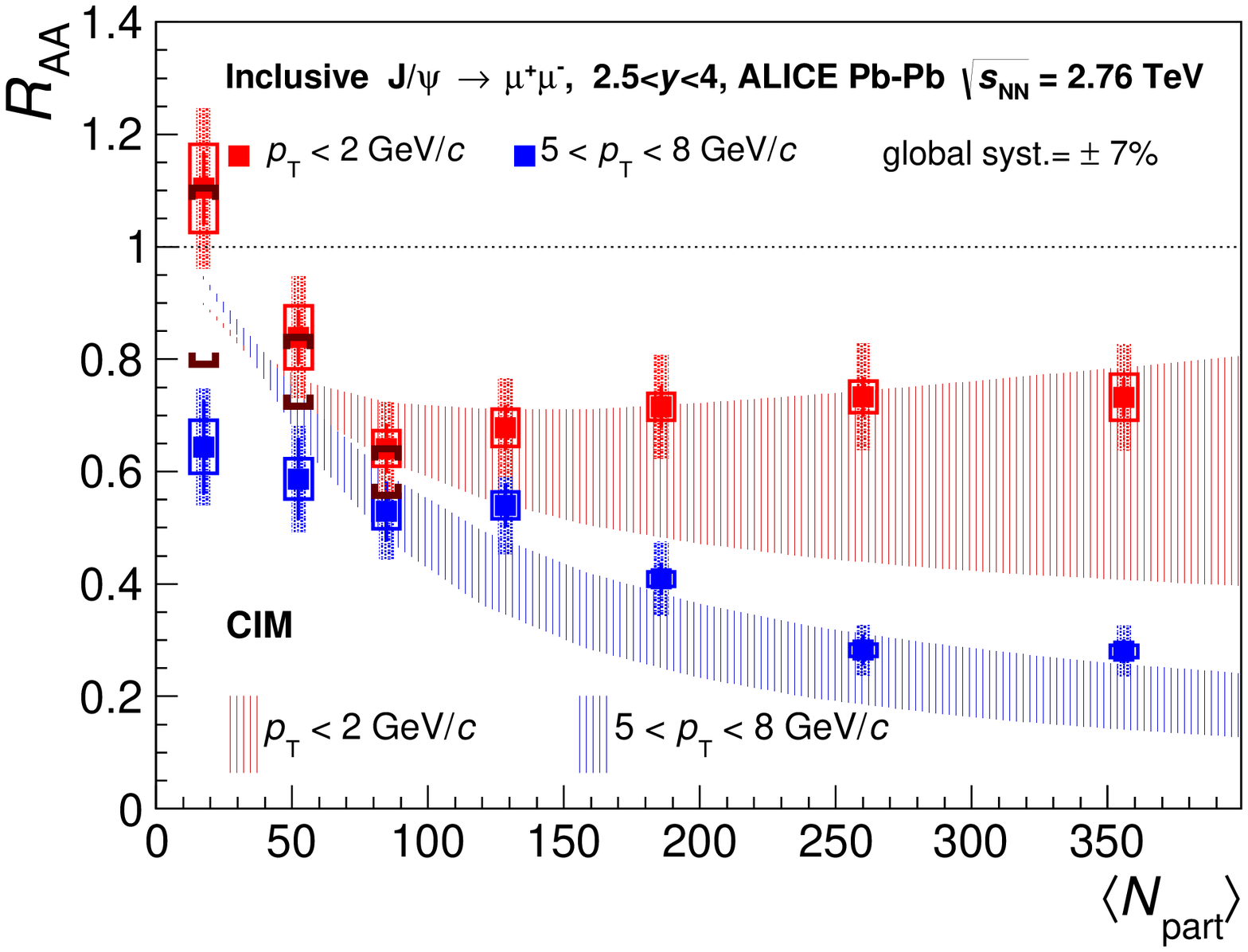}\\
\hspace{-1cm}
\includegraphics[width=0.54\linewidth,keepaspectratio]{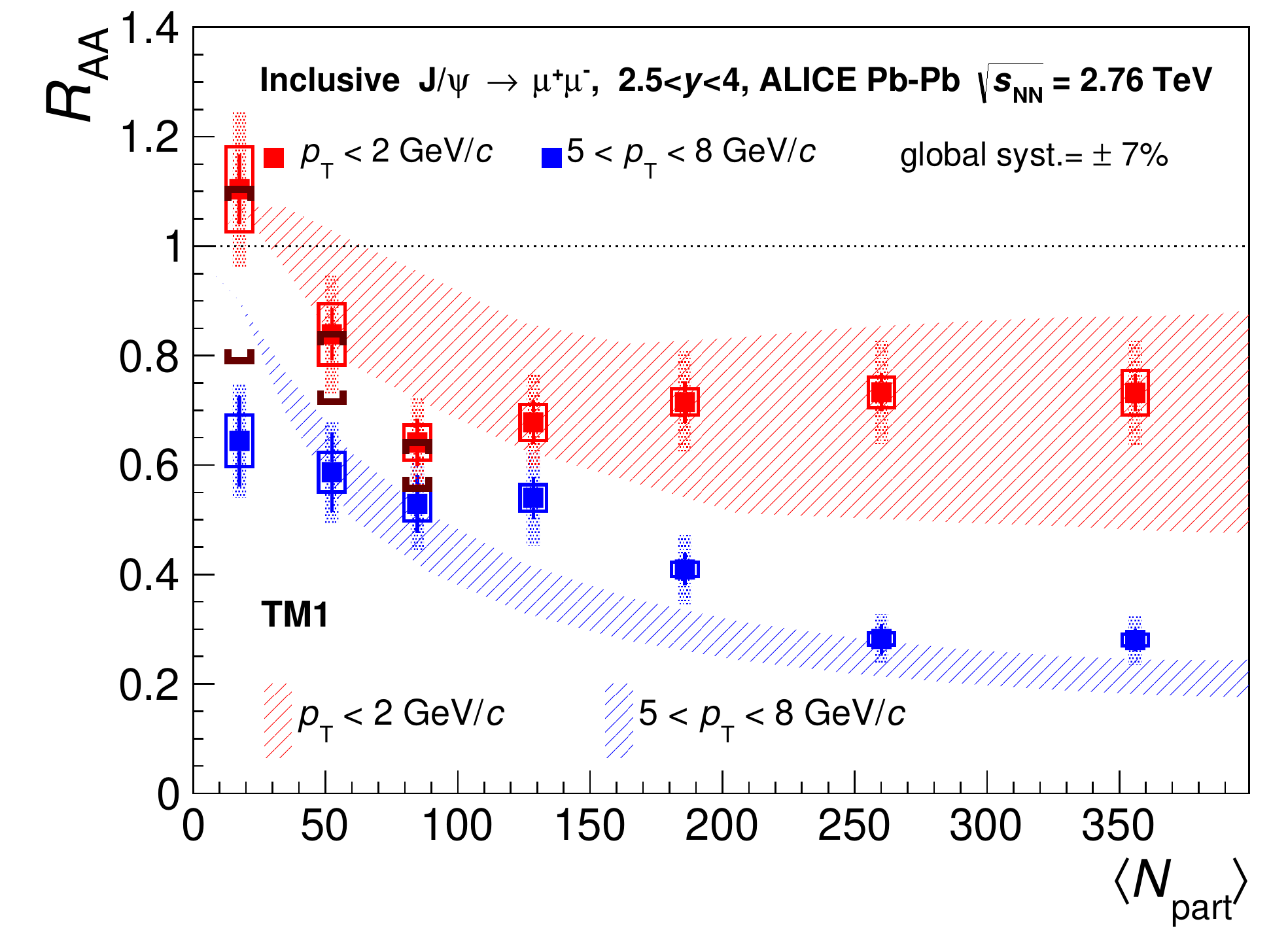}
\hspace{-0.4cm}
\includegraphics[width=0.54\linewidth,keepaspectratio]{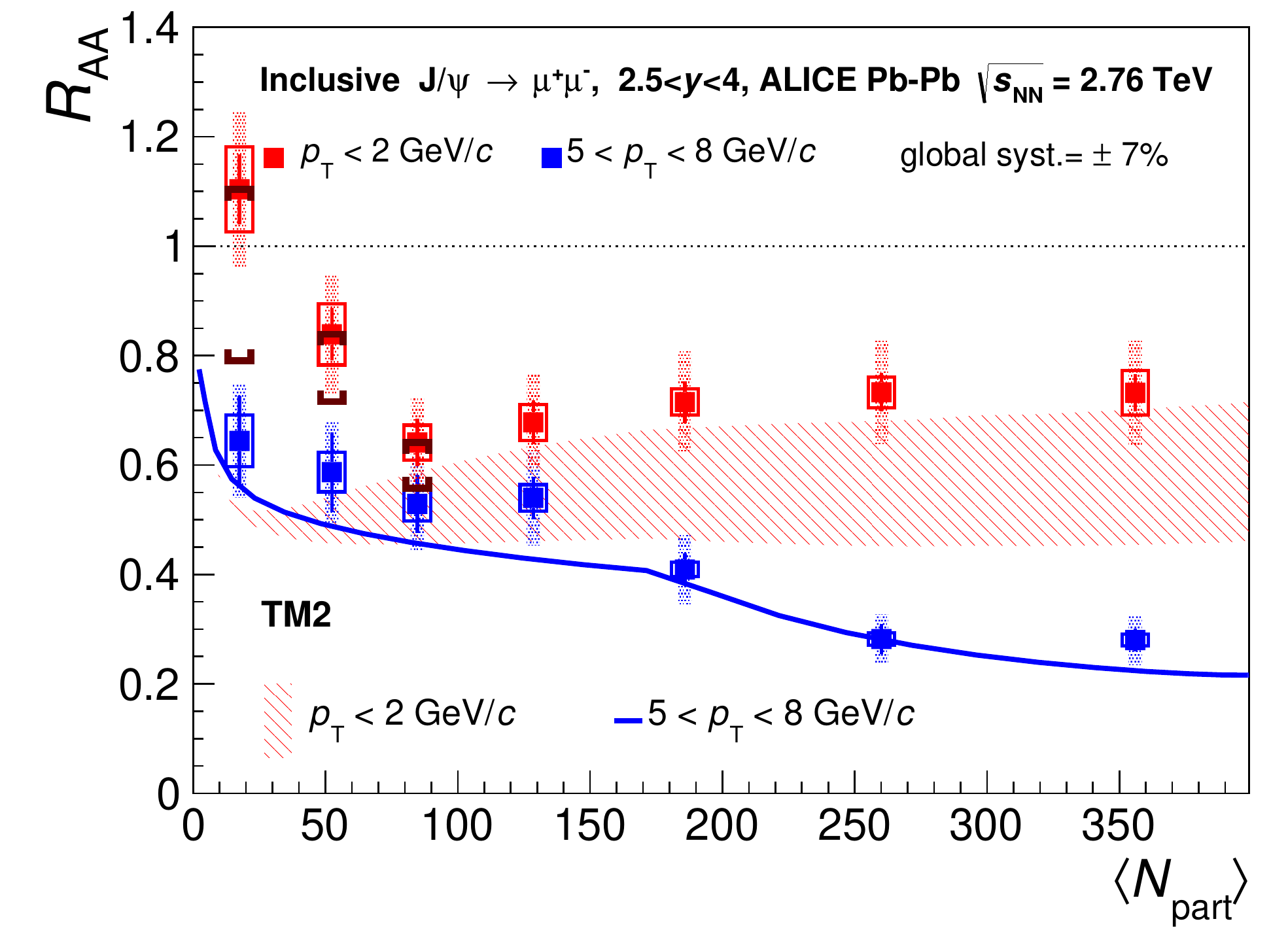}
\end{tabular}
\end{center}
\caption{\label{fig:raacentptranges} Inclusive \jpsi\ \Raa\ as a function of the number of participant nucleons measured in \PbPb\ collisions at \sqrtSnnE{2.76} for three \pt\ ranges  (0--2, 2--5 and 5--8\gevc) and comparisons of the  lowest and highest \pt\ range to the transport and to the comover interaction models~\cite{Zhao:2011cv,Zhou:2014kka,Ferreiro:2012rq}. The brackets quantify the possible range of variation of the hadronic \jpsi\ \Raa\ for two extreme hypotheses on the photo-production contamination in the inclusive measurement.}
\end{figure}

The impact  of non-prompt  \jpsi\ on the inclusive \Raa\ analysis was studied.
The \Raa\ of prompt \jpsi\ is estimated (see Eq.~\ref{eq:promptraa}) to be about 7\% larger than the inclusive \jpsi\ \Raa\ if the beauty component is fully suppressed. 
In the other extreme case, where the B-meson production is not affected by the medium and scales with the number of binary collisions, i.e. $\Raa^{\mathrm{non\mbox{-}prompt}}  = 1$, the \Raa\ of prompt \jpsi\ would be about 6\% smaller in central  collisions  and about 1\% smaller in peripheral collisions.
The excess in the inclusive \jpsi\ yield observed at very low \pt~\cite{Adam:2015gba} also influences the \Raa\ in the most peripheral collisions. 
A large fraction of this contribution (about 75\% as explained in section~\ref{sec:observables}) can be removed by selecting \jpsi\ with a \pt\ higher than $0.3 \gevc$. Assuming that the hadronic \jpsi\ \Raa\ in the ranges $ 0 < \pt < 0.3\gevc $ and  $ 0.3 < \pt < 8\gevc$ are the same, it becomes possible to estimate the impact of the \jpsi\ photo-production  on the inclusive \Raa. In the centrality classes \cent{60}{70}, \cent{70}{80}\ and \cent{80}{90}, the hadronic \jpsi\ \Raa\ would be about 5\%, 11\% and  25\%  lower, respectively. 
Extreme hypotheses were  made to define upper and lower limits, represented with brackets on the   Figs.~\ref{fig:raacent}, \ref{fig:raacentptranges} and \ref{fig:raapt}.
The upper limit calculation assumes no \jpsi\ from photo-production thus the inclusive measurement only contains hadronic production. The lower limit assumes that i) all \jpsi\ produced with a \pt\ smaller than 0.3\gevc\ originate from photo-production and ii) the efficiency of the 0.3\gevc\ \pt\ selection is reduced from 75\% to 60\% (corresponding to an increase by a factor two of the \jpsi\ photo-production above 0.3\gevc).

The comparison with theoretical models, shown on the right-hand side of Fig.~\ref{fig:raacent}, helps in the interpretation of the large difference observed between the PHENIX and the ALICE results.

The Statistical Hadronization Model (SHM)~\cite{Andronic:2011yq}  assumes deconfinement and thermal equilibration of the bulk of the \ccbar\ pairs. 
Charmonium production occurs at the phase boundary via the statistical hadronization of charm quarks. 
The prediction is given for two values of the charm cross section $\mathrm{d} \sigma_{\ccbar} / \mathrm{d}y = 0.15$ and 0.25  mb at forward rapidity.
These values are derived from the measured charm cross section in \pp\ collisions at \sqrtSE{2.76\ \mathrm{and}\ 7}~\cite{Abelev:2012vra} bracketing the expectation for gluon shadowing in the Pb-nucleus between 0.6 and 1.0.
Production of non-prompt \jpsi\ from decays of B-mesons is not considered.

The two transport  models from Zhao (TM1)~\cite{Zhao:2011cv}  and Zhou (TM2)~\cite{Zhou:2014kka}   
 mainly differ in the rate  equation controlling the \jpsi\ dissociation and regeneration.  
In TM1, shadowing is  implemented  via a simple parametrization, leading to a 30\% suppression in the most central \PbPb\ collisions. 
The charm cross section  is assumed to be $ \mathrm{d} \sigma_{\ccbar}/ \mathrm{d} y \approx 0.5$ mb at forward rapidity, the fraction of \jpsi\ from beauty hadrons to be  10\% and no b-quenching is introduced in the calculation. 
This model is presented as a band connecting the results obtained with (lower limit) and without (upper limit) shadowing and is interpreted by the authors as the uncertainty of the prediction.
In TM2, the shadowing is given by the  EKS98 parametrization~\cite{Eskola:1998df}. The  charm cross section is taken in the range  d$\sigma_{\ccbar}/$d$y \approx 0.4-0.5$ mb at forward rapidity; the calculations for these two values provide  the lower and upper limits of the band displayed in the figure. 
The fraction of \jpsi\ from beauty hadrons is assumed to be 10\% with a b-quenching of 0.8, increased to 0.4 for \pt\ above 5\gevc.

The Comover Interaction Model (CIM)~\cite{Ferreiro:2012rq} implements shadowing, interaction with a co-moving dense partonic medium and recombination effects.  
The shadowing is calculated within the Glauber-Gribov theory  making use of the generalized Schwimmer model of multiple scattering. 
The \jpsi\ dissociation cross section due to comover interaction is taken as $\sigma_{\rm co} = $ 0.65 mb from low-energy data.
Recombination effects are included by adding a gain term proportional to $\sigma_{\rm co}$ and to the number of \qc\ and \qcbar\ quarks, thus no additional parameter is added to the model. 
The charm cross section $ \mathrm{d} \sigma_{\ccbar}/ \mathrm{d} y$ at forward rapidity is taken in the range 0.4 to 0.6 mb, which gives respectively the lower and upper limits of the calculation. 
Production of non-prompt \jpsi\  is not considered.

To match our \jpsi\ \Raa\ results, all models above need to include in their calculation a  sizeable \jpsi\ production from deconfined \qc\ and \qcbar\ quarks.  

A different test of these models was carried out by studying the \jpsi\ \Raa\ centrality dependence in \pt\ intervals. 
Figure~\ref{fig:raacentptranges} displays the  measurement of the inclusive \jpsi\ \Raa\ as a function of the number of participant nucleons measured in \PbPb\ collisions at \sqrtSnnE{2.76} for the three \pt\ ranges 0--2, 2--5 and 5--8\gevc.
The uncorrelated systematic uncertainties shown at each point were separated into
uncorrelated as a function of centrality (open boxes) and fully correlated as a function of centrality but uncorrelated as a function of \pt\ (shaded areas). 
For $\Npart \gtrsim 150$,  the low \pt\ \jpsi\ \Raa\ is significantly larger than the mid and high \pt\ ones.   
In the most central bin, the \Raa\ values corresponding to the lowest and the highest  \pt\   are separated by 3.9$\sigma$.
For $\Npart \lesssim 150$,
the centrality dependence exhibits similar trends  for the 2--5 and 5--8\gevc\ ranges, while the most peripheral ($\Npart \sim 20 $) \Raa\ measurement in the low \pt\ (0--2 \gevc) range appears to deviate from the others. 
However, the \jpsi\ yield excess observed at very low \pt\ may have a sizable effect in the 0--2 \gevc\ interval.  In the centrality classes \cent{40}{50}, \cent{50}{60}\ and \cent{60}{90}, based on the same assumptions made for the $ 0 < \pt < 8\gevc$ case, the hadronic \jpsi\ \Raa\ would be about 5\%, 6\% and  18\%  lower, respectively.
Due to the increase of the non-prompt \jpsi\ component at large \pt, the difference between the measured inclusive \jpsi\ \Raa\ and the prompt \jpsi\ \Raa\ increases with \pt. 
If the beauty contribution is fully (not) suppressed, \Raa\ of prompt \jpsi\ is estimated to be 6\%, 8\% and 11\% larger (0--3\%, 3--10\%  and 7--30\% smaller, depending on centrality) for the  \pt\ ranges 0--2, 2--5 and 5--8\gevc, respectively.

Calculations from the transport  models and the comover interaction model are plotted on top of the 
results shown in Fig.~\ref{fig:raacentptranges}.
For the most peripheral collisions ($\Npart \lesssim 100$), the models cannot correctly reproduce the \Raa\ centrality dependence for  both the low and high \pt\ ranges. 
For the most central collisions ($\Npart \gtrsim 100$), the \Raa\ centrality dependence for high \pt\  \jpsi\ is reasonably reproduced by all models.
Concerning the low \pt\ range in the most central events,  the measurement is compatible with the upper side of the theoretical uncertainty band from the CIM and TM2 models.  
For these models, it corresponds to  the highest value for $ \mathrm{d} \sigma_{\ccbar}/ \mathrm{d} y$, 0.6 and 0.5 mb respectively.

\subsection{Transverse momentum  dependence of \Raa}
\begin{figure}[h!]
\begin{center}
\begin{tabular}{cc} 
\hspace{-1cm}
\includegraphics[width=0.54\linewidth,keepaspectratio]{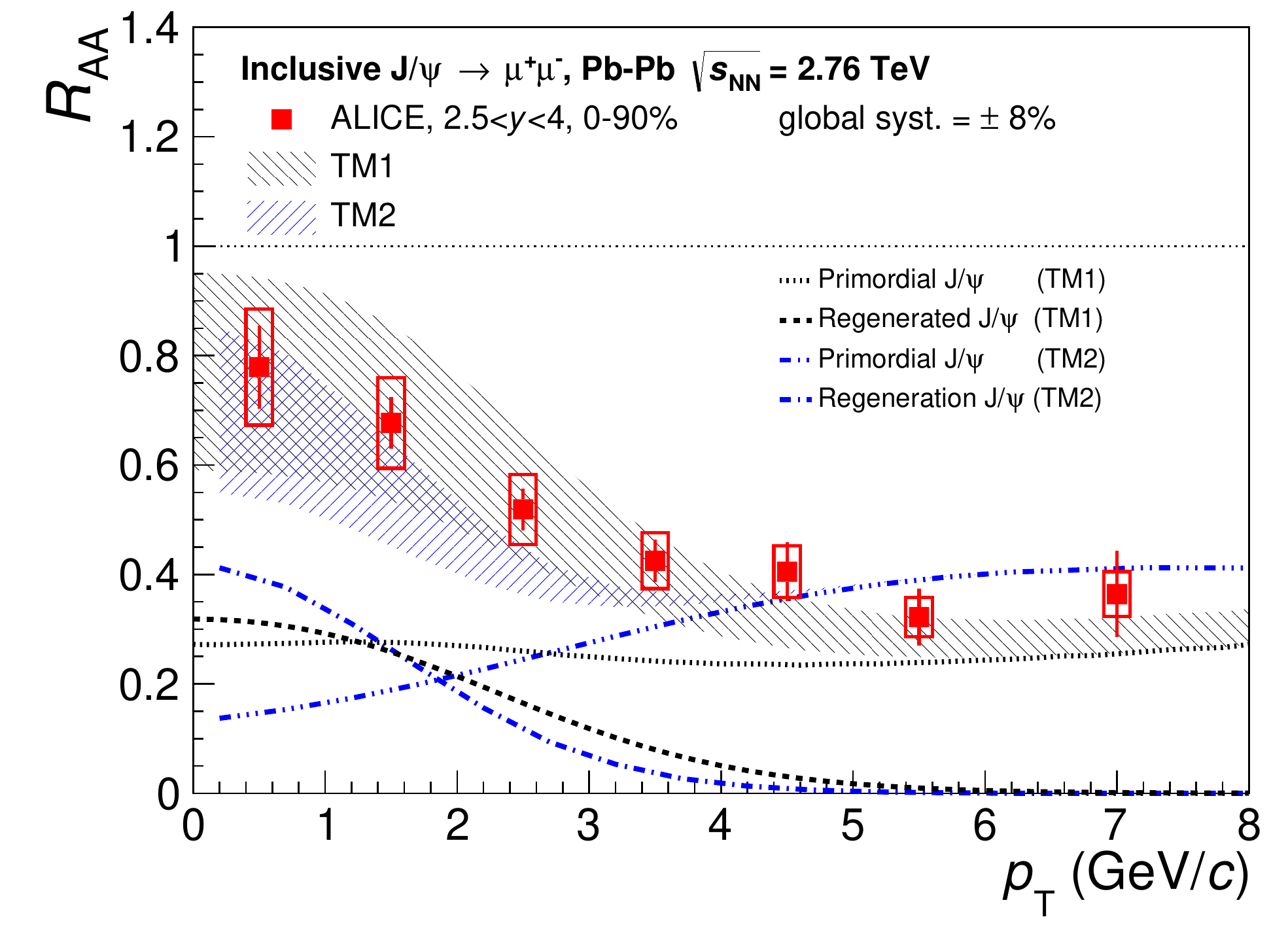}
\hspace{-0.38cm}
\includegraphics[width=0.54\linewidth,keepaspectratio]{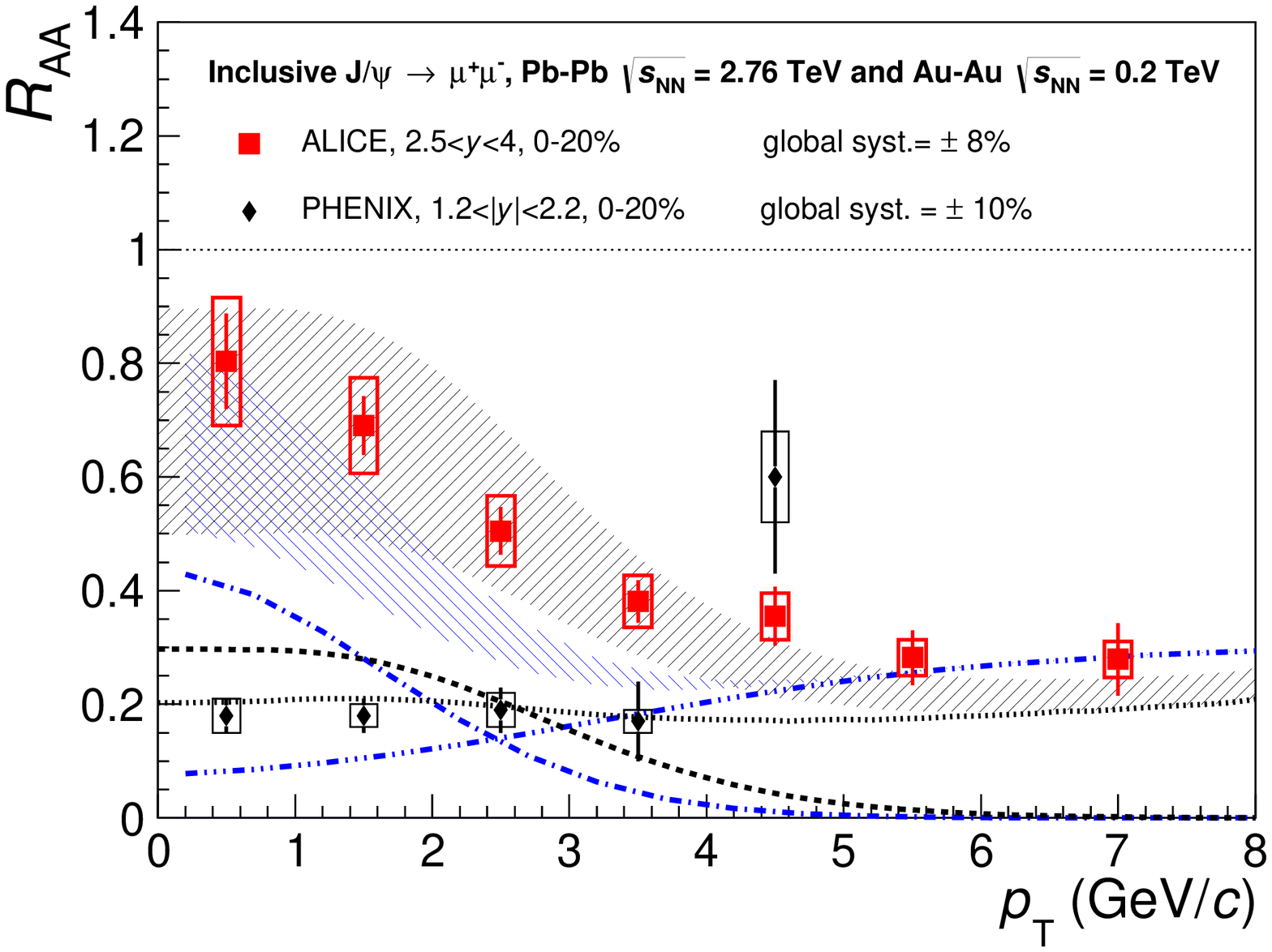} \\
\hspace{-1cm}
\includegraphics[width=0.54\linewidth,keepaspectratio]{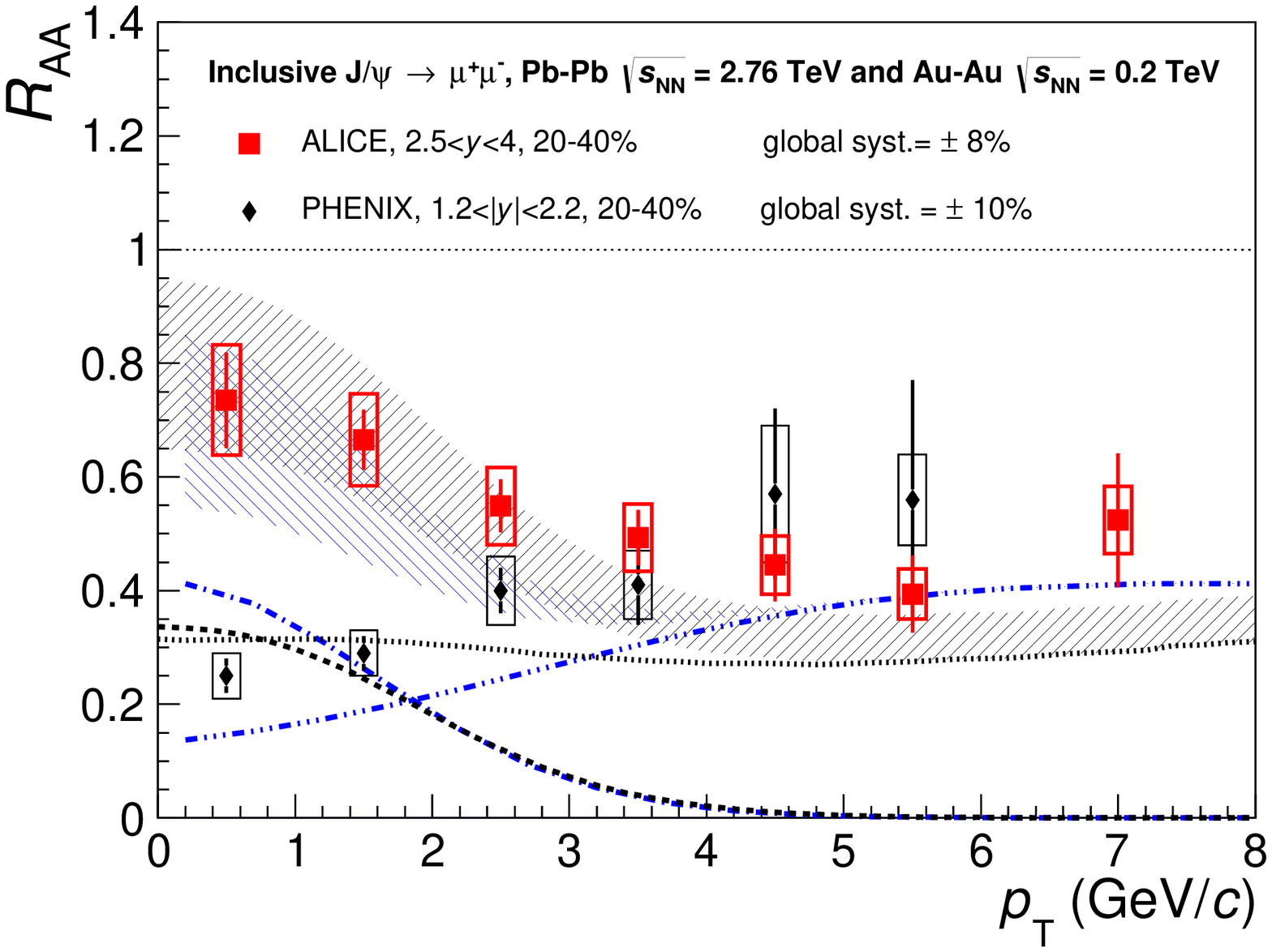}
\hspace{-0.38cm}
\includegraphics[width=0.54\linewidth,keepaspectratio]{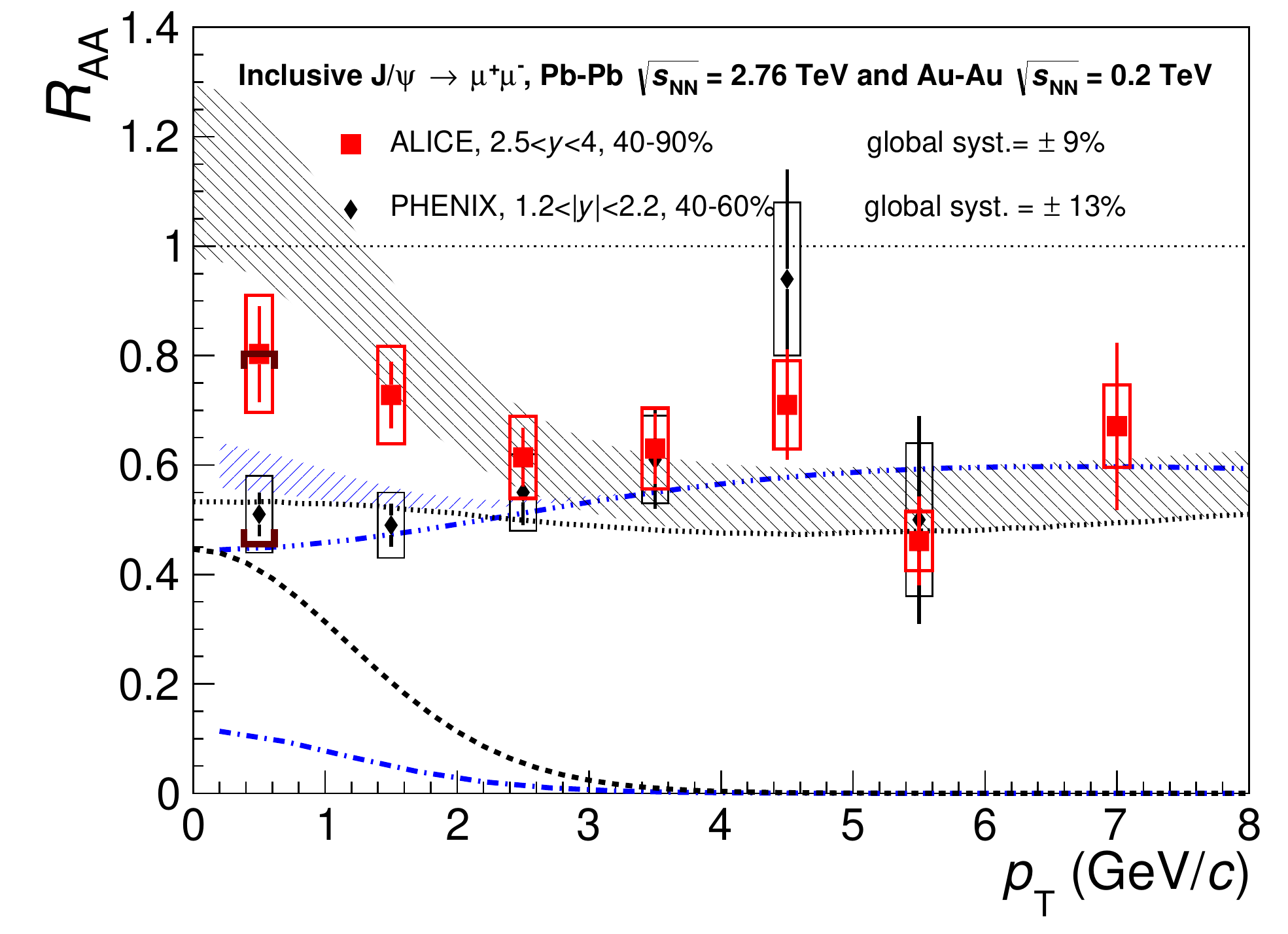}
\end{tabular}
\end{center}
\caption{\label{fig:raapt} Inclusive \jpsi\ \Raa\ as a function of the \jpsi\ \pt\ for $ 2.5 < \y < 4$  in the centrality class \cent{0}{90}~\cite{Abelev:2013ila} compared  to transport models~\cite{Zhao:2011cv,Zhou:2014kka} (top left). 
The comparison is done with PHENIX results~\cite{Adare:2011yf}  and transport models in the \cent{0}{20}~\cite{Abelev:2013ila} (top right),  \cent{20}{40}\ (bottom left) and \cent{40}{90}\ (bottom right) centrality classes.  
The brackets shown in the lowest \pt\ interval for the centrality class \cent{40}{90}\  quantify the possible range of variation of the hadronic \jpsi\ \Raa\ for two extreme hypotheses on the photo-production contamination in the inclusive measurement.
Upper limits  from PHENIX at high \pt\ are not represented. }
\end{figure}

The \pt\ dependence of the inclusive \jpsi\ \Raa\  in the rapidity range $2.5 <  \y < 4$  is shown in Fig.~\ref{fig:raapt} for the full centrality range \cent{0}{90} and  for three centrality classes  \cent{0}{20}~\cite{Abelev:2013ila}, \cent{20}{40}\ and \cent{40}{90}.  In Fig.~\ref{fig:raapt} top left corner,  the  inclusive \jpsi\ \Raa\ in the centrality class \cent{0}{90}\  shows a decrease of about $50\%$ from low to high \pt.
At low \pt, the measurement is close to 0.8 showing very little suppression. 
At high \pt, our \Raa\ value  is similar to that of CMS~\cite{Chatrchyan:2012np}. They measured, in the different rapidity range $1.6 < |\y| < 2.4$, an inclusive \jpsi\  $ \Raa =  0.41 \pm 0.05  \pm 0.04$ for $ 3 < \pt <  30\gevc$. The corresponding mean \pt\ is $6.27\,\gevc$.
When beauty contribution is fully (not) suppressed the prompt \jpsi\ \Raa\ is expected to be  5\% larger (2\% smaller) for  $ \pt  < 1\gevc $ and 17\% larger (30\% smaller) for $6 < \pt  < 8\gevc$.  

The transport model calculations  TM1~\cite{Zhao:2011cv}  and TM2~\cite{Zhou:2014kka}  are also shown in  Fig.~\ref{fig:raapt}. 
Both models reproduce reasonably well the \cent{0}{90}\ centrality measurement at high \pt. 
At low \pt, TM1 reproduces rather well our measurement,  while the data points sit on the upper limit of the TM2 calculation. 
One can also appreciate the relative contributions of the primordial (from the initial hard parton scattering) and regenerated  (from coalescence of \qc\ and \qcbar\ quarks in the deconfined medium)  components in these two calculations. 
The contribution of regenerated \jpsi\ is concentrated at low \pt\ and its relative fraction with respect to the initial production differs between the models. 
In TM1,  it is of the same order of the  primordial  \jpsi\ production, which is about constant  over the full \pt\ range.
In TM2, the regenerated \jpsi\ contribution  is almost three times larger than the primordial one in the lowest  \pt\ interval.  
For $\pt > 5 \gevc$, only the primordial production remains.

In the other panels of Fig.~\ref{fig:raapt}, the ALICE measurements are compared to those from PHENIX in \AuAu\ collisions at \sqrtSnnE{0.2} for the \cent{0}{20}, \cent{20}{40}\ and \cent{40}{60}\ centrality classes~\cite{Adare:2011yf}.
For $\pt  < 1\gevc $,  for all centrality ranges, the prompt \jpsi\ \Raa\ is expected to be  5\% larger  (2\% smaller) when the beauty contribution is fully (not) suppressed.  
For $6 < \pt  < 8\gevc $ the effect is much larger: 
if the beauty contribution is fully  suppressed,  the prompt \jpsi\ \Raa\  would be 17\%  larger in  all centrality ranges. If the beauty contribution is not  suppressed,  the prompt \jpsi\ \Raa\  would be 44\%, 15\% and 8\% smaller in  the centrality ranges \cent{0}{20}, \cent{20}{40}\ and  \cent{40}{90}, respectively. 
The very low \pt\ excess in the inclusive \jpsi\ yield mentioned before has a non-negligible impact in the $0 < \pt < 1 \gevc$ range in the most peripheral centrality class \cent{40}{90}: following the same assumptions made for the $ 0 < \pt < 8\gevc$ case, the estimated hadronic \jpsi\ \Raa\ would be about 22\% lower.
In the most central collisions (\cent{0}{20}), the inclusive \jpsi\  \Raa\ at low \pt\  is almost four times larger in \PbPb\ collisions at  \sqrtSnnE{2.76} than in \AuAu\ collisions at  \sqrtSnnE{0.2}. 
This difference cannot be explained only by the possible change in the size of the CNM effects that can be expected due to the different rapidity coverage and collision energy between the two measurements.
Such a behavior, on the other hand, is expected by all the recombination models described in the previous section.  
The same trend is observed in the centrality class \cent{20}{40}, where the large difference between the PHENIX and ALICE results observed at low \pt\ vanishes at high \pt. 
Concerning the most peripheral collisions, the inclusive \jpsi\ \Raa\ is still slightly larger 
for ALICE results at low \pt. 
However, here the  comparison between the two experiments is done with different centrality  classes,  \cent{40}{90}\ (ALICE) and  \cent{40}{60}\ (PHENIX), so that a firm conclusion, also because of the uncertainty size, cannot be drawn.
Transport model calculations for \PbPb\ collisions at \sqrtSnnE{2.76} are also presented for the \cent{0}{20}, \cent{20}{40}\ and \cent{40}{90}\  centrality classes. 
TM1 shows a good agreement with the measurements in  the \cent{0}{20}\ and  \cent{20}{40}\  centrality classes, while TM2 tends to underestimate the data for $\pt < 5 \gevc$.
In the most peripheral centrality class (\cent{40}{90}), the two models follow  significantly different trends, but  the uncertainties from the measurement are too large to discriminate them.
However, if the very low \pt\ excess is taken into account, a rather flat \pt\ dependence of the \Raa\ is expected, pushing our measurement aside from TM1 calculations in this specific range. For the high \pt\ region, both models reproduce well the experimental results in all the centrality classes.

\subsection{Rapidity dependence of \Raa}
\begin{figure}[h!]
\begin{center}
\begin{tabular}{c}
\includegraphics[width=0.54\linewidth,keepaspectratio]{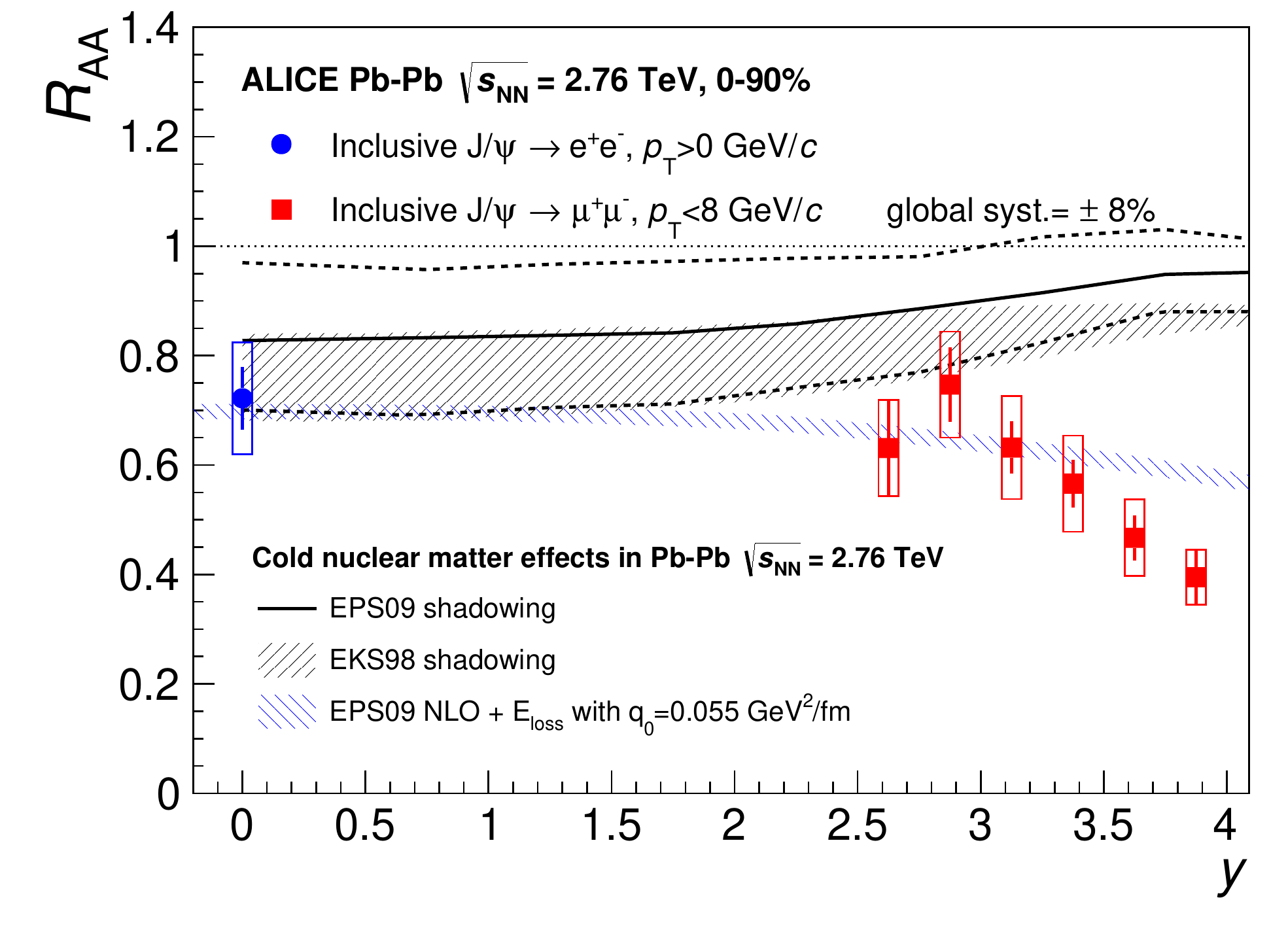}
\end{tabular}
\end{center}
\caption{\label{fig:raarap}  Inclusive \jpsi\ \Raa\ as a function of the \jpsi\ rapidity measured in \PbPb\ collisions at \sqrtSnnE{2.76}~\cite{Abelev:2013ila},  compared to theoretical calculations of CNM effects due to shadowing and/or coherent energy loss~\cite{Rakotozafindrabe:2011rw,Vogt:2010aa,Arleo:2012rs}.}
\end{figure}

The rapidity dependence of the inclusive \jpsi\ \Raa\ in \PbPb\ collisions~\cite{Abelev:2013ila} is shown in Fig.~\ref{fig:raarap}. 
The inclusive \jpsi\ \Raa\ measured in the rapidity range $ |\y| < 0.8 $  is about 0.7, consistent with the value measured at  $\y \sim 3$.    
From $y \sim 3$ to $y \sim  4$,  the \jpsi\ \Raa\ shows a decreasing trend leading to a drop of about  40\%. 
The influence of non-prompt \jpsi\ on this result is small, as the prompt \jpsi\  \Raa\  is expected to be only 8\% larger (5\% smaller) for  $2.5 < \y < 2.75$ and 6\% larger (9\% smaller) for $3.75 < \y  < 4$  if the beauty contribution is fully (not) suppressed.

The \PbPb\ measurements are compared to theoretical calculations, which only consider shadowing and coherent energy loss. The break-up of the \ccbar\ pair and nuclear absorption are not taken into account in any of the models. 
Shadowing only predictions are made within the Color Singlet Model at Leading  Order~\cite{Rakotozafindrabe:2011rw} and the Color Evaporation Model at Next to Leading Order~\cite{Vogt:2010aa}, with the EKS98~\cite{Eskola:1998df}	and the EPS09~\cite{Eskola:2009uj} parametrizations of the nPDF, respectively. 
For EKS98 (EPS09) the upper and lower limits correspond to the uncertainty in the factorization scale (uncertainty of the nPDF).
Finally, a theoretical prediction,  which includes a contribution from coherent parton energy loss processes in addition to EPS09 shadowing~\cite{Arleo:2012rs} is also shown. 
All models show a fair agreement with our measurements over a wide rapidity range,  $|\y| \lesssim 3$.
If the amplitude of CNM effects is correctly given by the calculations shown in Fig.~\ref{fig:raarap}, the observed \jpsi\ suppression due to CNM effects could be as large as 40\%. 
Moreover, if an additional \jpsi\ suppression occurs in the hot nuclear matter (as expected from lower energy measurement and observed at high \pt\ by both CMS and ALICE), other mechanisms compensating this suppression are needed to explain the \Raa\ measurements.
Figure~\ref{fig:raarap} supports this scenario, where suppression effects in hot matter are qualitatively counterbalanced by recombination. 
This is indeed what is expected from all models featuring recombination discussed in this paper. 
At higher rapidity, for $|\y| \gtrsim 3$, the models implementing only CNM effects tend to deviate from the data, although the one combining shadowing with coherent energy loss seems to match the decreasing trend of the \Raa\ better. 
Such a decrease of the \Raa\ values can also be explained by recombination models, where a reduction of the recombination effects is  expected with increasing rapidity, due to the decrease of $ \mathrm{d} \sigma_{\ccbar}/ \mathrm{d} y$.

\section{\psiratio\ ratio}

\begin{figure}[h!]
\begin{center}
\begin{tabular}{cc}
\hspace{-1cm}
\includegraphics[width=0.54\linewidth,keepaspectratio]{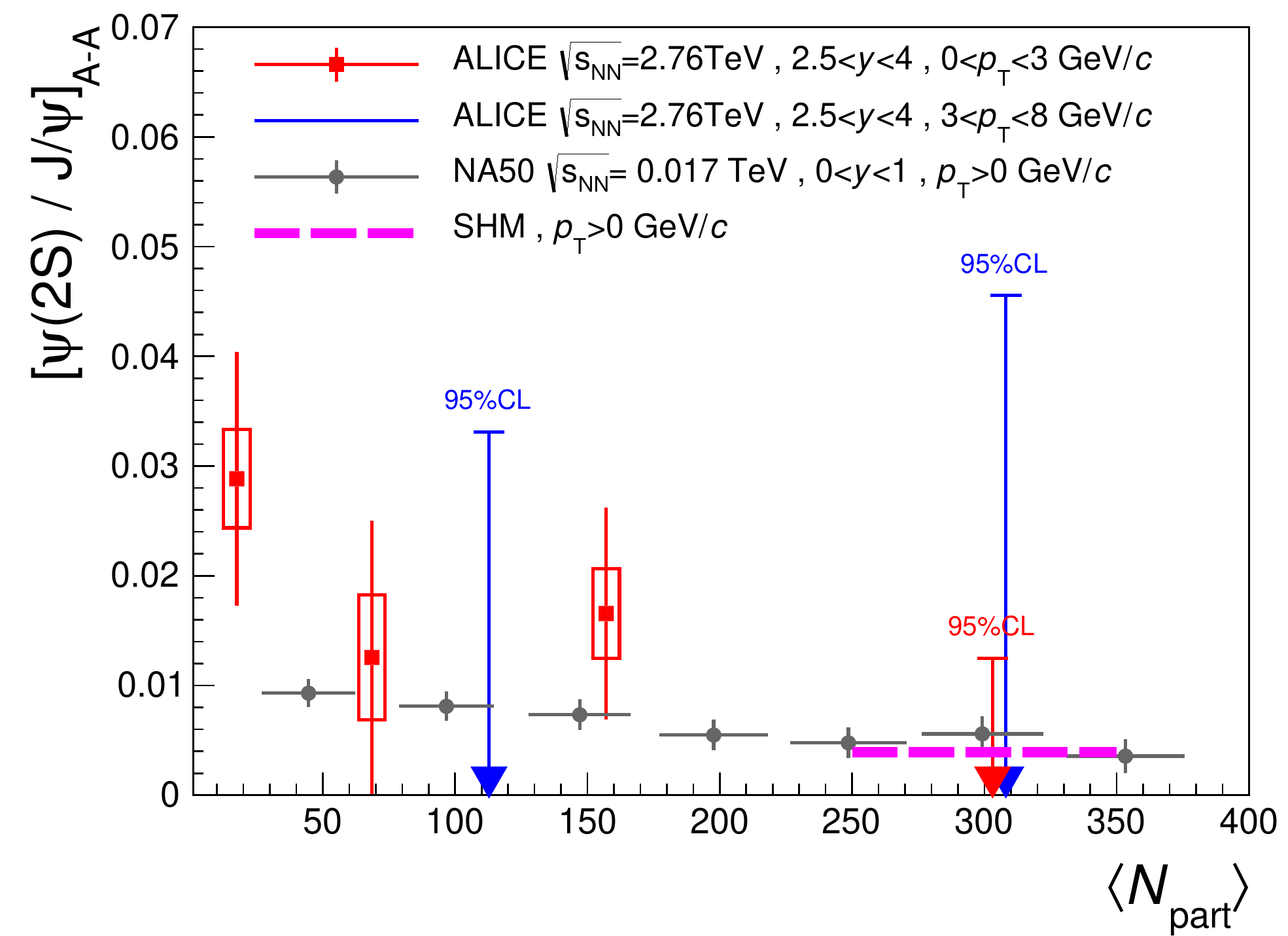}
\hspace{-.2cm}
\includegraphics[width=0.54\linewidth,keepaspectratio]{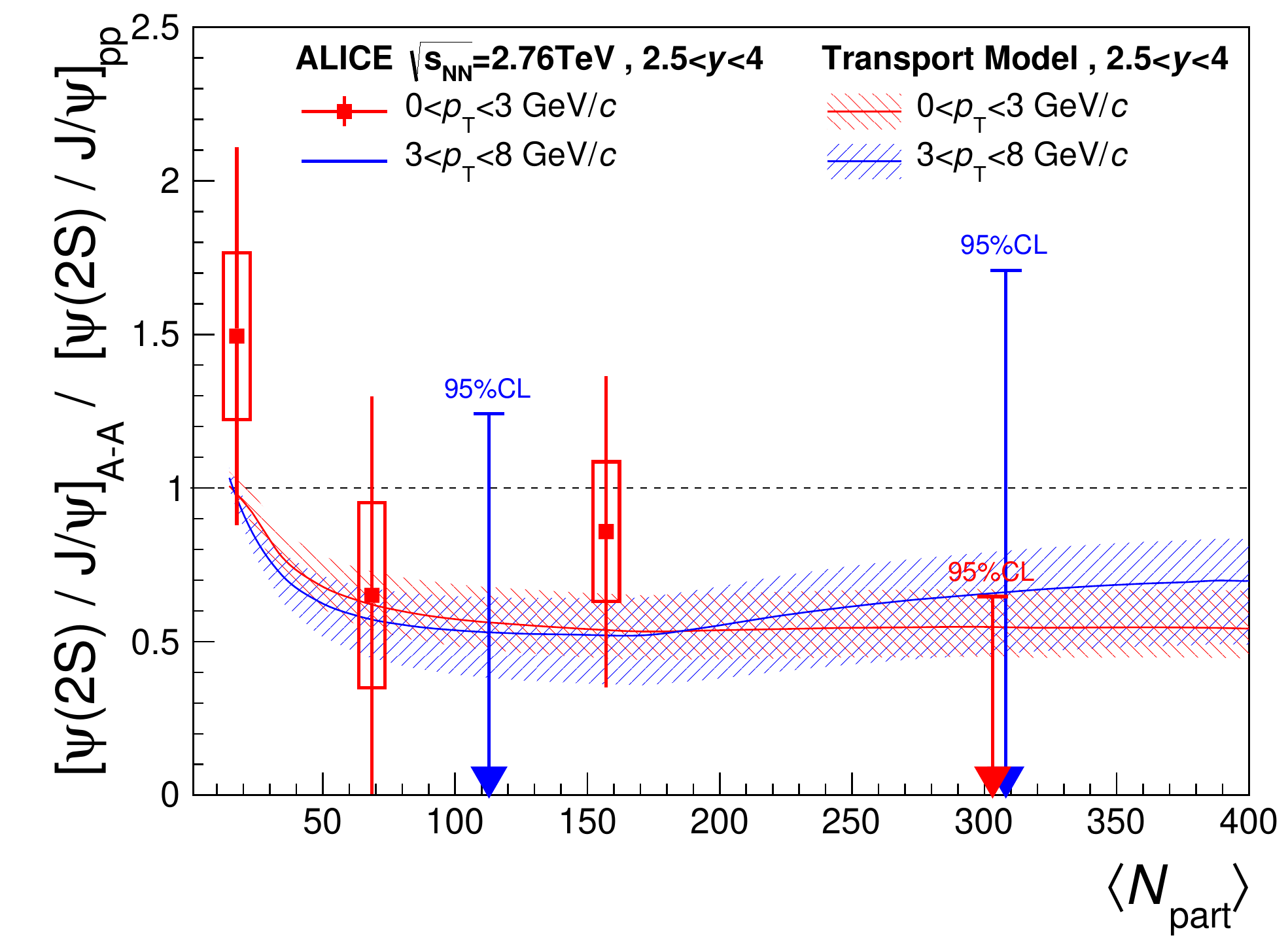}
\end{tabular}
\end{center}
\caption{\label{fig:psiprimecomparison} 
Inclusive \psiratio\ ratio  measured as a function of \Npart\  in \PbPb\ collisions at \sqrtSnnE{2.76} for two \pt\ intervals,  compared to NA50 results~\cite{Alessandro:2006ju} and to a theoretical calculation~\cite{Andronic:2009sv} (left). 
Double ratio, as a function of \Npart,  between the \psip\ and \jpsi\ measured in \PbPb\ at \sqrtSnnE{2.76} and \pp\ collisions  at \sqrtSE{7},   compared to  theoretical calculations~\cite{Chen:2013wmr} (right). }
\end{figure}

The ratio between  inclusive \psip\ and  \jpsi\  yields  measured in \PbPb\ collisions at \sqrtSnnE{2.76} is shown in the left side of Fig.~\ref{fig:psiprimecomparison}  as a function of \Npart.
In the interval $ \pt < 3\gevc$, the \psip\ signal was  extracted in three centrality classes (\cent{20}{40}, \cent{40}{60}\ and \cent{60}{90}) while only the 95\% confidence level upper limit was  established  for the centrality class \cent{0}{20}. 
At higher \pt, in the interval  $ 3 < \pt < 8\gevc $, the yield of \psip\ could not be extracted and the  95\% confidence level upper limit is shown for the \cent{0}{20}\ and \cent{20}{60}\ most central collisions. 

Our results are compared to the corresponding measurement at SPS energy (\sqrtSnnE{0.017}), performed in a region close to mid-rapidity ($0 < \y < 1 $)~\cite{Alessandro:2006ju}. Within the rather large uncertainties of our measurement, no clear \sqrtSnn\ or \y-dependence can be seen, in agreement with expectations from the SHM~\cite{Andronic:2009sv}. Prediction from the SHM for the prompt $\psiratio_{\PbPb}$ ratio at \sqrtSnnE{2.76} in our rapidity domain is also reported in Fig.~\ref{fig:psiprimecomparison}.

The double ratio $\psiratio_{\PbPb}/\psiratio_{\pp}$ is shown as a function of \Npart~in the right-hand side of Fig.~\ref{fig:psiprimecomparison}. 
Statistical uncertainties (including those coming from \PbPb\ and from the normalization to \pp) are shown as vertical bars, while systematic uncertainties are shown as open boxes. 
The results do not allow a firm conclusion since statistical fluctuations inside one standard deviation allow our data points to range between very low double ratios (strong \psip\ suppression with respect to \jpsi) to values higher than unity (less \psip\ suppression with respect to \jpsi). 
Nevertheless, the limit set for the lowest \pt\ bin for the \cent{0}{20}\ most central collisions points to a larger suppression of the \psip\ in that region.
A transport model calculation~\cite{Chen:2013wmr} for inclusive \psip~and \jpsi~production is  shown for the two \pt\ intervals considered.
The theoretical uncertainty band is due to  different choices of the quenching factor for the b-quark. 
A qualitative agreement  can be appreciated for both \pt\ intervals.

CMS  has measured the double ratio $\psiratio_{\PbPb}/\psiratio_{\pp}$  dependence on centrality \cite{Khachatryan:2014bva} for prompt \psip\ and \jpsi.
In the rapidity and transverse momentum intervals  $1.6 < |\y| < 2.4 $ and $ 3 < \pt < 30\gevc$ and for the \cent{0}{20}\ most central collisions,  a double ratio of  $2.31 \pm 0.53 (\mathrm{stat.})  \pm 0.37 (\mathrm{stat.}) \pm 0.15 (\mathrm{pp})$ is obtained.
This result sits at the upper edge of our  confidence limit estimated in the same centrality range for $2.5 < \y < 4$  and $ 3 < \pt < 8\gev/c$.
In more peripheral collisions, CMS results fall inside  the limits given by this analysis.

It is worth underlying that our result is for inclusive \psip~and \jpsi~production, while SHM predictions and CMS results are for prompt charmonia production.
The impact of the B-mesons feed-down  on the ratio was  extensively studied in~\cite{Chen:2013wmr}, showing a very strong influence of the non-prompt 
\psip\ component on the final result. According to this study, removing this non-prompt contribution would lead to a significantly lower double ratio at high 
\Npart: in the $ 0 < \pt < 3\gevc $ bin a 60\% decrease is expected, while in the $ 3 < \pt < 8\gevc $ bin the effect could be even stronger, leading to a 80\% decrease.

\section{Conclusions}

We have presented a study of \jpsi\ and \psip\ production in \PbPb\ collisions at \sqrtSnnE{2.76} in the transverse momentum and rapidity ranges $  \pt < 8\gevc$ and $2.5 < \y < 4$. 
This analysis was carried out in the muon spectrometer system, whose tracking and triggering capabilities were described in detail. 

The $\psiratio_{\PbPb}$ ratio   was measured in two \pt\ ranges as a function of centrality.
In some  intervals, only the  95\% confidence level upper limits could be obtained.   
The suppression pattern of the \psip\ is compatible with that of the  \jpsi\ in most of the centrality and \pt\ intervals studied. 
The large uncertainties leave open the possibility of  strong enhancement or suppression factors.
An accurate \psip\ measurement in \PbPb\ would require significantly more statistics
than the one presented in this analysis.

The \jpsi\ signal was  extracted as a function  of \pt, \y\ and the collision centrality. 
We have computed the \jpsi\  \meanpt\ and \meanptsq. 
The \jpsi\  \meanpt\ in  \PbPb\ collisions  decreases significantly (5$\sigma$ effect) from peripheral to central collisions.  
In addition we have studied \raa\ defined as the ratio of the   \jpsi\ \meanptsq\ measured in \PbPb\ and  \pp\ collisions at the same energy. 
The \raa\  exhibits a clear decrease as a function of centrality for  \PbPb\ collisions. 

The nuclear modification factor, \Raa, of inclusive \jpsi\ was  measured as a function of centrality.  
A constant suppression of about 40\% was observed for \Npart\ larger than 70~\cite{Abelev:2013ila}. 
New studies of the \jpsi\ suppression pattern as a function of centrality for three \pt\ ranges were presented. 
Above $\Npart \sim 150 $, the low \pt\  \jpsi\ \Raa\  clearly differs from the high \pt\  \jpsi\ \Raa\ and is about three times larger for $\Npart > 250 $, corresponding to a 3.9$\sigma$ separation. 
Complementary to this, the \pt\ dependence of the suppression pattern was  analysed for the different centrality classes. 
An increase of the inclusive \jpsi\ \Raa\ with decreasing \pt\ is observed below 5 \gevc\ in the most central \PbPb\ collisions (\cent{0}{20}), while no significant \pt\ dependence is seen in the most peripheral collisions (\cent{40}{90}).
As a function of rapidity, the results published in~\cite{Abelev:2013ila} 
show compatible \Raa\ values for $|\y| <  0.9 $  and $2.5<\y<~3$. 
For larger rapidity, a decreasing trend is visible. 

Comparisons of the \raa\ and \Raa\ measured in ALICE with lower energy experiments show significant differences. 
The decreasing trend of \raa\ observed as a function of centrality is opposite to NA50 and PHENIX measurements.   
The \Raa\ in the most central collisions is three times larger than the one measured by PHENIX, and the difference reaches a factor four in the \pt\ region below 1\gevc.   
If the suppression sources observed at lower energies, which were related to color screening in hot nuclear matter on top of CNM effects, are still present at the LHC, then other mechanisms compensating the \jpsi\ suppression are needed to explain the ALICE measurements.  This conclusion is further substantiated, in the region $ |\y| < 3$, by the comparison of the inclusive \jpsi\ \Raa\ measurements  as a function of \y\  to models implementing only CNM effects, which shows a qualitative agreement.

The inclusive \jpsi\ \raa\ and \Raa\ measurements  were also  compared to various theoretical calculations including  hot and cold  nuclear matter effects.
The hadronic part of the \jpsi\ \Raa\ was estimated when needed to allow for a direct comparison to models, which do not implement the \jpsi\ production mechanism at the origin  of the observed very low \pt\ excess~\cite{Adam:2015gba}.
All these models feature a full or partial \jpsi\ production from charm quarks recombination and are in fair agreement with the experimental results.
The transport models considered in this paper are also able to generate an amount of  \jpsi\ elliptic flow comparable to the one measured in ALICE~\cite{ALICE:2013xna}.  
The double differential studies of the inclusive \jpsi\ \Raa\ as a function of centrality and \pt\ brings new constraints to the models. 
Reproducing the suppression pattern in peripheral collisions for both low and high \pt\  \jpsi\  is challenging for all models. 
Some tensions also appear in describing the \Raa\ evolution at low \pt\ for all centrality classes. 
However, the uncertainties on the measurements on one side, and on the CNM and $ \mathrm{d} \sigma_{\ccbar}/ \mathrm{d} y$  in the theoretical calculations on the other side, do not allow for drawing a firm conclusion. 
The large uncertainties on the model predictions also show the limit of using the \Raa\ as an observable to measure the \jpsi\ suppression due to hot medium effects. 
Ideally one should, in \PbPb\ collisions, compare  the \jpsi\ production to the charm production to cancel out the  cold nuclear matter effects affecting the initial \ccbar\ dynamics.  
However, the measurement of the charm cross section in \PbPb\ collisions is very ambitious and still remains to be done at the LHC.  
  
To summarize, the \jpsi\  \raa\ and \Raa\  measured in \PbPb\ collisions at \sqrtSnnE{2.76} show  a new behavior with respect to measurements made at lower energies. 
In addition to the strong  \jpsi\ suppression observed at  high \pt,  ALICE results show that at low \pt\ a new contribution is necessary to explain the data. In  all available model calculations, this contribution is related to a recombination mechanism of  charm quarks.

\newenvironment{acknowledgement}{\relax}{\relax}
\begin{acknowledgement}
\section*{Acknowledgments}
The ALICE Collaboration would like to thank all its engineers and technicians for their invaluable contributions to the construction of the experiment and the CERN accelerator teams for the outstanding performance of the LHC complex.
The ALICE Collaboration gratefully acknowledges the resources and support provided by all Grid centres and the Worldwide LHC Computing Grid (WLCG) collaboration.
The ALICE Collaboration acknowledges the following funding agencies for their support in building and
running the ALICE detector:
State Committee of Science,  World Federation of Scientists (WFS)
and Swiss Fonds Kidagan, Armenia;
Conselho Nacional de Desenvolvimento Cient\'{\i}fico e Tecnol\'{o}gico (CNPq), Financiadora de Estudos e Projetos (FINEP),
Funda\c{c}\~{a}o de Amparo \`{a} Pesquisa do Estado de S\~{a}o Paulo (FAPESP);
National Natural Science Foundation of China (NSFC), the Chinese Ministry of Education (CMOE)
and the Ministry of Science and Technology of China (MSTC);
Ministry of Education and Youth of the Czech Republic;
Danish Natural Science Research Council, the Carlsberg Foundation and the Danish National Research Foundation;
The European Research Council under the European Community's Seventh Framework Programme;
Helsinki Institute of Physics and the Academy of Finland;
French CNRS-IN2P3, the `Region Pays de Loire', `Region Alsace', `Region Auvergne' and CEA, France;
German Bundesministerium fur Bildung, Wissenschaft, Forschung und Technologie (BMBF) and the Helmholtz Association;
General Secretariat for Research and Technology, Ministry of
Development, Greece;
Hungarian Orszagos Tudomanyos Kutatasi Alappgrammok (OTKA) and National Office for Research and Technology (NKTH);
Department of Atomic Energy and Department of Science and Technology of the Government of India;
Istituto Nazionale di Fisica Nucleare (INFN) and Centro Fermi -
Museo Storico della Fisica e Centro Studi e Ricerche "Enrico
Fermi", Italy;
MEXT Grant-in-Aid for Specially Promoted Research, Ja\-pan;
Joint Institute for Nuclear Research, Dubna;
National Research Foundation of Korea (NRF);
Consejo Nacional de Cienca y Tecnologia (CONACYT), Direccion General de Asuntos del Personal Academico (DGAPA), M\'{e}xico, Amerique Latine Formation academique – European Commission (ALFA-EC) and the EPLANET Program (European Particle Physics Latin American Network);
Stichting voor Fundamenteel Onderzoek der Materie (FOM) and the Nederlandse Organisatie voor Wetenschappelijk Onderzoek (NWO), Netherlands;
Research Council of Norway (NFR);
National Science Centre, Poland;
Ministry of National Education/Institute for Atomic Physics and Consiliul Naţional al Cercetării Ştiinţifice - Executive Agency for Higher Education Research Development and Innovation Funding (CNCS-UEFISCDI) - Romania;
Ministry of Education and Science of Russian Federation, Russian
Academy of Sciences, Russian Federal Agency of Atomic Energy,
Russian Federal Agency for Science and Innovations and The Russian
Foundation for Basic Research;
Ministry of Education of Slovakia;
Department of Science and Technology, South Africa;
Centro de Investigaciones Energeticas, Medioambientales y Tecnologicas (CIEMAT), E-Infrastructure shared between Europe and Latin America (EELA), Ministerio de Econom\'{i}a y Competitividad (MINECO) of Spain, Xunta de Galicia (Conseller\'{\i}a de Educaci\'{o}n),
Centro de Aplicaciones Tecnológicas y Desarrollo Nuclear (CEA\-DEN), Cubaenerg\'{\i}a, Cuba, and IAEA (International Atomic Energy Agency);
Swedish Research Council (VR) and Knut $\&$ Alice Wallenberg
Foundation (KAW);
Ukraine Ministry of Education and Science;
United Kingdom Science and Technology Facilities Council (STFC);
The United States Department of Energy, the United States National
Science Foundation, the State of Texas, and the State of Ohio;
Ministry of Science, Education and Sports of Croatia and  Unity through Knowledge Fund, Croatia;
Council of Scientific and Industrial Research (CSIR), New Delhi, India

\end{acknowledgement}

\bibliographystyle{utphys}
\bibliography{paper}

\appendix
\section{Data tables}
\label{app:datatables}
This appendix provides all the numerical values obtained in this analysis.

The inclusive \jpsi\ differential \pt\ yields in \PbPb\ in centrality classes are given in Tab.~\ref{tab:yieldptcent}. 
Tables~\ref{tab:raavscent} to \ref{tab:raavscent_pt58} present the inclusive \jpsi\ \Raa\ and associated \PbPb\ yields as a function of centrality for $ 2.5 < \y < 4.0 $ and  four \pt\ ranges,   $ \pt < 8\gevc$, $  \pt \leq 2\gevc$,  $ 2 < \pt < 5\gevc$  and $ 5 < \pt < 8\gevc$.
Tables~\ref{tab:raavspt020} to \ref{tab:raavspt090} show the  \pt\ dependence of the inclusive \jpsi\ \Raa\ and associated \PbPb\ yields for the centrality classes \cent{0}{20}, \cent{20}{40}, \cent{0}{40}, \cent{40}{90} and \cent{0}{90}. 
Table~\ref{tab:raavsy} shows the  \y\ dependence of the inclusive \jpsi\ \Raa\ and associated \PbPb\ yields for the centrality class \cent{0}{90} in the  \pt\ range   $ \pt < 8\gevc$.
Then, the inclusive \jpsi\ \Raa\ results with a low \pt\ cut at 0.3 \gevc\ are presented. The reference \pp\ cross section needed to build the \Raa\ was extracted with the method described in~\cite{Adam:2015gba}.
The inclusive \jpsi\ \Raa\  centrality dependence for $ 2.5 < y < 4$ in the \pt\ ranges $0.3 < \pt < 8\gevc$ and $0.3 < \pt < 2\gevc$ is shown in Tab.~\ref{tab:raavscent_300mevcut}. 
The inclusive \jpsi\ \Raa\ in the \pt\ range $0.3 < \pt < 1\gevc$  for $ 2.5 < y < 4$ in four  centrality classes \cent{0}{90}, \cent{0}{20}, \cent{20}{40} and \cent{40}{90} is given in Tab.~\ref{tab:raavspt_300mevcut}.
Finally, Tab.~\ref{tab:psi2res} presents the inclusive $\psiratio_{\PbPb}$ and $\psiratio_{\PbPb} \; / \; \psiratio_{\pp}$ ratios as a function of centrality for the \pt\ intervals $ \pt < 3\gevc$ and $ 3 < \pt < 8\gevc$.

\begin{table}[h!]
\begin{center}
\begin{tabular}{c|c|c|c} 
\hline
\hline 
  \multicolumn{4}{c}{ $ \mathrm{d}^{2} Y_{\jpsi} / \mathrm{d} \y \mathrm{d} \pt (\gevc)^{-1} \, \times 10^{3}$ } \\
\hline
    \pt\ (\gevc) &    \cent{0}{20}    &       \cent{20}{40}      &     \cent{40}{90}         \\ 
\hline
0.0--0.5   &  $3.253 \pm 0.386 \pm 0.446$ 			  &  $1.366 \pm 0.081 \pm 0.165$ 		      &  $0.257 \pm 0.017 \pm 0.031$ \\
0.5--1.0   &  $8.012 \pm 0.487 \pm 1.087$ 			  &  $2.571 \pm 0.199 \pm 0.310$ 			  &  $0.346 \pm 0.024 \pm 0.042$ \\ 
1.0--1.5   &  $9.909 \pm 0.603 \pm 1.149$ 			  &  $3.494 \pm 0.255 \pm 0.388$ 			  &  $0.533 \pm 0.030 \pm 0.061$ \\ 
1.5--2.0   &  $8.193 \pm 0.505 \pm 0.907$ 			  &  $2.907 \pm 0.194 \pm 0.320$ 			  &  $0.493 \pm 0.037 \pm 0.053$ \\ 
2.0--2.5   &  $6.342 \pm 0.401 \pm 0.701$ 			  &  $2.371 \pm 0.164 \pm 0.260$ 			  &  $0.441 \pm 0.034 \pm 0.049$ \\ 
2.5--3.0   &  $4.759 \pm 0.316 \pm 0.542$ 			  &  $1.997 \pm 0.134 \pm 0.227$ 			  &  $0.270 \pm 0.020 \pm 0.029$ \\ 
3.0--3.5   &  $2.735 \pm 0.183 \pm 0.290$ 			  &  $1.313 \pm 0.087 \pm 0.151$ 			  &  $0.222 \pm 0.016 \pm 0.023$ \\ 
3.5--4.0   &  $1.876 \pm 0.134 \pm 0.201$ 			  &  $0.874 \pm 0.068 \pm 0.092$ 			  &  $0.174 \pm 0.013 \pm 0.018$ \\ 
4.0--4.5   &  $1.075 \pm 0.098 \pm 0.109$ 			  &  $0.483 \pm 0.037 \pm 0.048$ 			  &  $0.108 \pm 0.009 \pm 0.011$ \\ 
4.5--5.0   &  $0.731 \pm 0.069 \pm 0.073$ 			  &  $0.339 \pm 0.030 \pm 0.033$ 			  &  $0.076 \pm 0.007 \pm 0.007$ \\ 
5.0--5.5   &  $0.453 \pm 0.047 \pm 0.045$ 			  &  $0.263 \pm 0.023 \pm 0.026$ 			  &  $0.042 \pm 0.005 \pm 0.004$ \\ 
5.5--6.0   &  $0.345 \pm 0.039 \pm 0.046$ 			  &  $0.132 \pm 0.016 \pm 0.014$ 			  &  $0.028 \pm 0.004 \pm 0.003$ \\ 
6.0--8.0   &  $0.099 \pm 0.009 \pm 0.010$ 			  &  $0.068 \pm 0.005 \pm 0.007$ 			  &  $0.012 \pm 0.001 \pm 0.001$ \\ 
\hline    
\end{tabular}
\end{center}
\caption{\label{tab:yieldptcent} Inclusive \jpsi\ yields (as defined by Eq.~\ref{eq:jpsiyield})  in  \pt\  intervals for the \cent{0}{20}, \cent{20}{40}  and \cent{40}{90} most central \PbPb\ collisions.  
The rapidity range is $2.5 < \y < 4$. 
Statistical and systematic uncertainties are also reported as $ \mathrm{d}^{2} Y_{\jpsi} / \mathrm{d} \y \mathrm{d} \pt  \pm {\rm statistical~uncertainty} \pm {\rm systematic~uncertainty} $.
A global systematic uncertainty of 4\% affects all the values. 
A 2\%, 1\% and 2\% systematic uncertainty, independent of \pt,  affects the centrality  classes  \cent{0}{20}, \cent{20}{40}  and \cent{40}{90}, respectively. 
}
\end{table}

\begin{table}[h!]
\centering
\begin{tabular}{c|c|c}
\hline
Centrality   &   $\Raa \pm {\rm (stat.)} \pm {\rm (syst.)} $~\cite{Abelev:2013ila}&  $Y_{\jpsi} \pm {\rm (stat.)} \pm {\rm (syst.)} \times 10^{3}$\\
\hline
\cent{0}{10} 	  & $0.557 \pm 0.019 \pm 0.024$   & $43.095 \pm 1.454 \pm 1.049$ \\
\cent{10}{20} 	  & $0.573 \pm 0.020 \pm 0.022$   & $27.212 \pm 0.979 \pm 0.501$ \\
\cent{20}{30} 	  & $0.598 \pm 0.022 \pm 0.020$   & $17.409 \pm 0.638 \pm 0.188$ \\ 
\cent{30}{40} 	  & $0.577 \pm 0.024 \pm 0.025$   & $9.671 \pm 0.406 \pm 0.211$ \\ 
\cent{40}{50} 	  & $0.609 \pm 0.028 \pm 0.030$   & $5.413 \pm 0.247 \pm 0.041$ \\ 
\cent{50}{60} 	  & $0.725 \pm 0.036 \pm 0.043$   & $3.246 \pm 0.160 \pm 0.050$ \\ 
\cent{60}{70} 	  & $0.839 \pm 0.041 \pm 0.058$   & $1.677 \pm 0.083 \pm 0.024$ \\ 
\cent{70}{80} 	  & $0.849 \pm 0.063 \pm 0.068$   & $0.701 \pm 0.051 \pm 0.014$ \\ 
\cent{80}{90}     & $1.094 \pm 0.106 \pm 0.104$   & $0.362 \pm 0.033 \pm 0.008$ \\ 
\hline
\end{tabular}
\caption{\label{tab:raavscent} Inclusive  \jpsi\ \Raa\ and \PbPb\ yields as a function of centrality,  for $ \pt < 8\gevc$ and $ 2.5 < \y < 4.0$.
Statistical and systematic uncertainties are also reported.
A global systematic uncertainty of 15\%  (12\%)  affects all the  \Raa\ (yields) values.}  		
\end{table} 

\begin{table}[h!]
\centering
\begin{tabular}{c|c|c}
\hline
Centrality   &   $\Raa \pm {\rm (stat.)} \pm {\rm (syst.)} $  & $Y_{\jpsi} \pm {\rm (stat.)} \pm {\rm (syst.)} \times 10^{3}$\\
\hline
\cent{0}{10} 	  & $0.732 \pm 0.034 \pm 0.041$   & $27.932 \pm 1.302 \pm 1.282$ \\ 
\cent{10}{20} 	  & $0.733 \pm 0.035 \pm 0.028$   & $17.159 \pm 0.824 \pm 0.383$ \\ 
\cent{20}{30} 	  & $0.715 \pm 0.038 \pm 0.024$   & $10.113 \pm 0.541 \pm 0.115$ \\ 
\cent{30}{40} 	  & $0.678 \pm 0.040 \pm 0.033$   & $5.516 \pm 0.322 \pm 0.182$ \\ 
\cent{40}{50} 	  & $0.641 \pm 0.044 \pm 0.032$   & $2.789 \pm 0.190 \pm 0.064$ \\ 
\cent{50}{60} 	  & $0.839 \pm 0.048 \pm 0.056$   & $1.799 \pm 0.103 \pm 0.070$ \\ 
\cent{60}{90} 	  & $1.104 \pm 0.064 \pm 0.078$   & $0.559 \pm 0.032 \pm 0.016$ \\ 
\hline
\end{tabular}
\caption{\label{tab:raavscent_pt02} Inclusive  \jpsi\ \Raa\ and \PbPb\ yields as a function of centrality, for  $ \pt < 2\gevc$ and $ 2.5 < \y < 4.0$.
Statistical and systematic uncertainties are also reported.
A global systematic uncertainty of 15\%  (12\%)  affects all the \Raa\ (yields) values.
}
\end{table} 

\begin{table}[h!]
\centering
\begin{tabular}{c|c|c}
\hline
Centrality   &   $\Raa \pm {\rm (stat.)} \pm {\rm (syst.)} $  & $Y_{\jpsi} \pm {\rm (stat.)} \pm {\rm (syst.)} \times 10^{3}$\\
\hline
\cent{0}{10} 	  & $0.425 \pm 0.019 \pm 0.017$   & $15.540 \pm 0.681 \pm 0.379$ \\ 
\cent{10}{20} 	  & $0.461 \pm 0.019 \pm 0.016$   & $10.336 \pm 0.431 \pm 0.168$ \\ 
\cent{20}{30} 	  & $0.529 \pm 0.022 \pm 0.018$   & $7.164 \pm 0.293 \pm 0.106$ \\ 
\cent{30}{40} 	  & $0.498 \pm 0.025 \pm 0.027$   & $3.879 \pm 0.194 \pm 0.153$ \\ 
\cent{40}{50} 	  & $0.595 \pm 0.030 \pm 0.029$   & $2.481 \pm 0.126 \pm 0.049$ \\ 
\cent{50}{60} 	  & $0.675 \pm 0.042 \pm 0.041$   & $1.386 \pm 0.085 \pm 0.037$ \\ 
\cent{60}{90} 	  & $0.722 \pm 0.044 \pm 0.050$   & $0.350 \pm 0.021 \pm 0.009$ \\
\hline
\end{tabular}
\caption{\label{tab:raavscent_pt25} Inclusive  \jpsi\ \Raa\ and \PbPb\ yields as a function of centrality,  for $ 2 < \pt < 5\gevc$ and $ 2.5 < \y < 4.0 $.
Statistical and systematic uncertainties are also reported.
A global systematic uncertainty of 14\%  (11\%)  affects all the \Raa\ (yields) values.
}
\end{table} 

\begin{table}[h!]
\centering
\begin{tabular}{c|c|c}
\hline
Centrality   &   $\Raa \pm {\rm (stat.)} \pm {\rm (syst.)} $  &  $Y_{\jpsi} \pm {\rm (stat.)} \pm {\rm (syst.)} \times 10^{3}$\\
\hline
\cent{0}{10} 	  & $0.280 \pm 0.021 \pm 0.011$   & $1.093 \pm 0.081 \pm 0.027$ \\ 
\cent{10}{20} 	  & $0.282 \pm 0.027 \pm 0.011$   & $0.677 \pm 0.064 \pm 0.016$ \\ 
\cent{20}{30} 	  & $0.410 \pm 0.029 \pm 0.013$   & $0.594 \pm 0.042 \pm 0.006$ \\ 
\cent{30}{40} 	  & $0.540 \pm 0.039 \pm 0.024$   & $0.449 \pm 0.033 \pm 0.012$ \\ 
\cent{40}{50} 	  & $0.529 \pm 0.053 \pm 0.031$   & $0.236 \pm 0.024 \pm 0.009$ \\ 
\cent{50}{60} 	  & $0.587 \pm 0.073 \pm 0.036$   & $0.129 \pm 0.016 \pm 0.004$ \\ 
\cent{60}{90} 	  & $0.644 \pm 0.083 \pm 0.047$   & $0.033 \pm 0.004 \pm 0.001$ \\ 
\hline
\end{tabular}
\caption{\label{tab:raavscent_pt58} Inclusive  \jpsi\ \Raa\ and \PbPb\ yields as a function of centrality, for  $ 5 < \pt < 8\gevc$ and $ 2.5 < \y < 4.0 $.
Statistical and systematic uncertainties are also reported.
A global systematic uncertainty of 18\%  (10\%)  affects all the \Raa\ (yields) values.
}
\end{table} 

\begin{table}[h!]
\centering
\begin{tabular}{c|c|c} 
\hline
\pt\ (\gevc)  &   $\Raa \pm {\rm (stat.)} \pm {\rm (syst.)} $ ~\cite{Abelev:2013ila}&  $ \mathrm{d}^{2} Y_{\jpsi} / \mathrm{d} \y \mathrm{d} \pt  \pm {\rm (stat.)} \pm {\rm (syst.) }   (\gevc)^{-1} \, \times 10^{3}$\\
\hline
0--1	  & $0.803 \pm 0.084 \pm 0.113$   & $5.771 \pm 0.345 \pm 0.748$ \\ 
1--2	  & $0.690 \pm 0.052 \pm 0.084$   & $9.134 \pm 0.411 \pm 0.987$ \\ 
2--3	  & $0.505 \pm 0.042 \pm 0.062$   & $5.539 \pm 0.284 \pm 0.604$ \\ 
3--4	  & $0.381 \pm 0.037 \pm 0.046$   & $2.305 \pm 0.116 \pm 0.247$ \\ 
4--5	  & $0.355 \pm 0.052 \pm 0.041$   & $0.905 \pm 0.068 \pm 0.090$ \\ 
5--6	  & $0.282 \pm 0.048 \pm 0.032$   & $0.388 \pm 0.030 \pm 0.038$ \\ 
6--8	  & $0.279 \pm 0.064 \pm 0.032$   & $0.100 \pm 0.009 \pm 0.010$ \\ 
\hline
\end{tabular}	
\caption{\label{tab:raavspt020} Inclusive  \jpsi\ \Raa\  and \PbPb\ yields as a function of \pt\ for the \cent{0}{20} centrality class and $ 2.5 < \y < 4.0 $.
Statistical and systematic uncertainties are also reported.
A global systematic uncertainty of 8\%  (4\%)  affects all the \Raa\ (yields) values.
}	
\end{table} 

\begin{table}[h!]
\centering
\begin{tabular}{c|c|c}
\hline
 \pt\ (\gevc)  &   $\Raa \pm {\rm (stat.)} \pm {\rm (syst.)} $  &  $ \mathrm{d}^{2} Y_{\jpsi} / \mathrm{d} \y \mathrm{d} \pt  \pm {\rm (stat.)} \pm {\rm (syst.)}   (\gevc)^{-1} \, \times 10^{3}$\\
\hline
0--1	  & $0.733 \pm 0.080 \pm 0.097$   & $1.909 \pm 0.128 \pm 0.229$ \\ 
1--2	  & $0.660 \pm 0.051 \pm 0.080$   & $3.189 \pm 0.154 \pm 0.344$ \\ 
2--3	  & $0.543 \pm 0.044 \pm 0.067$   & $2.167 \pm 0.106 \pm 0.238$ \\ 
3--4	  & $0.493 \pm 0.048 \pm 0.060$   & $1.084 \pm 0.055 \pm 0.117$ \\ 
4--5	  & $0.444 \pm 0.063 \pm 0.051$   & $0.411 \pm 0.027 \pm 0.040$ \\ 
5--6	  & $0.399 \pm 0.067 \pm 0.045$   & $0.200 \pm 0.014 \pm 0.020$ \\ 
6--8	  & $0.523 \pm 0.116 \pm 0.059$   & $0.068 \pm 0.005 \pm 0.007$ \\ 
\hline
\end{tabular}
\caption{\label{tab:raavspt2040} Inclusive  \jpsi\ \Raa\  and \PbPb\ yields as a function of \pt\ for the \cent{20}{40} centrality class and $ 2.5 < \y < 4.0 $.
Statistical and systematic uncertainties are also reported.
A global systematic uncertainty of 8\%  (4\%)  affects all the \Raa\ (yields) values.
}
\end{table} 

\begin{table}[h!]
\centering
\begin{tabular}{c|c|c} 
\hline
\pt\ (\gevc)  &   $\Raa \pm {\rm (stat.)} \pm {\rm (syst.)} $  &  $ \mathrm{d}^{2} Y_{\jpsi} / \mathrm{d} \y \mathrm{d} \pt  \pm {\rm (stat.)} \pm {\rm (syst.)}   (\gevc)^{-1} \, \times 10^{3}$\\
\hline
0--1	  & $0.767 \pm 0.074 \pm 0.105$   & $3.754 \pm 0.163 \pm 0.472$ \\ 
1--2	  & $0.672 \pm 0.046 \pm 0.082$   & $6.103 \pm 0.212 \pm 0.662$ \\ 
2--3	  & $0.515 \pm 0.038 \pm 0.064$   & $3.865 \pm 0.134 \pm 0.428$ \\ 
3--4	  & $0.411 \pm 0.038 \pm 0.049$   & $1.698 \pm 0.063 \pm 0.178$ \\ 
4--5	  & $0.376 \pm 0.051 \pm 0.043$   & $0.655 \pm 0.033 \pm 0.064$ \\ 
5--6	  & $0.315 \pm 0.050 \pm 0.036$   & $0.296 \pm 0.016 \pm 0.029$ \\ 
6--8	  & $0.340 \pm 0.075 \pm 0.038$   & $0.083 \pm 0.005 \pm 0.008$ \\ 
\hline
\end{tabular}
\caption{\label{tab:raavspt040} Inclusive  \jpsi\ \Raa\  and \PbPb\ yields as a function of \pt\ for the \cent{0}{40}  centrality class and $ 2.5 < \y < 4.0 $.
Statistical and systematic uncertainties are also reported.
A global systematic uncertainty of 8\%  (4\%)  affects all the \Raa\ (yields) values.
}	
\end{table} 

\begin{table}[h!]
\centering
\begin{tabular}{c|c|c}
\hline
\pt\ (\gevc)  &   $\Raa \pm {\rm (stat.)} \pm {\rm (syst.)} $  &  $ \mathrm{d}^{2} Y_{\jpsi} / \mathrm{d} \y \mathrm{d} \pt  \pm {\rm (stat.)} \pm {\rm (syst.)}   (\gevc)^{-1} \, \times 10^{3}$\\
\hline
0--1	  & $0.815 \pm 0.081 \pm 0.107$   & $0.305 \pm 0.015 \pm 0.036$ \\ 
1--2	  & $0.732 \pm 0.059 \pm 0.090$   & $0.508 \pm 0.028 \pm 0.055$ \\ 
2--3	  & $0.617 \pm 0.053 \pm 0.076$   & $0.354 \pm 0.020 \pm 0.038$ \\ 
3--4	  & $0.627 \pm 0.062 \pm 0.074$   & $0.198 \pm 0.010 \pm 0.020$ \\ 
4--5	  & $0.693 \pm 0.097 \pm 0.079$   & $0.092 \pm 0.006 \pm 0.009$ \\ 
5--6	  & $0.489 \pm 0.087 \pm 0.055$   & $0.035 \pm 0.003 \pm 0.003$ \\ 
6--8	  & $0.646 \pm 0.150 \pm 0.072$   & $0.012 \pm 0.001 \pm 0.001$ \\ 
\hline
\end{tabular}
\caption{\label{tab:raavspt4090} Inclusive  \jpsi\ \Raa\  and \PbPb\ yields as a function of \pt\ for the \cent{40}{90} centrality class and $ 2.5 < \y < 4.0 $.
Statistical and systematic uncertainties are also reported.
A global systematic uncertainty of 9\%  (4\%)  affects all the \Raa\ (yields) values.
}
\end{table} 

\begin{table}[h!]
\centering
\begin{tabular}{c|c|c}
\hline
\pt\ (\gevc)  &   $\Raa \pm {\rm (stat.)} \pm {\rm (syst.)} $~\cite{Abelev:2013ila}  &  $ \mathrm{d}^{2} Y_{\jpsi} / \mathrm{d} \y \mathrm{d} \pt  \pm {\rm (stat.)} \pm {\rm (syst.)}   (\gevc)^{-1} \, \times 10^{3}$\\
\hline
0--1	  & $0.779 \pm 0.076 \pm 0.106$   & $1.857 \pm 0.081 \pm 0.230$ \\ 
1--2	  & $0.677 \pm 0.047 \pm 0.083$   & $2.993 \pm 0.104 \pm 0.323$ \\ 
2--3	  & $0.519 \pm 0.038 \pm 0.064$   & $1.896 \pm 0.064 \pm 0.206$ \\ 
3--4	  & $0.425 \pm 0.039 \pm 0.051$   & $0.855 \pm 0.029 \pm 0.089$ \\ 
4--5	  & $0.405 \pm 0.054 \pm 0.047$   & $0.343 \pm 0.015 \pm 0.033$ \\ 
5--6	  & $0.322 \pm 0.052 \pm 0.036$   & $0.147 \pm 0.007 \pm 0.015$ \\ 
6--8	  & $0.364 \pm 0.079 \pm 0.041$   & $0.043 \pm 0.002 \pm 0.004$ \\ 
\hline
\end{tabular}
\caption{\label{tab:raavspt090} Inclusive  \jpsi\ \Raa\  and \PbPb\ yields as a function of \pt\ for the \cent{0}{90}  centrality class and $ 2.5 < \y < 4.0 $.
Statistical and systematic uncertainties are also reported.
A global systematic uncertainty of 8\%  (4\%)  affects all the \Raa\ (yields) values.
} 	
\end{table} 

\begin{table}[h!]
\centering
\begin{tabular}{c|c|c}
\hline
 \y    &   $\Raa \pm {\rm (stat.)} \pm {\rm (syst.)} $~\cite{Abelev:2013ila}  & $ \mathrm{d}^{2} Y_{\jpsi} / \mathrm{d} \y \mathrm{d} \pt  \pm {\rm (stat.)} \pm {\rm (syst.)}   (\gevc)^{-1} \, \times 10^{3}$\\
\hline
2.50--2.75	  & $0.631 \pm 0.087 \pm 0.088$   & $1.509 \pm 0.114 \pm 0.191$ \\ 
2.75--3.00	  & $0.747 \pm 0.068 \pm 0.097$   & $1.387 \pm 0.058 \pm 0.162$ \\ 
3.00--3.25	  & $0.632 \pm 0.048 \pm 0.094$   & $1.120 \pm 0.039 \pm 0.154$ \\ 
3.25--3.50	  & $0.566 \pm 0.044 \pm 0.088$   & $0.891 \pm 0.032 \pm 0.130$ \\ 
3.50--3.75	  & $0.467 \pm 0.041 \pm 0.070$   & $0.733 \pm 0.025 \pm 0.101$ \\ 
3.75--4.00	  & $0.395 \pm 0.050 \pm 0.050$   & $0.528 \pm 0.029 \pm 0.058$ \\
\hline
\end{tabular}
\caption{\label{tab:raavsy} Inclusive  \jpsi\ \Raa\  and \PbPb\ yields as a function of \y\ for the \cent{0}{90}  centrality class and $  \pt < 8\gevc$.
Statistical and systematic uncertainties are also reported.
A global systematic uncertainty of 8\%  (4\%)  affects all the \Raa\ (yields) values.
} 	
\end{table} 

\begin{table}[h!]
\centering
\begin{tabular}{c|c|c}
\hline
\hline
             &  \multicolumn{2}{c}{ $\Raa \pm {\rm (stat.)} \pm {\rm (syst.)}$ } \\
\hline
Centrality   &  $ 0.3 < \pt < 8\gevc$ &  $ 0.3 < \pt < 2\gevc$\\
\hline
\cent{0}{10} 	  & $0.545 \pm 0.017 \pm 0.026$   & $0.745 \pm 0.041 \pm 0.042$ \\
\cent{10}{20} 	  & $0.560 \pm 0.018 \pm 0.021$   & $0.736 \pm 0.036 \pm 0.028$ \\
\cent{20}{30} 	  & $0.594 \pm 0.020 \pm 0.020$   & $0.716 \pm 0.038 \pm 0.025$ \\ 
\cent{30}{40} 	  & $0.570 \pm 0.021 \pm 0.025$   & $0.671 \pm 0.040 \pm 0.032$ \\ 
\cent{40}{50} 	  & $0.592 \pm 0.025 \pm 0.029$   & $0.619 \pm 0.045 \pm 0.032$ \\ 
\cent{50}{60} 	  & $0.715 \pm 0.033 \pm 0.044$   & $0.801 \pm 0.049 \pm 0.054$ \\ 
\cent{60}{70} 	  & $0.805 \pm 0.043 \pm 0.057$   & \multirow{3}{*}{$\Biggr\rbrace  0.959 \pm 0.057 \pm 0.067$} \\ 
\cent{70}{80} 	  & $0.778 \pm 0.062 \pm 0.064$   & \\ 
\cent{80}{90}     & $0.887 \pm 0.097 \pm 0.088$   & \\ 
\hline
\end{tabular}
\caption{\label{tab:raavscent_300mevcut} 
Inclusive \jpsi\ \Raa\ as a function of centrality, for  $ 0.3 < \pt < 8\gevc$  and $ 0.3 < \pt < 2\gevc$ in the rapidity range $ 2.5 < \y < 4.0 $.
Statistical and systematic uncertainties are also reported.
A global systematic uncertainty of 15\%  affects all the  \Raa\  values.
}
\end{table} 

\begin{table}[h!]
\centering
\begin{tabular}{c|c} 
\hline
\hline
Centrality      &   $\Raa \pm {\rm (stat.)} \pm {\rm (syst.)}$ for  $0.3 < \pt < 1\gevc$ \\
\hline
\cent{0}{90}    &   $0.775 \pm 0.057 \pm 0.113$   \\
\hline
\cent{0}{20}    &   $0.803 \pm 0.066 \pm 0.123$   \\
\cent{20}{40}   &   $0.733 \pm 0.067 \pm 0.103$   \\
\cent{40}{90}   &   $0.688 \pm 0.057 \pm 0.098$   \\
\hline
\end{tabular}	
\caption{\label{tab:raavspt_300mevcut} 
Inclusive  \jpsi\ \Raa\  for $ 2.5 < \y < 4.0 $ in the centrality classes  \cent{0}{90}, \cent{0}{20}, \cent{20}{40} and \cent{40}{90} for the lowest \pt\ range when the $0.3\gevc$ \pt\ cut is applied.
Statistical and systematic uncertainties are also reported.
A global systematic uncertainty of 8\%, 8\%,  8\% and 9\%  affect the \Raa\ values, respectively.
}
\end{table} 

\begin{table}[h!]
\begin{center}
\begin{tabular}{c|c|c|c}
\hline
\hline
 \pt\ (\gevc)	&  Centrality	& $\psiratio_{\PbPb}$ & $\psiratio_{\PbPb} \; / \; \psiratio_{\pp}$         \\
 \hline
 0--3  	&  \cent{0}{20}  	&  $ < 0.012$ (95\% CL)          & $ < 0.65 $ (95\% CL) \\
 0--3  	&  \cent{20}{40}  	&  $ 0.017 \pm 0.010 \pm 0.004$  & $ 0.86 \pm 0.51 \pm 0.23 $ \\
 0--3  	&  \cent{40}{60}  	&  $ 0.013 \pm 0.012 \pm 0.006$  & $ 0.65 \pm 0.65 \pm 0.30 $ \\
 0--3  	&  \cent{60}{90}  	&  $ 0.029 \pm 0.012 \pm 0.004$  & $ 1.49 \pm 0.62 \pm 0.27 $ \\
 3--8 	&  \cent{0}{20} 	&  $ < 0.046$ (95\% CL)          & $ < 1.71 $ (95\% CL) \\
 3--8  	&  \cent{20}{60}  	&  $ < 0.033$ (95\% CL)          & $ < 1.24 $ (95\% CL) \\
\hline    
\end{tabular}
\end{center}
\caption{\label{tab:psi2res} Inclusive $\psiratio_{\PbPb}$ and $\psiratio_{\PbPb} \; / \; \psiratio_{\pp}$ ratios as a function of centrality for two \pt\ intervals.
Statistical and systematic uncertainties are reported when the value is not given as an upper limit.
}
\end{table}

\clearpage
\newpage
\section{The ALICE Collaboration}
\label{app:collab}



\begingroup
\small
\begin{flushleft}
J.~Adam\Irefn{org40}\And
D.~Adamov\'{a}\Irefn{org83}\And
M.M.~Aggarwal\Irefn{org87}\And
G.~Aglieri Rinella\Irefn{org36}\And
M.~Agnello\Irefn{org111}\And
N.~Agrawal\Irefn{org48}\And
Z.~Ahammed\Irefn{org132}\And
S.U.~Ahn\Irefn{org68}\And
I.~Aimo\Irefn{org94}\textsuperscript{,}\Irefn{org111}\And
S.~Aiola\Irefn{org137}
% ...
\renewcommand\labelenumi{\textsuperscript{\theenumi}~}

\section*{Affiliation notes}
\renewcommand\theenumi{\roman{enumi}}
\begin{Authlist}
\item \Adef{0}Deceased
\item \Adef{idp3817696}{Also at: M.V. Lomonosov Moscow State University, D.V. Skobeltsyn Institute of Nuclear, Physics, Moscow, Russia}
\item \Adef{idp5959776}{Also at: University of Kansas, Lawrence, Kansas, United States}
\end{Authlist}

\section*{Collaboration Institutes}
\renewcommand\theenumi{\arabic{enumi}~}
\begin{Authlist}

\item \Idef{org1}A.I. Alikhanyan National Science Laboratory (Yerevan Physics Institute) Foundation, Yerevan, Armenia
\item \Idef{org2}Benem\'{e}rita Universidad Aut\'{o}noma de Puebla, Puebla, Mexico
\item \Idef{org3}Bogolyubov Institute for Theoretical Physics, Kiev, Ukraine
\item \Idef{org4}Bose Institute, Department of Physics and Centre for Astroparticle Physics and Space Science (CAPSS), Kolkata, India
\item \Idef{org5}Budker Institute for Nuclear Physics, Novosibirsk, Russia
\item \Idef{org6}California Polytechnic State University, San Luis Obispo, California, United States
\item \Idef{org7}Central China Normal University, Wuhan, China
\item \Idef{org8}Centre de Calcul de l'IN2P3, Villeurbanne, France
\item \Idef{org9}Centro de Aplicaciones Tecnol\'{o}gicas y Desarrollo Nuclear (CEADEN), Havana, Cuba
\item \Idef{org10}Centro de Investigaciones Energ\'{e}ticas Medioambientales y Tecnol\'{o}gicas (CIEMAT), Madrid, Spain
\item \Idef{org11}Centro de Investigaci\'{o}n y de Estudios Avanzados (CINVESTAV), Mexico City and M\'{e}rida, Mexico
\item \Idef{org12}Centro Fermi - Museo Storico della Fisica e Centro Studi e Ricerche ``Enrico Fermi'', Rome, Italy
\item \Idef{org13}Chicago State University, Chicago, Illinois, USA
\item \Idef{org14}China Institute of Atomic Energy, Beijing, China
\item \Idef{org15}Commissariat \`{a} l'Energie Atomique, IRFU, Saclay, France
\item \Idef{org16}COMSATS Institute of Information Technology (CIIT), Islamabad, Pakistan
\item \Idef{org17}Departamento de F\'{\i}sica de Part\'{\i}culas and IGFAE, Universidad de Santiago de Compostela, Santiago de Compostela, Spain
\item \Idef{org18}Department of Physics and Technology, University of Bergen, Bergen, Norway
\item \Idef{org19}Department of Physics, Aligarh Muslim University, Aligarh, India
\item \Idef{org20}Department of Physics, Ohio State University, Columbus, Ohio, United States
\item \Idef{org21}Department of Physics, Sejong University, Seoul, South Korea
\item \Idef{org22}Department of Physics, University of Oslo, Oslo, Norway
\item \Idef{org23}Dipartimento di Elettrotecnica ed Elettronica del Politecnico, Bari, Italy
\item \Idef{org24}Dipartimento di Fisica dell'Universit\`{a} 'La Sapienza' and Sezione INFN Rome, Italy
\item \Idef{org25}Dipartimento di Fisica dell'Universit\`{a} and Sezione INFN, Cagliari, Italy
\item \Idef{org26}Dipartimento di Fisica dell'Universit\`{a} and Sezione INFN, Trieste, Italy
\item \Idef{org27}Dipartimento di Fisica dell'Universit\`{a} and Sezione INFN, Turin, Italy
\item \Idef{org28}Dipartimento di Fisica e Astronomia dell'Universit\`{a} and Sezione INFN, Bologna, Italy
\item \Idef{org29}Dipartimento di Fisica e Astronomia dell'Universit\`{a} and Sezione INFN, Catania, Italy
\item \Idef{org30}Dipartimento di Fisica e Astronomia dell'Universit\`{a} and Sezione INFN, Padova, Italy
\item \Idef{org31}Dipartimento di Fisica `E.R.~Caianiello' dell'Universit\`{a} and Gruppo Collegato INFN, Salerno, Italy
\item \Idef{org32}Dipartimento di Scienze e Innovazione Tecnologica dell'Universit\`{a} del  Piemonte Orientale and Gruppo Collegato INFN, Alessandria, Italy
\item \Idef{org33}Dipartimento Interateneo di Fisica `M.~Merlin' and Sezione INFN, Bari, Italy
\item \Idef{org34}Division of Experimental High Energy Physics, University of Lund, Lund, Sweden
\item \Idef{org35}Eberhard Karls Universit\"{a}t T\"{u}bingen, T\"{u}bingen, Germany
\item \Idef{org36}European Organization for Nuclear Research (CERN), Geneva, Switzerland
\item \Idef{org37}Excellence Cluster Universe, Technische Universit\"{a}t M\"{u}nchen, Munich, Germany
\item \Idef{org38}Faculty of Engineering, Bergen University College, Bergen, Norway
\item \Idef{org39}Faculty of Mathematics, Physics and Informatics, Comenius University, Bratislava, Slovakia
\item \Idef{org40}Faculty of Nuclear Sciences and Physical Engineering, Czech Technical University in Prague, Prague, Czech Republic
\item \Idef{org41}Faculty of Science, P.J.~\v{S}af\'{a}rik University, Ko\v{s}ice, Slovakia
\item \Idef{org42}Faculty of Technology, Buskerud and Vestfold University College, Vestfold, Norway
\item \Idef{org43}Frankfurt Institute for Advanced Studies, Johann Wolfgang Goethe-Universit\"{a}t Frankfurt, Frankfurt, Germany
\item \Idef{org44}Gangneung-Wonju National University, Gangneung, South Korea
\item \Idef{org45}Gauhati University, Department of Physics, Guwahati, India
\item \Idef{org46}Helsinki Institute of Physics (HIP), Helsinki, Finland
\item \Idef{org47}Hiroshima University, Hiroshima, Japan
\item \Idef{org48}Indian Institute of Technology Bombay (IIT), Mumbai, India
\item \Idef{org49}Indian Institute of Technology Indore, Indore (IITI), India
\item \Idef{org50}Inha University, Incheon, South Korea
\item \Idef{org51}Institut de Physique Nucl\'eaire d'Orsay (IPNO), Universit\'e Paris-Sud, CNRS-IN2P3, Orsay, France
\item \Idef{org52}Institut f\"{u}r Informatik, Johann Wolfgang Goethe-Universit\"{a}t Frankfurt, Frankfurt, Germany
\item \Idef{org53}Institut f\"{u}r Kernphysik, Johann Wolfgang Goethe-Universit\"{a}t Frankfurt, Frankfurt, Germany
\item \Idef{org54}Institut f\"{u}r Kernphysik, Westf\"{a}lische Wilhelms-Universit\"{a}t M\"{u}nster, M\"{u}nster, Germany
\item \Idef{org55}Institut Pluridisciplinaire Hubert Curien (IPHC), Universit\'{e} de Strasbourg, CNRS-IN2P3, Strasbourg, France
\item \Idef{org56}Institute for Nuclear Research, Academy of Sciences, Moscow, Russia
\item \Idef{org57}Institute for Subatomic Physics of Utrecht University, Utrecht, Netherlands
\item \Idef{org58}Institute for Theoretical and Experimental Physics, Moscow, Russia
\item \Idef{org59}Institute of Experimental Physics, Slovak Academy of Sciences, Ko\v{s}ice, Slovakia
\item \Idef{org60}Institute of Physics, Academy of Sciences of the Czech Republic, Prague, Czech Republic
\item \Idef{org61}Institute of Physics, Bhubaneswar, India
\item \Idef{org62}Institute of Space Science (ISS), Bucharest, Romania
\item \Idef{org63}Instituto de Ciencias Nucleares, Universidad Nacional Aut\'{o}noma de M\'{e}xico, Mexico City, Mexico
\item \Idef{org64}Instituto de F\'{\i}sica, Universidad Nacional Aut\'{o}noma de M\'{e}xico, Mexico City, Mexico
\item \Idef{org65}iThemba LABS, National Research Foundation, Somerset West, South Africa
\item \Idef{org66}Joint Institute for Nuclear Research (JINR), Dubna, Russia
\item \Idef{org67}Konkuk University, Seoul, South Korea
\item \Idef{org68}Korea Institute of Science and Technology Information, Daejeon, South Korea
\item \Idef{org69}KTO Karatay University, Konya, Turkey
\item \Idef{org70}Laboratoire de Physique Corpusculaire (LPC), Clermont Universit\'{e}, Universit\'{e} Blaise Pascal, CNRS--IN2P3, Clermont-Ferrand, France
\item \Idef{org71}Laboratoire de Physique Subatomique et de Cosmologie, Universit\'{e} Grenoble-Alpes, CNRS-IN2P3, Grenoble, France
\item \Idef{org72}Laboratori Nazionali di Frascati, INFN, Frascati, Italy
\item \Idef{org73}Laboratori Nazionali di Legnaro, INFN, Legnaro, Italy
\item \Idef{org74}Lawrence Berkeley National Laboratory, Berkeley, California, United States
\item \Idef{org75}Lawrence Livermore National Laboratory, Livermore, California, United States
\item \Idef{org76}Moscow Engineering Physics Institute, Moscow, Russia
\item \Idef{org77}National Centre for Nuclear Studies, Warsaw, Poland
\item \Idef{org78}National Institute for Physics and Nuclear Engineering, Bucharest, Romania
\item \Idef{org79}National Institute of Science Education and Research, Bhubaneswar, India
\item \Idef{org80}Niels Bohr Institute, University of Copenhagen, Copenhagen, Denmark
\item \Idef{org81}Nikhef, Nationaal instituut voor subatomaire fysica, Amsterdam, Netherlands
\item \Idef{org82}Nuclear Physics Group, STFC Daresbury Laboratory, Daresbury, United Kingdom
\item \Idef{org83}Nuclear Physics Institute, Academy of Sciences of the Czech Republic, \v{R}e\v{z} u Prahy, Czech Republic
\item \Idef{org84}Oak Ridge National Laboratory, Oak Ridge, Tennessee, United States
\item \Idef{org85}Petersburg Nuclear Physics Institute, Gatchina, Russia
\item \Idef{org86}Physics Department, Creighton University, Omaha, Nebraska, United States
\item \Idef{org87}Physics Department, Panjab University, Chandigarh, India
\item \Idef{org88}Physics Department, University of Athens, Athens, Greece
\item \Idef{org89}Physics Department, University of Cape Town, Cape Town, South Africa
\item \Idef{org90}Physics Department, University of Jammu, Jammu, India
\item \Idef{org91}Physics Department, University of Rajasthan, Jaipur, India
\item \Idef{org92}Physik Department, Technische Universit\"{a}t M\"{u}nchen, Munich, Germany
\item \Idef{org93}Physikalisches Institut, Ruprecht-Karls-Universit\"{a}t Heidelberg, Heidelberg, Germany
\item \Idef{org94}Politecnico di Torino, Turin, Italy
\item \Idef{org95}Purdue University, West Lafayette, Indiana, United States
\item \Idef{org96}Pusan National University, Pusan, South Korea
\item \Idef{org97}Research Division and ExtreMe Matter Institute EMMI, GSI Helmholtzzentrum f\"ur Schwerionenforschung, Darmstadt, Germany
\item \Idef{org98}Rudjer Bo\v{s}kovi\'{c} Institute, Zagreb, Croatia
\item \Idef{org99}Russian Federal Nuclear Center (VNIIEF), Sarov, Russia
\item \Idef{org100}Russian Research Centre Kurchatov Institute, Moscow, Russia
\item \Idef{org101}Saha Institute of Nuclear Physics, Kolkata, India
\item \Idef{org102}School of Physics and Astronomy, University of Birmingham, Birmingham, United Kingdom
\item \Idef{org103}Secci\'{o}n F\'{\i}sica, Departamento de Ciencias, Pontificia Universidad Cat\'{o}lica del Per\'{u}, Lima, Peru
\item \Idef{org104}Sezione INFN, Bari, Italy
\item \Idef{org105}Sezione INFN, Bologna, Italy
\item \Idef{org106}Sezione INFN, Cagliari, Italy
\item \Idef{org107}Sezione INFN, Catania, Italy
\item \Idef{org108}Sezione INFN, Padova, Italy
\item \Idef{org109}Sezione INFN, Rome, Italy
\item \Idef{org110}Sezione INFN, Trieste, Italy
\item \Idef{org111}Sezione INFN, Turin, Italy
\item \Idef{org112}SSC IHEP of NRC Kurchatov institute, Protvino, Russia
\item \Idef{org113}SUBATECH, Ecole des Mines de Nantes, Universit\'{e} de Nantes, CNRS-IN2P3, Nantes, France
\item \Idef{org114}Suranaree University of Technology, Nakhon Ratchasima, Thailand
\item \Idef{org115}Technical University of Ko\v{s}ice, Ko\v{s}ice, Slovakia
\item \Idef{org116}Technical University of Split FESB, Split, Croatia
\item \Idef{org117}The Henryk Niewodniczanski Institute of Nuclear Physics, Polish Academy of Sciences, Cracow, Poland
\item \Idef{org118}The University of Texas at Austin, Physics Department, Austin, Texas, USA
\item \Idef{org119}Universidad Aut\'{o}noma de Sinaloa, Culiac\'{a}n, Mexico
\item \Idef{org120}Universidade de S\~{a}o Paulo (USP), S\~{a}o Paulo, Brazil
\item \Idef{org121}Universidade Estadual de Campinas (UNICAMP), Campinas, Brazil
\item \Idef{org122}University of Houston, Houston, Texas, United States
\item \Idef{org123}University of Jyv\"{a}skyl\"{a}, Jyv\"{a}skyl\"{a}, Finland
\item \Idef{org124}University of Liverpool, Liverpool, United Kingdom
\item \Idef{org125}University of Tennessee, Knoxville, Tennessee, United States
\item \Idef{org126}University of the Witwatersrand, Johannesburg, South Africa
\item \Idef{org127}University of Tokyo, Tokyo, Japan
\item \Idef{org128}University of Tsukuba, Tsukuba, Japan
\item \Idef{org129}University of Zagreb, Zagreb, Croatia
\item \Idef{org130}Universit\'{e} de Lyon, Universit\'{e} Lyon 1, CNRS/IN2P3, IPN-Lyon, Villeurbanne, France
\item \Idef{org131}V.~Fock Institute for Physics, St. Petersburg State University, St. Petersburg, Russia
\item \Idef{org132}Variable Energy Cyclotron Centre, Kolkata, India
\item \Idef{org133}Vin\v{c}a Institute of Nuclear Sciences, Belgrade, Serbia
\item \Idef{org134}Warsaw University of Technology, Warsaw, Poland
\item \Idef{org135}Wayne State University, Detroit, Michigan, United States
\item \Idef{org136}Wigner Research Centre for Physics, Hungarian Academy of Sciences, Budapest, Hungary
\item \Idef{org137}Yale University, New Haven, Connecticut, United States
\item \Idef{org138}Yonsei University, Seoul, South Korea
\item \Idef{org139}Zentrum f\"{u}r Technologietransfer und Telekommunikation (ZTT), Fachhochschule Worms, Worms, Germany
\end{Authlist}
\endgroup

\end{document}